\documentclass[11pt,a4paper]{article}

\usepackage[T1]{fontenc}
\usepackage[latin1,utf8]{inputenc}

\pdfoutput=1

\usepackage{amsmath,amssymb,amsfonts}
\usepackage[normal,font=small,labelfont=bf,labelsep=period]{caption}
\usepackage[pdftex]{color,graphicx}
\usepackage[english]{babel}
\usepackage[compress]{cite}
\usepackage{multirow}
\usepackage{tikz}
\usepackage{tikz-cd}
\usepackage{graphicx}

\usepackage{afterpage}

\usepackage[multiple]{footmisc}

\usepackage{slashed}

\usepackage[linktoc=all,colorlinks=true,linkcolor=black,urlcolor=magenta,citecolor=black,hyperfootnotes=false]{hyperref}
\hypersetup{%
  colorlinks = true,
  linkcolor  = black
}

\newcommand{\eq}[1]{(\ref{#1})}

\newcommand*\diff{\mathop{}\!\mathrm{d}}

\newcommand{\beq}{\begin{equation}}
\newcommand{\eeq}{\end{equation}}
\newcommand{\bp}{\boldsymbol{p}}
\newcommand{\bx}{\boldsymbol{x}}
\newcommand{\bk}{\boldsymbol{k}}
\newcommand{\bn}{\boldsymbol{n}}
\def\bea{\begin{eqnarray}}
\def\eea{\end{eqnarray}}
\def\be{\begin{equation}}
\def\ee{\end{equation}}

\newcommand{\tncdm}{T_{\star}}
\newcommand{\hatmncdm}{m_\text{DM}}
\newcommand{\mncdm}{m_\text{DM}}
\newcommand{\ncdm}{}
\newcommand{\wdm}{\text{WDM}}
\newcommand{\Ydec}{\mathcal{R}}

\definecolor{gbcolor}{rgb}{.9,.22,.12} 
\definecolor{gbcolor2}{rgb}{.7,.1,.6} 
\definecolor{gbcolor3}{rgb}{.05,.2,.9}

\makeatletter
\newcommand{\Cen}[2]{%
  \ifmeasuring@
    #2%
  \else
    \makebox[\ifcase\expandafter #1\maxcolumn@widths\fi]{$\displaystyle#2$}%
  \fi
}
\makeatother

\numberwithin{equation}{section}

\usepackage[symbols,nogroupskip,numberedsection,
sort=use]{glossaries-extra}

\makenoidxglossaries

\glsxtrnewsymbol[description={\quad Comoving momentum: \quad}]{commov}{\ensuremath{q\,T_\star}}

\glsxtrnewsymbol[description={\quad Metric perturbations in synchronous gauge: \quad}]{handeta}{\ensuremath{h} and \ensuremath{\eta}}

\glsxtrnewsymbol[description={\quad NCDM ``temperature'': \quad }]{ftemp}{\ensuremath{ T_\star}}

\glsxtrnewsymbol[description={\quad Phase space distribution: \quad }]{fased}{\ensuremath{f}}

\glsxtrnewsymbol[description={\quad Equation of state: \quad }]{eqstate}{\ensuremath{w}}

\glsxtrnewsymbol[description={\quad Phase space distribution perturbation: \quad }]{psif}{\ensuremath{\Psi}}

\glsxtrnewsymbol[description={\quad Transfer function: \quad}]{trans}{\ensuremath{\mathcal{T}}}

\glsxtrnewsymbol[description={\quad WDM  mass: \quad }]{wdmass}{\ensuremath{ m_{\rm WDM}}}

\glsxtrnewsymbol[description={\quad NCDM mass: \quad}]{ncdmm}{\ensuremath{m_{\rm DM}}}

\glsxtrnewsymbol[description={Momentum (modulus)}]{p}{\ensuremath{p}}

\glsxtrnewsymbol[description={Momentum}]{4}{\ensuremath{\mathbf{p}}}

\glsxtrnewsymbol[description={Energy (of a particle)}]{5}{\ensuremath{p_0}}

\glsxtrnewsymbol[description={\quad  Relative density fluctuation: \quad }]{deltar}{\ensuremath{\delta}}

\glsxtrnewsymbol[description={\quad Pressure: \quad }]{Pp}{\ensuremath{P}}

\glsxtrnewsymbol[description={\quad  Energy density: \quad }]{rhop}{\ensuremath{\rho}}

\glsxtrnewsymbol[description={\quad  Velocity divergence: \quad }]{thetap}{\ensuremath{\theta}}

\glsxtrnewsymbol[description={\quad Anisotropic stress: \quad }]{sigmap}{\ensuremath{\sigma}}

\glsxtrnewsymbol[description={\quad Scale factor (FLRW):\quad }]{ap}{\ensuremath{a}}

\glsxtrnewsymbol[description={\quad Hubble parameter (function): \quad}]{Hp}{\ensuremath{H}}

\glsxtrnewsymbol[description={\quad Conformal Hubble parameter (function): \quad}]{mathcalH}{\ensuremath{\mathcal{H}}}

\glsxtrnewsymbol[description={\quad Comoving ``energy'':\quad }]{comene}{\ensuremath{\epsilon}}

\glsxtrnewsymbol[description={Conformal time'}]{13}{\ensuremath{\tau}}

\glsxtrnewsymbol[description={Cosmic time}]{14}{\ensuremath{t}}

\glsxtrnewsymbol[description={\quad Phase space distribution parametrization: \quad}]{fasedp}{\ensuremath{\alpha, \beta, \gamma}}

\glsxtrnewsymbol[description={\quad Reduced Planck mass: \quad}]{mplanck}{\ensuremath{M_P}}

\glsxtrnewsymbol[description={\quad {Collision term:} \quad}]{collisionterm}{\ensuremath{\mathcal{C}}}

\glsxtrnewsymbol[description={\quad {Number density of the species $\chi$:} \quad}]{nchi}{\ensuremath{n_\chi}}

\glsxtrnewsymbol[description={\quad {Normalization scale of the $2 \rightarrow 2$ cross section:} \quad}]{Lambda}{\ensuremath{\Lambda}}

\glsxtrnewsymbol[description={\quad {Rest-frame sound speed:} \quad}]{cs}{\ensuremath{\hat c_s}}

\glsxtrnewsymbol[description={\quad {Adiabatic sound speed:} \quad}]{ca}{\ensuremath{c_a}}

\glsxtrnewsymbol[description={\quad {Free-streaming scale:} \quad}]{kFS}{\ensuremath{k_\text{FS}}}

\glsxtrnewsymbol[description={\quad {Free-streaming horizon:} \quad}]{kH}{\ensuremath{k_\text{H}}}

\glsxtrnewsymbol[description={\quad {Reheating temperature:} \quad}]{Treh}{\ensuremath{T_{\rm reh}}}

\glsxtrnewsymbol[description={\quad {Highest temperature during reheating:} \quad}]{Tmax}{\ensuremath{T_{\rm max}}}

\glsxtrnewsymbol[description={\quad {Internal degrees of freedom of the species $\chi$:} \quad}]{gchi}{\ensuremath{g_\chi}}

\glsxtrnewsymbol[description={\quad {Reheating temperature parametrization:} \quad}]{b}{\ensuremath{b}}

\glsxtrnewsymbol[description={\quad {Effective number of relativistic species:} \quad}]{Neff}{\ensuremath{N_{\rm eff} }}

\glsxtrnewsymbol[description={\quad {Number of effective entropy degrees of freedom:} \quad}]{gstars}{\ensuremath{g_{\rm *s} }}

\glsxtrnewsymbol[description={\quad {Inflaton scalar field:} \quad}]{inflaton}{\ensuremath{\phi}}

\glsxtrnewsymbol[description={\quad {Inflaton decay width:} \quad}]{Gammaphi}{\ensuremath{\Gamma_{\phi}}}

\usepackage{glossary-mcols}

\usepackage{censor}

\begin{document}

\begin{flushright}
\footnotesize
{IFT-UAM/CSIC-20-135}

\end{flushright}
\vspace{1cm}

\begin{center}
{\LARGE\color{black}\bf How warm are non-thermal relics?\\ Lyman-$\alpha$ bounds on out-of-equilibrium dark matter \\[1mm] }

\medskip
\bigskip\color{black}\vspace{0.6cm}

{\large\bf Guillermo Ballesteros, Marcos A.\ G.\ Garcia, Mathias Pierre
}
\\[7mm]

{\it Instituto de F\'isica Te\'orica UAM/CSIC,\\
Calle Nicol\'as Cabrera 13-15, Cantoblanco E-28049 Madrid, Spain}\\
\vspace{0.2cm}
{\it Departamento de F\'isica Te\'orica, Universidad Aut\'onoma de Madrid (UAM)\\ Campus de Cantoblanco, 28049 Madrid, Spain}\\

\end{center}

\vspace{1cm}

\centerline{\large\bf Abstract}
\begin{quote}
\large 

We investigate the power spectrum of Non-Cold Dark Matter (NCDM) produced in a state out of thermal equilibrium. We consider dark matter production from the decay of scalar condensates (inflaton, moduli), the decay of thermalized and non-thermalized particles, and from thermal and non-thermal freeze-in. For each case, we compute the NCDM phase space distribution and the linear matter power spectrum, which features a cutoff analogous to that for Warm Dark Matter (WDM). This scale is solely determined by the equation of state of NCDM. We propose a mapping procedure that translates the WDM Lyman-$\alpha$ mass bound to NCDM scenarios. This procedure does not require expensive ad hoc numerical computations of the non-linear matter power spectrum. By applying it, we obtain bounds on several NCDM possibilities, ranging from $m_{\rm DM}\gtrsim {\rm EeV}$ for DM production from inflaton decay with a low reheating temperature, to sub-keV values for non-thermal freeze-in. We discuss the phenomenological implications of these results for specific examples which include strongly-stabilized and non-stabilized supersymmetric moduli, gravitino production from inflaton decay, $Z'$ and spin-2 mediated freeze-in, and non-supersymmetric spin-3/2 DM.

\end{quote}

\begin{center} 

\vfill\flushleft
\noindent\rule{6cm}{0.4pt}\\
{\small  \tt guillermo.ballesteros@uam.es, marcosa.garcia@uam.es, mathias.pierre@uam.es}

\end{center}

\newpage

\tableofcontents

\section{Introduction and results}
\label{sec:intro}

\subsection{Motivation}

After a few decades of remarkable improvement, dark matter (DM) direct detection experiments have reached a sensitivity on the nucleon-DM scattering cross section around 
$10^{-46}~\text{cm}^2$ for DM masses of the order of the electroweak scale~\cite{Aprile:2018dbl}. The absence
of any confirmed experimental signal (also in indirect detection and colliders) strongly constrains the viable parameter space of Weakly Interacting Massive Particle (WIMP) models of DM based on the vanilla freeze-out mechanism. This calls for a  reassessment of the attractiveness of this framework in the simplest models~\cite{Escudero:2016gzx,Arcadi:2017kky,Arcadi:2019lka}. In this context, exploring theoretically and experimentally other scenarios \cite{Bernal:2017kxu} that can achieve the correct DM abundance is necessary. A well-known example of such a scenario is the freeze-in mechanism~\cite{Hall:2009bx}. Other examples are a dark sector that thermalizes only with itself~\cite{Chu:2011be,Hambye:2019dwd} and a DM depletion process lead by cannibalization \cite{Kuflik:2015isi,Kuflik:2017iqs,Hochberg:2014dra}. These proposals assume feeble couplings between the SM and the DM,\footnote{A large energy scale, even larger than the reheating temperature, can be invoked to justify suppressed SM-DM interactions. Just to give some examples, this scale can be identified with the Planck mass in gravitino DM~\cite{ehnos,gravitino1,gravitino2,gravitino3,gravitino7,gravitino8,gravitino12}, with the mass of heavy gauge fields in Grand Unified Theories~\cite{Mambrini:2013iaa,mnoqz1,Bhattacharyya:2018evo} and with a new physics threshold in scenarios inspired by modified gravity~\cite{Garny:2017kha,Bernal:2018qlk,Anastasopoulos:2020gbu,Brax:2020gqg}.} helping them to satisfy current bouds. Consequently, this tends to reduce the chances of testing them by traditional means \cite{Bernal:2017kxu}. However, several phenomenological studies \cite{Chu:2011be,Hochberg:2015vrg,Choi:2017zww,Bernal:2015bla,Hochberg:2018rjs,Choi:2016hid,Choi:2019zeb} have highlighted various possibilities for observing such DM candidates.

Scenarios in which the DM is produced by non-standard mechanisms may feature an important DM self-interaction cross section or a free-streaming scale, affecting the large scale structure of the universe. These properties could allow to alleviate purported tensions in the $\Lambda$CDM model at  galactic and sub-galactic scales~\cite{de_Blok_2010,Viel:2013fqw,Bullock:2017xww}. Also, non-thermal DM models have been proposed to address  the tensions between early and late time determinations of the Hubble constant \cite{Blinov:2020uvz} and of the clustering of matter \cite{Heimersheim:2020aoc,Abellan:2020pmw}.

Indeed, in absence of thermodynamic equilibrium between the DM and the SM, the DM phase space distribution can differ significantly from the standard freeze-out case. This opens a possibility for discriminating between different DM models and production mechanisms. The DM component in the standard $\Lambda$CDM model of cosmology is assumed to be entirely pressureless. A non-vanishing DM kinetic energy
would then result in a cutoff  in the matter power spectrum on small wavelength Fourier modes (as compared to $\Lambda$CDM  prediction). 

An interesting possibility for testing these Non-Cold Dark Matter (NCDM) models --which do not conform to the standard freeze-out mechanism-- is to measure the Ly-$\alpha$ forest of absorption lines of light emitted by distant quasars around redshift $z=2-4$, which is produced due to the neutral hydrogen present in the intergalactic medium. This provides enough information on the matter power spectrum at sufficiently small scales for probing the aforementioned cutoff. 

\subsection{Ly-$\alpha$ constraints on out-of-equilibrium dark matter}

The well known Ly-$\alpha$ bound on the DM mass for Warm Dark Matter (WDM)~\cite{Narayanan:2000tp,Viel:2005qj,Viel:2013fqw,Baur:2015jsy,Irsic:2017ixq,Palanque-Delabrouille:2019iyz,Garzilli:2019qki},
\beq \label{bound1}
\gls{wdmass}\;\gtrsim\; (1.9-5.3)~\text{keV at 95\% C.L.} \,,
\eeq
can be mapped into constraints on various out-of-equilibrium NCDM production mechanisms. To do this, we compute (for the first time) the phase space distributions in several of these models by integrating the Boltzmann transport equation, numerically and/or analytically, depending on the production process. For the large majority of the scenarios that we consider, the resulting phase space distributions can be remarkably well described by a generalized distribution of the form 
\begin{equation}\label{eq:allfita}
f(q)  \, \propto \, q^\alpha\,\exp{ \left(-\beta \, q^\gamma \right)}\,, 
\end{equation}
where $q$ denotes the DM comoving momentum and \gls{fasedp} are model-dependent constants.  
We then use \texttt{CLASS}~\cite{Blas:2011rf,Lesgourgues:2011rh} to compute,  for each of the NCDM models we consider, the linear power spectrum $\mathcal{P}_{\text{NCDM}}(k)$, or, more precisely, the linear transfer function, defined in terms of the ratio to the $\Lambda$CDM spectrum as follows, 
\begin{equation} \label{transfunct}
    \gls{trans}(k)\, \equiv \, \left(\dfrac{\mathcal{P}_{\text{NCDM}}(k)}{\mathcal{P}_{\Lambda\text{CDM}}(k)}\right)^{1/2}\,.
\end{equation}
We assume that the DM is entirely composed of a single NCDM species (and is produced by only one mechanism in each scenario). By varying the DM mass, we match the transfer function to the one of a fermionic WDM scenario (for which the bound  \eq{bound1} applies). We perform this matching numerically and, also, with an approximate semi-analytical procedure, demonstrating their equivalence. We find that the matching can be done with great accuracy for all the models we consider (and for all the relevant ranges of their parameters).  In addition, by approximating the NCDM species as a perfect fluid, we show that the cutoff in the transfer function can be entirely characterized in terms of the equation of state parameter of NCDM, $w$, which allows to translate the WDM mass Ly-$\alpha$ limit  \eq{bound1} to the NCDM case. This can be done for each of the NCDM models, without having to run specifically tailored N-body simulations or doing a dedicated non-linear analysis of the NCDM perturbations at small scales. A general analytical expression relates these bounds to each other via only the knowledge of the first and second moments of the phase space distribution, 
\begin{equation}\label{eq:somelabel}
  \gls{ncdmm} \, = \, m_\text{WDM} \left( \dfrac{\tncdm}{T_{\text{WDM},0}} \right) \sqrt{\dfrac{\langle q^2 \rangle}{\langle q^2 \rangle_\text{WDM}}}\quad \quad \text{(Lyman-$\alpha$ bound)}\,,
\end{equation}
where $\gls{ftemp}  \propto \langle q\rangle$ is the present ``temperature'' of NCDM, understood as the energy scale that normalizes the typical momentum of the distribution. In all cases, we find a remarkable agreement between this approximation and the numerical computation of the linear power spectrum. Using this procedure, we achieve (for most of our scenarios and in the range of scales of interest) a $\lesssim 3\%$ error in the matching to the transfer function of WDM, see Figure \ref{fig:transfer_function_residual}. In the (very few) least precise examples that we consider the matching worsens to at most $10 \%$.

\begin{figure*}[t!]
\begin{center}
\includegraphics[width=0.9\textwidth]{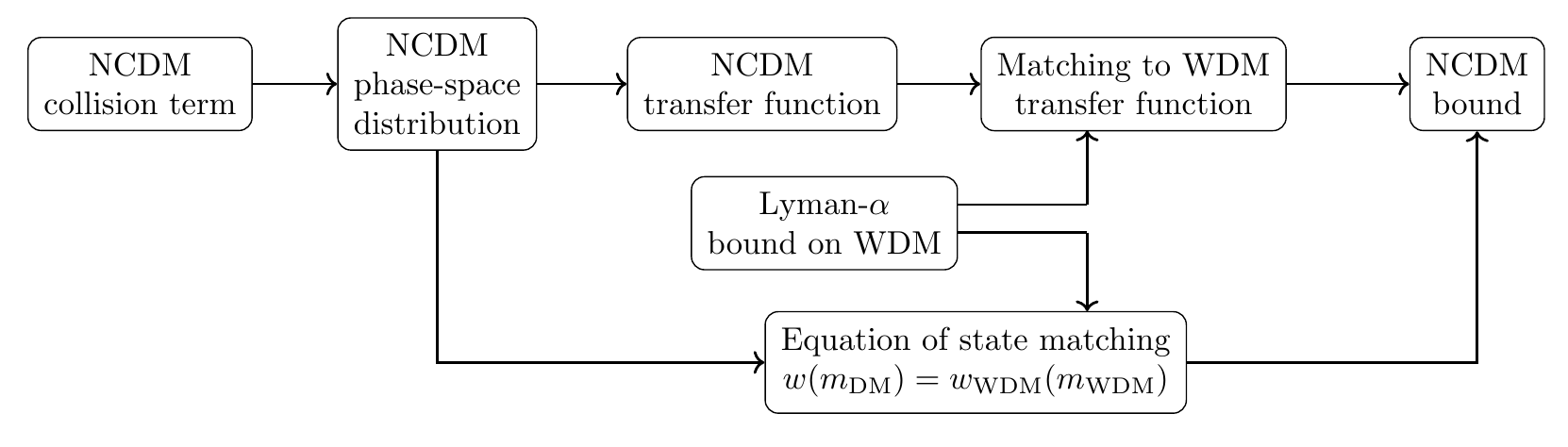}

\caption{Pipeline applied in this paper to derive bounds on a given NCDM model from the Lyman-$\alpha$ WDM mass limit. {The boxes represent the computational steps or inputs used to derive a bound on the NCDM mass.  Starting from the NCDM collision term and  Lyman-$\alpha$ bound on WDM, two possible paths allow to derive a bound on NCDM. Our matching procedure, going through the equation of state matching, allows us to obtain the NCDM bound without computing numerically the NCDM transfer function. 
}}
\label{fig:scheme}
\end{center}
 \end{figure*}

\begin{figure*}[t!]
\begin{center}

\includegraphics[width=0.99\textwidth]{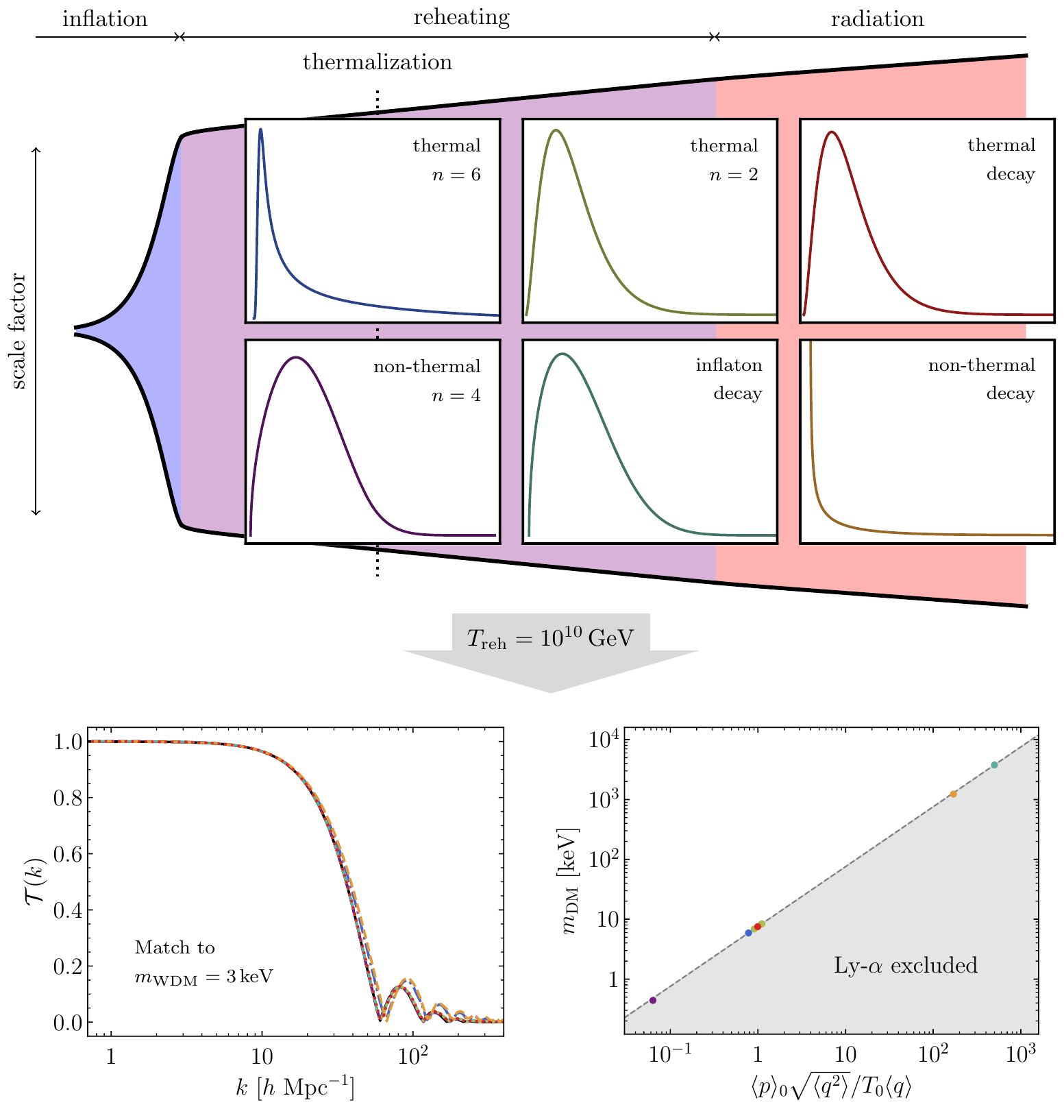}
\caption{ Out-of-equilibrium dark matter as a probe for early universe dynamics. The top figure depicts schematically the phase space distributions for the production scenarios considered in the present work, mapped over the history of the early universe. The relic distributions for the thermal freeze-in production with $n\geq 6$ and non-thermal freeze-in with $n>2$ are set at the thermalization time-scale, after inflation but well before the end of reheating. The relic abundance and the distribution of dark matter produced from inflaton decay or low-$n$ freeze-in are set at the matter-radiation transition at the end of reheating.  Dark matter can be also produced through decays of thermalized and non-thermalized fields during early radiation domination. Most distributions have the {approximate} form $f(q)\propto q^{\alpha}e^{-\beta q^{\gamma}}$. The bottom left figure shows that, in all cases, the transfer function of the linear power spectrum determines the lower bound on the dark matter mass, since $\mathcal{T}(k)$ can be matched to the corresponding WDM bound. The bottom right figure shows the range of masses that can be bound by Ly-$\alpha$ observations, depending on the dark matter production mechanism and the reheating temperature (here $T_{\rm reh}=10^{10}\,{\rm GeV}$).}
\label{fig:summary}
\end{center}
 \end{figure*}
This bound-mapping procedure  is summarized in Figs.\ \ref{fig:scheme} and \ref{fig:summary}. The top panel of the latter shows schematically the  different shapes of the distribution functions corresponding to six distinct NCDM production processes, active during or after reheating. We now proceed to enumerate these processes, and summarize our results for each of them.

\subsection{Dark matter production mechanisms}

We consider various non-equilibrium DM production mechanisms. They are all assumed to proceed perturbatively via the scattering or decay of particles, with time-scales
ranging from the very end of inflation, during the earliest stages of reheating, to the radiation dominated universe occurring after the end of reheating. We list these mechanisms below; see also Fig.~\ref{fig:summary}.\par\medskip

\noindent
{\bf Inflaton decay (Section~\ref{sec:perturbative_inflaton_decay}).} It is often assumed that the DM may
have been produced from the decay of the inflaton field, $\phi$. Even in the absence of tree-level inflaton-DM couplings, DM-SM interactions can generate a non-vanishing inflaton $\rightarrow$ DM decay channel at higher order in perturbation theory~\cite{Kaneta:2019zgw}. Assuming that this decay proceeds perturbatively through a two-body process, we find that the DM phase space distribution is of the form  $f(q) \propto q^{-3/2} e^{-0.74q^2}$, where the power-like behaviour at low $q$ arises from redshifting during the matter dominated reheating epoch, and the Gaussian tail comes from the depletion of the inflaton condensate at the end of reheating. The resulting Ly-$\alpha$ constraint on the DM mass is proportional to the ratio of the inflaton mass to the reheating temperature, being $m_{\rm DM}\gtrsim 3.8\,{\rm MeV}$ for $T_{\rm reh}=10^{10}\,{\rm GeV}$ and $m_{\phi}\simeq 3\times 10^{13}\,{\rm GeV}$, the later fixed by the measurement of the amplitude of the curvature power spectrum~\cite{Aghanim:2018eyx,Ellis:2015pla}. Higher reheating temperatures can reduce the limit down to the keV range, whereas lower reheating temperatures can increase it well beyond the TeV range.\par\medskip

\noindent
{\bf Moduli decay (Section \ref{sec:moduli_decay}).} In many SM extensions, in particular in supergravity and string constructions, there is a plethora of scalar fields with very weak couplings to the SM  (typically of gravitational strength) and masses that are typically of the order of the weak scale. These fields are known as moduli, and can have far reaching cosmological consequences if they are excited away from their vacuum values in the early Universe. We consider DM production from moduli decays in two scenarios: when the modulus dominates the energy of the Universe and decays at late times, and when the modulus is always subdominant to the inflaton/radiation background due to some stabilization mechanism. In the first case, the shape of the DM phase space distribution is identical to that for DM produced from inflaton decay, and the lower bound on $m_{\rm DM}$ is proportional to the ratio of the modulus mass ($m_Z$) to its reheating temperature, with $m_{\rm DM}\gtrsim 13\,{\rm GeV}$ for $m_Z=10\,{\rm TeV}$  and $T_{\rm reh}=1\,{\rm MeV}$. For the decay of stabilized moduli, we find non-thermal DM distributions of the form $f(q)\propto q^{-3/2}e^{-q^{3/2}}$ or $f(q)\propto q^{-1}e^{-q^2}$, depending on whether the modulus decays during or after reheating, respectively. In these cases, the limit on $m_{\rm DM}$ depends on the ratio of the modulus mass to the background temperature evaluated at the moment of its decay,  and on the ratio of the inflaton and modulus decay widths. \par\medskip

\noindent
{\bf Thermal and non-thermal decays (Section~\ref{sec:freezeindecay}).} DM could have been produced also from the decay of free particles. In such a case the DM phase space distribution and its present abundance depend strongly on the initial momentum distribution of the decaying particles. We consider here two possibilities: the decay of a thermalized particle species during radiation domination {\bf (Section \ref{sec:thermal_decay_PSD})}, and the decay of a particle with a non-equilibrium distribution, assumed to be produced from the decay of the inflaton {\bf (Section \ref{sec:nonthermaldecay})}. In both cases we assume that the decaying particle is much lighter than the inflaton, yet much heavier than DM. For the thermal decay case, we find that the DM  inherits a quasi-thermal distribution, $f(q)\sim q^{-1/2}e^{-q}$, and the bound on its mass is given by $m_{\rm DM}\gtrsim 7\,{\rm keV}$.

For the non-thermal decay, we find that the shape of the distribution is highly dependent on the momentum of the parent particle when it decays. If this initial state decays while it is relativistic,  the DM inherits the Gaussian tail of the parent unstable particle, $f(q)\sim q^{-5/2}e^{-0.74 q^2}$. The Ly-$\alpha$ constraint is identical to that for the direct decay of the inflaton to DM, reduced by a factor of $\sim 0.3$. If instead {the decaying particle is non-relativistic,} the DM phase space distribution is highly non-thermal, skewed towards large momenta, and not suitable for a fit of the form (\ref{eq:allfita}). The Ly-$\alpha$ constraint depends on the mass and width of the decaying particles; more specifically proportional to the ratio of the mass to the temperature $T_{\rm dec}\propto \Gamma^{1/2}$ at which the decay occurs.  \par\medskip

\noindent
{\bf Thermal freeze-in via scatterings (Section \ref{sec:thermalfreezein}).} We consider the possibility of a DM population generated via the freeze-in mechanism by annihilations of thermalized SM particles. We assume that the typical DM-SM scattering amplitude can be parametrized by
\beq\label{eq:amplitudesn}
|\mathcal{M}|^2 \;=\; 16\pi \frac{s^{\frac{n}{2}+1}}{\Lambda^{n+2}}\,,
\eeq
where $n$ is an integer, $\sqrt{s}$, the square root of the Mandelstam variable, is the center-of-mass energy (in the high-energy limit), and $\Lambda$ is some high-energy scale.\footnote{Small differences on the dependence on Maldestam variables in the high energy limit can be absorbed into the value of $\Lambda$.}

For $0 \leq n <6$, the DM is produced at the end of the reheating process, at the reheating temperature $T_{\rm reh}$. For $n=0,2$ the resulting DM momentum distribution is quasi-thermal with $\beta\sim 1$ and $\gamma=1$. Instead, for $n=4$ it has a nearly Gaussian tail. For these three scenarios, the matching of the power spectrum to WDM is excellent and the bound translates to $m_{\rm DM}\gtrsim 6-9\,{\rm keV}$, with the precise value depending on $n$ and the quantum statistics of the thermalized scatterers.

When $n\geq 6$, most of the DM is produced on the earliest stages of reheating, at the maximum temperature $T_{\rm max}$. When this is the case, the fitting expression (\ref{eq:allfita}) fails. In particular, for $n=6$ $f(q)$ interpolates between a $q^3$ behaviour at $q\ll 1$, and an exponential tail at $q\gg 1$, through a region where $f(q)\sim q^{-3}$. This relatively complicated form of the distribution translates to an imperfect match with the WDM power spectrum, which nevertheless leads to a bound of the form $m_{\rm DM}^2 \gtrsim 81\,{\rm keV^2}/ \ln (T_{\rm max}/T_{\rm reh})$.
 
\par\medskip

\noindent
{\bf Non-thermal freeze-in via scatterings (Section \ref{sec:thenonthfi}).} The delay between the end of inflation and the onset of thermal equilibrium in the primordial plasma can leave an imprint on the DM phase space distribution if the parent scatterers are produced directly from inflaton decays. Inflaton decay products are typically very energetic, with momenta of the order of the inflaton mass. Only after a process of soft radiation emission and energy transfer through scatterings these decay products 
reach thermal equilibrium. Thermalization occurs after the beginning of reheating, but well before it ends. As it turns out, if $n>2$ in (\ref{eq:amplitudesn}), most of the DM could have been produced non-thermally by the very first SM particles present in the universe~\cite{Garcia:2018wtq}. %

 As a proof of concept, we consider here annihilations with $n=4$. Notably, under the freeze-in assumption, the transport equation can be solved in a closed, albeit complicated, form. We find that the approximation $f(q)\sim q^{-3/2}e^{-2.5 q^{2.6}}$ adequately describes the DM phase space distribution. The power spectrum matching with WDM can be performed accurately, leading to a minimum DM mass of the form $m_{\rm DM} \propto m_{\phi}^{23/15} T_{\rm reh}^{-7/15}$, where $m_{\phi}$ is the inflaton mass, and $T_{\rm reh}$ the reheating temperature. For $T_{\rm reh}=10^{10}\,{\rm GeV}$, $m_{\rm DM}\gtrsim 0.4\,{\rm keV}$. \par\bigskip

Our paper is organized as follows. In Section~\ref{sec:NCDM} we review the treatment of NCDM relics in cosmological linear perturbation theory, and discuss the properties of the transfer function \eq{transfunct} and the re-scaling into NCDM of the Lyman-$\alpha$ bounds coming from WDM. In Sections \ref{sec:classdec} to \ref{sec:UVfreezein} we study the production mechanisms we just listed, their Ly-$\alpha$ bounds and the corresponding phenomenological implications. In Section \ref{sec:lightbntl} we discuss the implications for the effective number of relativistic species. We present our conclusions in Section~\ref{sec:conclusion}. 

Appendix~\ref{sec:computation_PSD} contains a brief review of the Boltzmann equation in the early universe, as well a detailed calculation of the generic form of the collision term for DM production via freeze-in (Appendix~\ref{app:figen}), and the integration of this collision term for the $n=4$ non-thermal freeze-in scenario (Appendix~\ref{app:nthfi}). A glossary of the main symbols used in this paper is provided in Appendix~\ref{symbols}. We use a natural system of units in which $k_B=\hbar=c=1$.

\section{Non-cold dark matter cosmology}
\label{sec:NCDM}

\subsection{Linear cosmological perturbation theory}

In the standard $\Lambda$CDM model of cosmology, the DM is assumed to be cold (CDM), i.e.\ presureless. Therefore, its equation of state parameter $w$ -- defined by the relation $\bar P=\gls{eqstate}\,\bar \rho$ where $\bar  \rho$ and $\bar  P$ are its (time-dependent) background energy density and pressure --  is exactly vanishing. However, DM particles produced in the early universe, of thermal or non-thermal origin, would  actually possess some momentum distribution with a non-vanishing averaged momentum $\langle p \rangle \neq 0$, which could manifests in a deviation from $w=0$ and, possibly, also through other moments of it. We will now discuss a set of approximations under which $w$ can be the sole function encoding the deviations from CDM, both at the level of the background dynamics and for linear perturbations analyses. In Sections \ref{sec:classdec}--\ref{sec:UVfreezein} we will study concrete examples of such DM creation processes.

The phase space distribution, $f$, of a general cosmological species is a function of position, momentum and (conformal) time, $\tau$, that characterizes its energy-momentum tensor. It is convenient to split it into a time-dependent homogeneous background part, $\bar{f}(|\bp|,\tau)$, plus a fluctuation quantified by a function $\gls{psif}\ll 1$, such that $f(\bx,\bp,\tau)=\bar{f}(|\bp|,\tau)[1+\Psi(\bx,\bp,\tau)]$; see e.g.\ \cite{Ma:1995ey}. The background energy density and pressure functions, $\bar  \rho$ and $\bar  P$, of a NCDM relic are then
\begin{equation}
    \bar \rho\ncdm=4\pi\left( \dfrac{\tncdm}{a} \right)^4 \int  q^2 \epsilon \bar{f}(q) \diff q \, ,  \qquad \bar P\ncdm= \dfrac{4\pi}{3}\left( \dfrac{\tncdm}{a} \right)^4 \int  q^2 \dfrac{q^2}{\epsilon} \bar{f}(q) \diff q \, ,
\end{equation}
where \gls{ap} is the scale factor of the Universe and  $\tncdm$ is a convenient energy scale that characterizes the DM density at the present time.
 
Following the conventions of~\cite{Lesgourgues:2011rh}\footnote{Our $\tncdm$ is a time-independent quantity denoted by $T_{\text{NCDM},0}$ in~\cite{Lesgourgues:2011rh}.} we define
\begin{equation}\label{eq:qdef}
    q=\dfrac{p\, a}{{\tncdm}} \quad \text{with}\,\quad \gls{comene} = \sqrt{q^2+\left( \dfrac{\mncdm\, a}{\tncdm} \right)^2}\,,
\end{equation}
where the product \gls{commov} is the comoving momentum and $ p=|\bp|$ is the (absolute value) of the momentum of individual NCDM particles. 
In Fourier space, the perturbation $\Psi$ of the NCDM phase space distribution can be expanded in Legendre polynomials $P_\ell$ as follows:
\begin{equation}
\Psi(\bk,\hat \bn,q,\tau)=\sum_{\ell=0}^\infty (-i)^\ell (2\ell +1)\Psi_\ell (\bk,q,\tau)P_\ell(\hat{\bk}\cdot\hat{\bn})    \, ,
\end{equation}
where $k$ is the comoving wavenumber of the perturbations in Fourier space, $ \bk = k\,\hat \bk$  and $p= \hat{\bn}\cdot \vec{\bp}$. 
The quantities defining the perturbed energy-momentum tensor are
\begin{align}
\begin{aligned}
      \delta \gls{rhop}\ncdm\,=\,&4 \pi \left( \dfrac{\tncdm}{a} \right)^4 \int  q^2 \epsilon \bar{f}(q)   \Psi_0 \,\diff q ,  \quad && \text{energy density fluctuation}\\
      \delta \gls{Pp}\ncdm\,=\,& \dfrac{4 \pi}{3}\left( \dfrac{\tncdm}{a} \right)^4 \int  q^2 \dfrac{q^2}{\epsilon} \bar{f}(q)   \Psi_0 \,  \diff q ,  \quad && \text{pressure (density) fluctuation}  \\
     (\bar \rho\ncdm +\bar P\ncdm )\gls{thetap}\ncdm\,=\,& 4 \pi k\left( \dfrac{\tncdm}{a} \right)^4 \int  q^3  \bar{f}(q)  \Psi_1 \, \diff q  , \quad && \text{velocity divergence}   \\
    (\bar \rho\ncdm +\bar P\ncdm )\gls{sigmap}\ncdm\,=\,& \dfrac{8 \pi k}{3}\left( \dfrac{\tncdm}{a} \right)^4 \int  q^2 \dfrac{q^2}{\epsilon} \bar{f}(q)  \Psi_2 \,\diff q  , \quad && \text{anisotropic stress.} 
\end{aligned}  
	\label{eq:ncdm_perturbed}
\end{align}
For decoupled NCDM, the phase space distribution satisfies the collisionless Boltzmann equation
\beq
\frac{\partial f}{\partial \tau} + \frac{\diff x^i}{\diff \tau}\frac{\partial f}{\partial x^i} + \frac{\diff q}{\diff\tau} \frac{\partial f}{\partial q} + \frac{\diff n_i}{\diff \tau}\frac{\partial f}{\partial n_i} \;=\;0\,,
\eeq
with $i=1,2,3$ and $\bn$ being a unitary 3-vector pointing in the direction of the momentum, as defined above. In the synchronous gauge, this equation leads to the following system for the quantities $\Psi_\ell$, 
    \begin{align}
    \begin{aligned}
        \dot \Psi_0 \, = \, & -\dfrac{qk}{\epsilon} \Psi_1 +\dfrac{1}{6} \dot h \dfrac{\diff \ln \bar{f}}{\diff \ln q} \, , \\
        \dot \Psi_1 \, = \, & \dfrac{qk}{3\epsilon}\Big( \Psi_0 -2 \Psi_2 \Big) \, ,  \\
        \dot \Psi_2 \, = \, & \dfrac{qk}{5\epsilon}\Big( 2 \Psi_1 -3 \Psi_3 \Big) - \Big( \dfrac{1}{15} \dot h +\dfrac{2}{5} \dot \eta \Big) \dfrac{\diff \ln \bar{f}}{\diff \ln q}\, ,   \\
        \dot \Psi_\ell \, = \, & \dfrac{qk}{(2 \ell +1)\epsilon}\Big( \ell \Psi_{\ell-1} -(\ell+1) \Psi_{\ell+1} \Big) \, , \quad [\ell \ge 3]
        \end{aligned}  
                \label{eq:ncdm-Boltzmann}
    \end{align}
where \gls{handeta} are the trace and traceless part of the metric perturbation \cite{Ma:1995ey}.

For a non-relativistic species, higher multipoles are typically suppressed by (positive) powers of $q/\epsilon\sim p/ \mncdm$, making any $\Psi_\ell $ with $\ell \geq 2$ much smaller than  $\Psi_0$ and  $\Psi_1$. In this case, the Boltzmann hierarchy can be truncated imposing $\Psi_\ell=0$ for $\ell> 1$,
as discussed in~\cite{Shoji:2010hm}, whose analysis shows the validity of this truncation. As argued also in~\cite{Kunz:2016yqy}, in this (non-relativistic) case $\Psi_0$ depends only mildly on the variable $q$; and the integrals in (\ref{eq:ncdm_perturbed}) are dominated by the low $q\ll \epsilon$ regime,\footnote{Notice that for heavy-tailed distributions, the later approximation is no longer valid and one cannot apply the analytical arguments presented in this section. For instance in the case where DM particles could have been produced from Primordial-Black-Hole evaporation~\cite{Lennon:2017tqq,Baldes:2020nuv}, where the distribution function behaves as $\bar{f}(q)\sim 1/q^5$ at large $q$. In that case the integral appearing in the background-pressure expression is always dominated by $q\gtrsim m_\text{DM}a/\tncdm$, and hence it cannot be restricted to $q\ll \epsilon$.}  so that we can identify $\delta P\ncdm/\delta \rho\ncdm \, \simeq \, \bar P\ncdm/  \bar \rho\ncdm = w\ncdm$. 

In this limit, the first two equations of \eq{eq:ncdm-Boltzmann}
 can be integrated over $q$, allowing us to describe the NCDM species with a coupled system of (continuity and Euler) equations:
\begin{align} \label{cont}
    \dot \delta \,& = \, -(1+w) \Big( \theta + \dfrac{\dot h}{2} \Big)-3 \mathcal{H} \left( \hat c_s^2-w \right) \delta+9 \mathcal{H}^2(1+w)\left( \hat c_s^2-c_a^2\right)\dfrac{\theta}{k^2} \, , \\
    \dot \theta \, &= \, - \mathcal{H}
    \left( 1 - 3  \hat c_s^2 \right) \theta+\dfrac{ \hat c_s^2}{1+w} k^2 \delta  \, , \label{Euler}
\end{align}
where $\gls{deltar}\,\equiv \, \delta \rho / \bar \rho$, 
 and $ \gls{mathcalH}=a\gls{Hp}$, where $H \equiv \dot a = \diff a/\diff \tau $. To first order in $w \ll 1$, the adiabatic sound speed is $  \gls{ca}^2 \equiv \dot{\bar P} / \dot{\bar \rho} \simeq 5w/3$. In addition, as shown in~\cite{Lesgourgues:2011rh}, for sufficiently non-relativistic species, the (rest frame) sound speed\footnote{See e.g.\ \cite{Ballesteros:2010ks} for its definition.} can be reasonably well approximated by the adiabatic sounds speed $  \gls{cs}^2\simeq c_a^2$.

Notice that by taking $w=0$ one recovers the usual CDM perturbation equation $\dot \delta =-1/2 \dot h$. In the NCDM domination era, from the perturbed Einstein equations, the trace of the metric fluctuation $h$ satisfies the equation
\begin{equation}
    \Ddot{h}+\mathcal{H}\dot h+3(1+3w)\mathcal{H}^2\delta\,=\,0 \, ,
    \label{eq:handdelta}
\end{equation}
allowing the system \eq{cont}--\eq{Euler} to be reduced to
\begin{equation}
       \Ddot \delta   +\mathcal{H}  \dot \delta -\frac{3}{2}\mathcal{H}^2\left(1-w\dfrac{10}{9} \frac{k^2}{\mathcal{H}^2} \right)\delta \, = \,0 \, .
       \label{eq:finaleqdeltaNCDM}
\end{equation}

In the limit where $w=0$ exactly, overdensities grow ``democratically'', i.e.~independently of $k$ (as in $\Lambda$CDM). However, for non-vanishing $w$, at a given time, there is a suppressed growth for modes larger than the free-streaming wavenumber $k>k_\text{FS}(a)$ with
\begin{equation}
    \gls{kFS}^{2}(a)\, = \, \dfrac{9}{10} \dfrac{\mathcal{H}^2}{w}\, = \, \dfrac{3}{2}\dfrac{\mathcal{H}^2}{c_g^2}\, .
    \label{eq:kcrit}
\end{equation}
Thus, a cutoff in the power spectrum can be observed at a given time for modes larger than the free-streaming horizon wavenumber $k_H(a)$, which can be expressed in term of $k_\text{FS}$ as~\cite{Heimersheim:2020aoc}
\begin{equation}
   \gls{kH}(a)\, \equiv \, \left[ \int_0^a \dfrac{1}{k_\text{FS}(\tilde{a})}  \dfrac{\diff \tilde{a}}{\tilde{a}} \right]^{-1} \, .
\end{equation}

From these equations, we see that $w$ is the only quantity (together with the current DM density) that controls in first approximation the suppression of the power spectrum at large $k$. In the non-relativistic limit, $w$ can be expressed in terms of the normalized second moment of the distribution function
\begin{equation}
 w\ncdm \,  \simeq  \dfrac{\delta P\ncdm}{\delta \rho\ncdm}\,= \, \dfrac{\tncdm^2}{3\mncdm^2}\frac{\langle q^2 \rangle\ncdm}{a^2}\, ,
    \label{eq:eos_ncdm}
\end{equation}
with the $n$-th moment being
\begin{equation}
     \langle q^n \rangle \, \equiv \, \dfrac{ \int  q^{n+2}  \bar{f}(q) \diff q }{ \int  q^2  \bar{f}(q) \diff q }\, .
     \label{eq:n-th_moment}
\end{equation}
As a result, given a phase space distribution for the DM, determination of its second moment is sufficient to estimate the cutoff of the matter power spectrum.

\subsection{Large scale structure} \label{keysec}
For a given NCDM cosmology, the cutoff can be described in terms of the transfer function $\mathcal{T}(k)$ defined as 
\begin{equation}
    \mathcal{T}(k)=\left(\dfrac{\mathcal{P}\ncdm(k)}{\mathcal{P}_{\Lambda\text{CDM}}(k)}\right)^{1/2}\, ,
    \label{eq:defTk}
\end{equation}
which compares (at the present time) the power spectrum for a given NCDM cosmology to the typical $\Lambda$CDM case. As we will now discuss, a small scale cutoff in the matter power spectrum 
may be one of the few possibilities at our disposal for distinguishing NCDM cosmologies from the paradigmatic $\Lambda$CDM model and thus probe the degree of DM ``warmness''. Light emitted by distant quasars and subsequently interacting with the neutral Hydrogen of the intergalactic medium around redshifts $z\sim2-6$ generates a pattern of absorption lines around $\sim 1000\, $\AA:  the Ly-$\alpha$ forest. This allows to probe the power spectrum on scales $k\sim (0.1-10)\, h\,\text{Mpc}^{-1}$ at the present time, by estimating the amount of  matter 
through a determination of  the Ly-$\alpha$ optical depth, thus providing one of the most stringent ways of testing NCDM models. 

Constraints from the Ly-$\alpha$ flux power spectrum on the DM properties are usually given as a lower bound on the 
WDM mass parameter, $m_\wdm$, used as a reference. Given $m_\wdm$, the WDM phase space is characterized by a single quantity: $T_\wdm$. In spite of our notation, this quantity, which decreases with time, is not a temperature, stricto sensu, since we assume the WDM species not to be in thermal equilibrium at recombination and later times. Such a DM candidate is assumed to have achieved a state of thermal equilibrium at some earlier time in the evolution of the Universe and would have subsequently decoupled later on, as it happens e.g.\  for neutrinos in the SM. Indeed, a good benchmark scenario for WDM, which we will assume from now on, is a fermionic DM candidate with two degrees of freedom having a Fermi-Dirac distribution. In this case the WDM relic density can be related to its mass and ``temperature'' $T_\wdm$ by
\begin{equation}
        \Omega_\wdm h^2\simeq \left( \dfrac{m_\wdm}{94~ \text{eV}} \right) \left( \dfrac{T_\wdm}{T_\nu} \right)^3\, ,
    \label{eq:densityWDM}
\end{equation}
where $T_\nu=(4/11)^{1/3}\,T
$ is the neutrino temperature as expected in the SM after $e^+e^-$ annihilations, assuming instantaneous decoupling, expressed as a function of the photon temperature $T$.  As usual, $h$ denotes the reduced Hubble constant, defined by the relation $H_0 \equiv 100 \, h \, \text{km}  \, \text{s}^{-1} \, \text{Mpc}^{-1} $.

Assuming that the WDM saturates
the DM density determined by Planck~\cite{Aghanim:2018eyx}, a numerical evaluation of the free-streaming horizon in Eq.~(\ref{eq:kcrit}) gives that the cutoff in the linear matter power spectrum occurs at
\begin{equation}
    k_H(a=1)\, \simeq \, 3.5 \, h\,\text{Mpc}^{-1} \, .
\end{equation}
for $m_\text{WDM}=1$ keV. As shown in~\cite{Bode:2000gq,Hansen:2001zv,Viel:2005qj}, an analytical fit for the transfer function in the WDM case is given by
\begin{equation} \label{fittinf}
    \mathcal{T}(k)=\left(1+(\alpha k)^{2 \nu}\right)^{-5/\nu} \, ,
\end{equation}
with the parameters
\begin{equation}
    \nu=1.12 \quad \text{and} \quad \alpha=0.24 \left( \dfrac{\gls{wdmass}}{1~\text{keV}} \dfrac{T_\wdm}{T_\nu}  \right)^{-0.83} \left( \dfrac{\Omega_\wdm h^2}{0.12} \right)^{-0.16} \text{Mpc}\,.
    \label{eq:Tkfit}
\end{equation} 	
Importantly, these fitting parameters are independent of the standard cosmological parameters (other than the DM abundance, $\Omega_\wdm$). The non-observation of a cutoff in actual data for the matter power spectrum can be translated into a constraint on the WDM mass. A recent analysis~\cite{Palanque-Delabrouille:2019iyz} gives a bound $m_\wdm>5.3$ keV at 95\% C.L., while the reference~\cite{Garzilli:2019qki} derived a less stringent bound $m_\wdm>1.9$ keV at 95\% C.L.,  by claiming a more conservative treatment of thermal history for the intergalactic medium. In the following we will take $m_\wdm>3$ keV as a reference but allow our results to be translated for a different value. 

The most up-to-date lower bounds on the WDM mass from Ly-$\alpha$ data \eq{bound1} have been obtained using the medium resolution X-shooter spectrographic observations of the intermediate redshift ($z: 3 - 4.2$) XQ-100 sample of quasars \cite{2016AA594A91L, Irsic:2017sop} and the higher-resolution, higher-redshift ($z: 4.2 - 5.4$) data from the HIRES/MIKE spectrographs \cite{bernstein2002,1994SPIE2198362V}. These data can be used in combination with probes of the matter power spectrum at smaller comoving scales ($k< (\rm{km/s})^{-1}$)  via Lyman-$\alpha$ data, in particular from the Baryon Oscillation Spectroscopic Survey (BOSS) of the Sloan Digital Sky Survey (SDSS-III) \cite{Palanque-Delabrouille:2013gaa,Chabanier:2018rga}.  For future prospects (including higher redshifts), potentially allowing an enhanced sensitivity to the cutoff of the spectrum see \cite{Witstok:2019sxp}.

In principle, it may seem reasonable to assume that in order to compare the expected matter power spectrum for a (more general, non Fermi-Dirac) NCDM cosmology to the WDM case, it should be essential to take into account the non-linear behaviour of the DM density field on the small cosmological distances ($1-100$~Mpc) probed by Ly-$\alpha$ data. Performing such a comparison requires costly N-body simulations for each possible NCDM case of interest. Nevertheless, the authors of Ref.~\cite{Murgia:2017lwo,Murgia:2018now} have performed a large set of N-body simulations of models featuring an ample variety of transfer functions, confronting the resulting power spectra to Ly-$\alpha$ data, and concluding that all the models that are ruled out can also be rejected by doing a simpler, linear analysis.

As we will show in the next sections, the shape of the linear power spectra of the various NCDM models we consider turns out to be very similar to the one for WDM, in spite of having, in some cases, notable differences at the level of the phase space. Therefore, we can translate directly the WDM Ly-$\alpha$ bounds by computing, numerically, the linear transfer functions for our NCDM models using a Boltzmann code, such as  \texttt{CLASS}~\cite{Blas:2011rf,Lesgourgues:2011rh}, and comparing the result with the linear transfer function in the WDM case.

The shape of the transfer function at the scales relevant for the change induced in the matter power spectrum by WDM free-streaming can also be probed by comparing the number of satellite galaxies of the Milky Way with N-body simulations \cite{Kennedy:2013uta, Newton:2020cog}. This method gives a bound on the WDM mass that is complementary and  comparable to those obtained from Ly-$\alpha$ data. The initial conditions for these N-body simulations were set in \cite{Kennedy:2013uta, Newton:2020cog} to mimic the (linear) transfer function \eq{fittinf}. Assuming that the formation of satellite galaxies only depends on the nature of the DM through the linear transfer function, we can also map these WDM mass bounds into constraints on NCDM models that feature different distribution functions, just as we do with Ly-$\alpha$ bounds.

\subsection{Analytical rescaling and generalized phase space distribution}
\label{sec:analytical_rescaling}
 
Let us now consider a NCDM model for which, by assumption, $w\ll 1$ is the only quantity needed to characterize the cutoff in the linear transfer function. Then, we can estimate the bound on $w$ from Ly-$\alpha$ by finding the value of $\hatmncdm$ such that
\begin{align}
w\ncdm(\hatmncdm)=w_\wdm(m_\wdm)\,.
\label{eq:equalitiesw}
\end{align}
The bound is obtained by assuming that the cutoff scale of the linear matter power spectrum for WDM can be translated to that of NCDM equating the equations of state. A correspondence between two NCDM scenarios (a sterile neutrino and a particle that decouples while being relativistic) was proposed for the first time (to our knowledge) in Ref.~\cite{Colombi:1995ze}. By equating the power spectra, the authors found a relation between these two scenarios, which possess distribution functions with the same analytical expression but with different parameters. A similar matching procedure using the mean square of the DM velocity was proposed in~\cite{Kamada:2013sh} and extended in~\cite{Bae:2019sby,Kamada:2019kpe} for several freeze-in models. 
As we show below, our (generalized) matching relation can be applied to a wide variety of NCDM scenarios, even for those in which thermal equilibrium is not established before DM decoupling.
From Eq.~(\ref{eq:eos_ncdm}) we can write $w_\wdm$ as
\begin{equation}
    w_\wdm(a)\, \simeq \, 6 \times 10^{-15} \,a^{-2} \, \left( \dfrac{\text{keV}}{m_\wdm}\right)^{8/3}\, ,
\end{equation}
implying that the bound on $m_\wdm\sim $ keV from Ly-$\alpha$~\cite{Viel:2005qj,Palanque-Delabrouille:2019iyz,Garzilli:2019qki} translates into
\begin{align} 
w_\wdm(a=1)\lesssim 10^{-15}\,,
\label{eq:constraintw}
\end{align}
showing that DM is indeed very cold. It is worth emphasizing that the constraint from Eq.~(\ref{eq:constraintw}) corresponds to a constraint at recombination of $w_\wdm(a \sim 10^{-3})\lesssim 10^{-9}$  whereas analyses based on CMB data constrain this value only at the level of $w_\wdm(a \sim 10^{-3})\lesssim 10^{-4}$  \cite{Kunz:2016yqy,Ilic:2020onu}.

\par
For instance, a typical WIMP with a mass $m_{\rm DM}=100~\text{GeV}$ that decoupled at a freeze-out temperature $T_F \simeq m_{\rm DM}/20$ inherits a Maxwell-Boltzmann distribution after decoupling of the form
\begin{equation}
    f(p,t)\,=\,\dfrac{g_{\rm DM}}{(2 \pi)^3} \exp  \left[\dfrac{-p^2 a(t)^2}{2 m_{\rm DM}a_F^2  T_F} \right] \,.
\end{equation}
In the non-relativistic limit, the energy and pressure densities can be evaluated analytically and $w$ can be expressed as
\begin{equation}
    w(a)\,\simeq \,\dfrac{a_F^2}{a^2}\dfrac{ T_F}{ m_{\rm DM}}\,\simeq\, 10^{-29} \left(\dfrac{1}{a^2}\right) \left( \dfrac{ 20\, T_F}{ m_{\rm DM}}\right) \left( \dfrac{100~\text{GeV}}{m_{\rm DM}} \right)^2 \left (\dfrac{100}{g_*^F}\right)^{2/3}\,,
\end{equation}
{where $g_*^F$ denotes the effective number of degrees of freedom.} %
This value for $w$ is several orders of magnitude lower than the typical value constrained by Ly-$\alpha$. For our NCDM case, Eq.~(\ref{eq:equalitiesw}) leads to
\begin{equation}
    \hatmncdm \, = \, m_\wdm \left( \dfrac{\tncdm}{T_{\wdm}} \right) \sqrt{\dfrac{\langle q^2 \rangle\ncdm}{\langle q^2 \rangle_\wdm}}  \simeq \, 7.56~\text{keV}  \,  \left( \dfrac{m_\wdm}{3~\text{keV}} \right)^{4/3} \left( \dfrac{\tncdm}{T} \right) \sqrt{\langle q^2 \rangle\ncdm}\,.
    \label{eq:mncdm_fromeos}
\end{equation}
Alternatively, the bound can be expressed in terms of the mean momentum at the present time, $\langle p \rangle_{\ncdm 0}= \langle q \rangle\ncdm \, \tncdm$ 
where $\langle q \rangle$ is defined in Eq.~(\ref{eq:n-th_moment}), giving
\begin{equation}
    \hatmncdm \, \simeq \, 7.56~\text{keV}  \,  \left( \dfrac{m_\wdm}{3~\text{keV}} \right)^{4/3} \left( \dfrac{\langle p \rangle_{0}}{T_{0}} \right) \dfrac{\sqrt{\langle q^2 \rangle\ncdm}}{\langle q \rangle\ncdm}\,.
   \label{eq:mncdm_fromeos_4} 
\end{equation}
As we will show, most of the NCDM cases discussed in this paper can be well described with a generalized phase space distribution of the form
\begin{equation}\label{eq:allfit}
f(q) \, \propto \, q^\alpha\,\exp{ \left(-\beta \, q^\gamma \right)}\,,
\end{equation}
with constant $\alpha>-3$ and $\beta,\gamma>0$ as required for the DM number density to be finite. For this distribution the normalized $n$-th moment (\ref{eq:n-th_moment}) is
\begin{equation}
\langle q^n \rangle \, = \, \beta^{\frac{2-n}{\gamma}} \, \dfrac{\Gamma \left(\frac{1+n+\alpha}{\gamma} \right)}{\Gamma \left(\frac{3+\alpha}{\gamma} \right)}\,.
\end{equation}
The rescaling of the mass reproducing the same cutoff as the WDM case then gives

\begin{equation}
    \hatmncdm \, \simeq \, 7.56~\text{keV}  \,  \left( \dfrac{m_\wdm}{3~\text{keV}} \right)^{4/3} \left( \dfrac{\langle p \rangle_{0}}{T_{0}} \right) \, \sqrt{\dfrac{\Gamma \left(\frac{3+\alpha}{\gamma} \right) \, \Gamma \left(\frac{5+\alpha}{\gamma} \right)}{\Gamma^2 \left(\frac{4+\alpha}{\gamma} \right)}}\,,
   \label{eq:mncdm_fromeos_4_GPSD} 
\end{equation}
which does not depend explicitly on $\beta$.\footnote{However, the mean momentum $\langle p \rangle_{0}$ depends actually on this quantity.} As  we will show later for the various examples we consider, this allows to translate any bound from Ly-$\alpha$ on $m_\wdm$ to a bound for a given NCDM model on $\mncdm$ provided that the DM phase space distribution can be well described by \eq{eq:allfit}, with good precision and  without requiring a numerical computation of the power spectrum.\footnote{Notably, in the case we consider for which (\ref{eq:allfit}) does not apply ($n=6$ thermal freeze-in, for example), we can still find an analytical form for the constraint using the more general form of Eq.~(\ref{eq:mncdm_fromeos}).}

Provided that the NCDM equation-of-state parameter evolves as $w(a) \simeq w_0 \,a^{-2}$ at redshift $z< 10^{6}$, for which the wavenumbers $k$ relevant for Lyman-$\alpha$ data enters the horizon, the NCDM power spectrum should exhibit the same features as the WDM power spectrum at first order in $w$. A different $a$-dependence of $w(a)$ would affect our matching procedure. For instance, in cannibalistic dark matter scenarios, the equation-of-state parameter behaves as $w(a)\propto 1/\log a$ when number-changing processes are active and $w(a)\propto a^{-2}$ in the non-relativistic regime at later times $z<10^3$\cite{Heimersheim:2020aoc}. In this case it has been shown that the NCDM power spectrum can be matched to a WDM one with a good precision\cite{Heimersheim:2020aoc}, by introducing a similar matching procedure up to a correction factor of order one\cite{Garny:2018byk}.

\begin{figure}[!t]
\centering
    \includegraphics[width=1\textwidth]{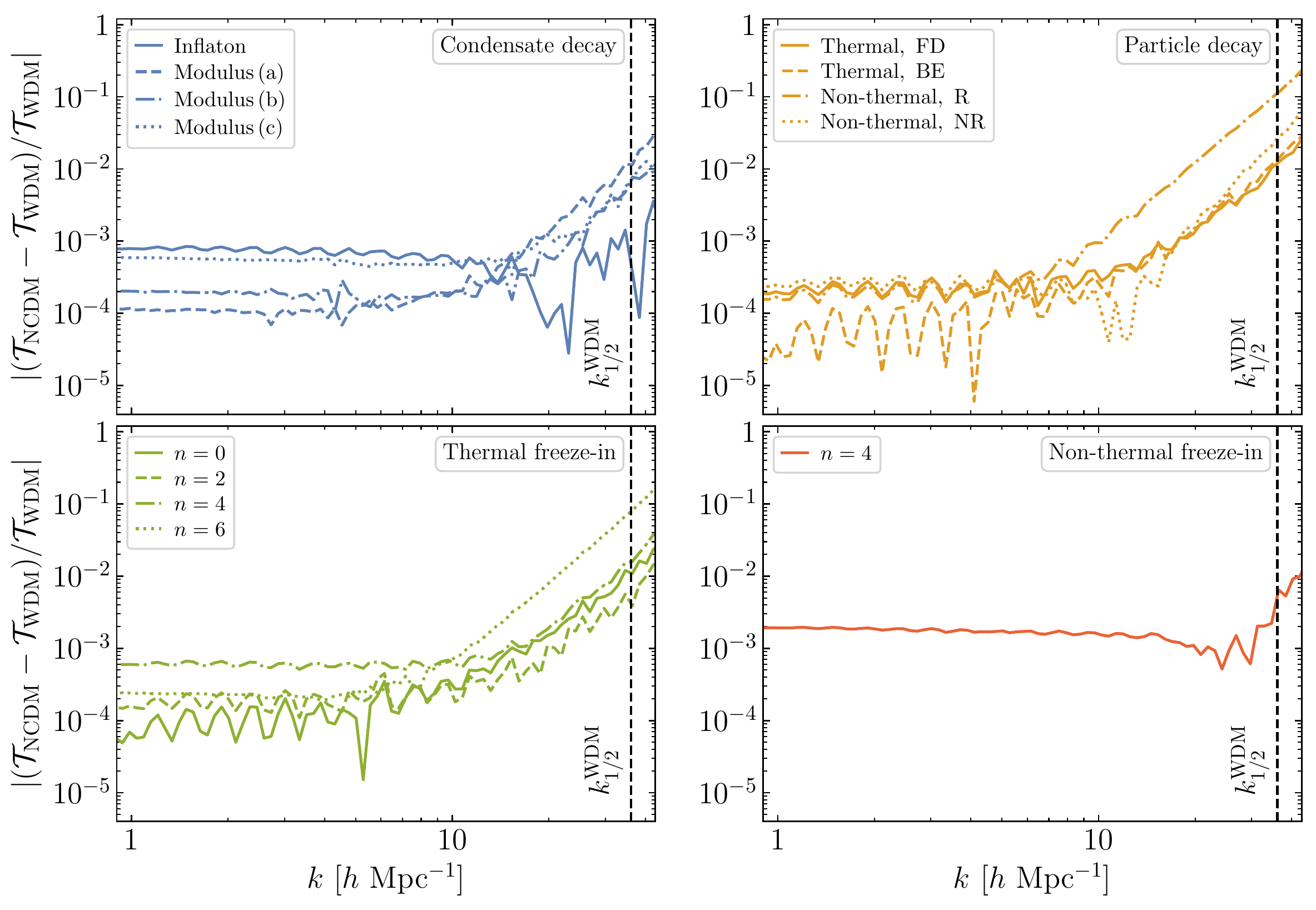}
    \caption{Relative difference between the transfer functions of the various NCDM models considered in this work and the WDM case (assuming $m_\text{WDM}=3$~keV). The scale $k_{1/2}^\text{WDM}$ defined in Section~\ref{sec:analytical_rescaling} is represented as a black vertical dashed line. Our matching procedure performs at the level of $\sim 3\%$ or better at this scale for most models. The notation for the modulus decay cases is introduced in Section~\ref{sec:moduli_PSD} (see Fig.~\ref{fig:TF_MD}). For thermal particle decays, FD and BE stand for the Fermi-Dirac and Bose-Einstein distributions for the decaying particle, respectively (see Fig.~\ref{fig:TF_TD}). For non-thermal decays, R and NR denote a relativistic or a non-relativistic decaying particle, respectively (see Eq.~\ref{eq:RorNR}). For thermal and non-thermal freeze-in, the parameter $n$ has been introduced in Eq.~(\ref{eq:amplitudesn}).}
    \label{fig:transfer_function_residual}
\end{figure}
The numerical precision achieved using our rescaling procedure is represented in Fig.~\ref{fig:transfer_function_residual}, which shows the relative difference between the various NCDM transfer functions considered in this work and the one for the WDM case. The transfer functions are computed numerically with \texttt{CLASS} and using our rescaling procedure to for a given WDM mass, which we assume to be $m_\text{WDM}=3$ keV in Fig.~\ref{fig:transfer_function_residual}. The NCDM transfer functions match accurately, mostly with a precision below the percent level, the WDM transfer function for $k <20\,h \, \text{Mpc}^{-1}$. The precision decreases for larger modes $k$. 

In order to estimate the difference on the cutoff scale expected between NCDM and WDM using our procedure, with a quantity more relevant for observational data, we represent in Fig.~\ref{fig:transfer_function_residual} the scale $k_{1/2}^\text{WDM}$ defined such that
$\mathcal{P}_\text{WDM}(k_{1/2}^\text{WDM})=(1/2)\mathcal{P}_{\Lambda\text{CDM}}(k_{1/2}^\text{WDM})$ for a given WDM mass. Fig.~\ref{fig:transfer_function_residual} shows that at $k_{1/2}^\text{WDM}$, our rescaling procedure allows to achieve a $3 \%$ difference or better on the NCDM transfer function, relative to the WDM case, for most of our scenarios. The least precise cases achieve a $\sim 10 \%$ difference at $k_{1/2}^\text{WDM}$. These correspond to DM production from the decay of a non-thermal relativistic particle and DM production via thermal freeze-in with $n=6$. 

An accurate estimate of the transfer function is numerically more challenging due to the specific shape of the phase space distributions in these cases. For this reason, we believe that the larger relative difference displayed in Fig.~\ref{fig:transfer_function_residual} for these cases can be partially attributed to the requirement of having a reasonable computation time, at the price of a limited precision.

\section{Decay of a classical condensate}
\label{sec:classdec}

We consider as the first application of our formalism the production of DM from the perturbative decay of a classical, spatially homogeneous, oscillating condensate. As a first example, we study the decay of the inflaton field into DM during reheating, assuming all other DM interactions can be neglected. We then consider the decay of a modulus field, a scalar field present in the early Universe with a non-vanishing vacuum misalignment: a displacement from its post-inflationary global minimum, which leads to a subsequent epoch of oscillations about this minimum. We explore the scenario in which the oscillations of the modulus dominate the energy density of the Universe at late times and, also, the case in which they are subdominant to the inflaton or radiation background. In all cases we find the non-thermal DM phase space distributions, and the corresponding Ly-$\alpha$ bounds on the DM mass.

\subsection{Perturbative inflaton decay}
\label{sec:perturbative_inflaton_decay}

Let us assume that the production of a DM particle $\chi$ proceeds through the two-body decay of the inflaton field \gls{inflaton} during reheating, i.e.~through a process of the form $\phi \rightarrow \chi+\psi$. The 
rest frame decay rate for this process is given by $\Gamma_{\phi\rightarrow \chi\psi}= {\rm Br}_{\chi} \gls{Gammaphi}$, where ${\rm Br}_{\chi}$ denotes the branching ratio to $\chi$, and \gls{Gammaphi} is the total decay rate of the inflaton. We assume the coupling of $\chi$ with $\phi$ is sufficiently weak to disregard the re-population of $\phi$ from inverse decays.\footnote{This is an assumption which is justified \textit{a posterori} by requiring the generated DM density to match the observed relic abundance. 
} Moreover, as in all other cases, we assume that the couplings of $\chi$ to the visible sector or to itself are not strong enough to bring it to kinetic and/or chemical equilibrium, and may therefore be disregarded. It must be emphasized that, for simplicity, in each of the cases discussed in this paper we assume that $100 \%$ of the DM relic abundance is produced by a single mechanism. In addition, we assume that the DM particles do not have significant interactions between them.\footnote{As highlighted recently in \cite{Heimersheim:2020aoc,Dvorkin:2020xga} self-interactions could affect the power spectrum, in particular for cases with light DM masses. Moreover, we neglect thermal effects that would give rise to a subdominant contribution for UV freeze-in but have been shown to alter the produced DM phase space distribution in IR-dominated freeze-in specific scenarios~\cite{Dvorkin:2019zdi}.}

In order to apply the procedure described in Section~\ref{sec:NCDM} for mapping the WDM bound on $m_\text{WDM}$ into a bound on the mass of $\chi$, we must first determine the form of the phase space distribution $f_\chi$ generated from decays of the inflaton field, by solving the Boltzmann transport equation

\beq\label{eq:boltzmanneq}
\frac{\partial f_{\chi}}{\partial t} - H|\bp|\frac{\partial f_{\chi}}{\partial |\bp|}\,=\,\mathcal{C}[f_{\chi}(|\bp|,t)]\,,
\eeq
where $\gls{collisionterm}$$[f_{\chi}]$ denotes the collision term, determined by the inflaton-DM interaction. In Appendix~\ref{sec:computation_PSD} we provide the general form of this collision term, as well as the general solution of (\ref{eq:boltzmanneq}) in the absence of inverse processes and in the free-streaming limit.

\subsubsection{DM phase space distribution}
\label{sec:inflaton_PSD}

Under the assumptions discussed above, the decay of the inflaton to $\chi$ will be perturbative. If this is true for all its decay channels, then $\phi$ is, on average, spatially homogeneous, and the phase space distribution 
may be written as $f_{\phi}(k,t)=(2\pi)^3n_{\phi}(t)\delta^{(3)}(\boldsymbol{k})$, with $n_{\phi}$ the instantaneous inflaton number density. Disregarding inverse decays, the collision term for the transport equation that determines the distribution function for $\chi$ takes the form
\begin{align} \label{eq:Cdecay} \notag
\mathcal{C}[f_\chi(p,t)] \;&=\; \frac{1}{2p_0} \int \frac{\diff ^3 {\boldsymbol{k}}}{(2\pi)^3 2k_0} \frac{g_{\psi} \diff ^3 {\bp}_{\psi}}{(2\pi)^3 2p_{\psi 0}} (2\pi)^4 \delta^{(4)}(k-p-p_{\psi})\\ 
&\hspace{145pt}\times |\mathcal{M}|^2_{\phi\rightarrow \chi\psi} f_{\phi}(k) \left(1\pm f_{\chi}(p) \pm f_{\psi}(p_{\psi}) \right)\\ \notag
&=\; \frac{\pi n_{\phi}}{4m_{\phi} p_0} \int_{\rm RF} \frac{g_{\psi} \diff ^3 {\bp}_{\psi}}{p_{\psi 0}} \delta(m_{\phi}-p_0 - p_{\psi 0}) \delta^{(3)}(\bp + \bp_{\psi}) |\mathcal{M}|^2_{\phi\rightarrow \chi\psi} \left(1\pm f_{\chi}(p) \pm f_{\psi}(p_{\psi}) \right) \\ \label{eq:BEPT}
&=\; \frac{2\pi^2}{g_{\chi} \varepsilon^2_{\psi}}n_{\phi} \Gamma_{\phi\rightarrow \chi\psi}(1\pm f_{\chi}(p_0)  \pm f_{\psi}( \varepsilon_{\psi})) \delta(p_0-\varepsilon_{\psi})\,,
\end{align}
where notations and conventions are detailed in the appendix. Here $\varepsilon_{\psi}= (m_{\phi}^2+m_{\psi}^2 - \mncdm^2) /2m_{\phi} $ denotes the energy of the daughter particle. The collision term can be further simplified in the limit when $\mncdm,m_{\psi}\ll m_{\phi}$, so that $p_0\simeq |\bp|=p$, and if the quantum statistics of the decay products can be neglected.\footnote{This is ensured provided that the effective coupling $y_{\chi}\equiv (8\pi \Gamma_{\phi\rightarrow\chi\psi}/m_{\phi})^{1/2}\ll 10^{-5}$.} If this is the case we can simply write 
\beq\label{eq:Cinfdec}
\mathcal{C}[f_\chi(p,t)]\;=\; \frac{8\pi^2}{g_{\chi} m_{\phi}^2}n_{\phi} \Gamma_{\phi\rightarrow \chi\psi} \delta(p-m_{\phi}/2)\,,
\eeq
Substitution of this collision term into the transport equation (\ref{eq:boltzmanneq}) yields an equation that has an exact solution in terms of the Hubble parameter $H$ and the inflaton occupation number~\cite{Garcia:2018wtq,Moroi:2020has},
\begin{align} \label{eq:fnonthermal}
f_{\chi}(p,t) \;&=\;  \frac{16\pi^2 \Gamma_{\phi\rightarrow \chi\psi} n_{\phi}(\hat t)}{g_{\chi} m_{\phi}^3 H(\hat t)} \theta(t - \hat t)\,,
\end{align}
where $\hat t$ is the solution to the equation
\beq\label{eq:thateq}
\frac{a(t)}{a(\hat t)}=\frac{m_{\phi}}{2p}\,.
\eeq

In order to obtain a closed form for $f_{\chi}$ we need to solve for the inflaton number density and the expansion rate. This can be achieved by integrating the Friedmann-Boltzmann system of equations
\begin{align}\label{eq:FB1}
\dot{\rho}_{\phi} + 3H\rho_{\phi} + \Gamma_{\phi}\rho_{\phi} \;&=\;0\,,\\ \label{eq:FB2}
\dot{\rho}_{r} + 4H \rho_{r} - \Gamma_{\phi} \rho_{\phi}\;&=\;0 \,,\\ \label{eq:FB3}
\rho_{\phi} + \rho_{r} \;=\; 3H^2 &M_P^2\,,
\end{align}
where the reduced Planck is $\gls{mplanck}=1/\sqrt{8\pi\,G}$ (being $G$ Newton's gravitational constant) and where we denote by $\rho_{\phi}$ and $\rho_{r}$ the energy densities of the inflaton condensate and that of its relativistic decay products, respectively. Note that (\ref{eq:FB2}) is nothing but the integrated version of the transport equation (\ref{eq:boltzmanneq}) for an ultrarelativistic species with ${\rm Br}_r=1$. Straightforward integration gives~\cite{Turner:1983he}
\beq\label{eq:nphi1}
n_{\phi}(t) \;=\; \frac{\rho_{\phi}(t)}{m_{\phi}} \;=\; \frac{\rho_{\rm end}}{m_{\phi}} \left(\frac{a(t)}{a_{\rm end}}\right)^{-3} e^{-\Gamma_{\phi}(t-t_{\rm end})}\,,
\eeq
where the sub-index ``end'' denotes quantities at the end of inflation. For $t_{\rm end} \ll t \ll \Gamma_{\phi}^{-1}$ the exponential in the previous expression can be disregarded: the Universe is dominated by the matter-like oscillations of $\phi$. Therefore, we may also approximate $a\propto t^{2/3}$, and $\hat t\simeq (2p/m_{\phi})^{3/2}t$. Substitution into (\ref{eq:fnonthermal}) yields the following expression for the phase space distribution of $\chi$ well before the end of reheating at $t_{\rm reh}\simeq \Gamma_{\phi}^{-1}$,
\begin{flalign}\label{eq:fpreth}
& \text{($t\ll t_{\rm reh}$)} & \Cen{3}{\begin{aligned} f_{\chi}(p,t) \;&=\; \frac{24\pi^2 {\rm Br}_{\chi} \Gamma_{\phi} }{g_{\chi} m_{\phi}^3} \left(\frac{m_{\phi}}{2p}\right)^{3/2} t\, n_{\phi}(t)\, \theta(m_{\phi}/2-p)\\
&\simeq\; \frac{24\pi^2 n_{\chi}(t) }{g_{\chi} m_{\phi}^3} \left(\frac{m_{\phi}}{2p}\right)^{3/2} \, \theta(m_{\phi}/2-p)\,. \end{aligned}}      &&  
\end{flalign}
Here we have approximated the number density of decay products as
\beq\label{eq:ndecprod}
n_{\chi}(t) \;\simeq\; {\rm Br}_{\chi}\frac{\rho_{\rm end}}{m_{\phi}} \left(1- e^{-\Gamma_{\phi}(t-t_{\rm end})} \right) \left(\frac{a(t)}{a_{\rm end}}\right)^{-3} \,.
\eeq
obtained by counting the quanta produced from inflaton decay. Note the consistency of (\ref{eq:fpreth}) with the defining relation (\ref{eq:nfrel}) between $f_{\chi}$ and $n_{\chi}$. \par\bigskip

The distribution (\ref{eq:fpreth}) will come handy for our study of non-thermal freeze-in in Section~\ref{sec:thenonthfi}. For our present purposes, though, this distribution is incomplete, as it lacks the high momentum tail that will be generated when the inflaton energy density begins to get exhausted. We must therefore extend (\ref{eq:fpreth}) beyond the end of reheating. As a first approximation, we evaluate (\ref{eq:fnonthermal}) at $t_{\rm reh} = \Gamma_{\phi}^{-1}$, the moment of time at which the energy density in $\phi$ is approximately equal to that in radiation, $\rho_{\phi}\simeq \rho_r$, and where $H_{\rm reh}\simeq 2\Gamma_{\phi}/3$. With the reheating temperature given by
\beq
T_{\rm reh} \;=\; \left(\frac{30 \rho_{\rm rad}}{\pi^2 g_{*s}^{\rm reh}}\right)^{1/4}\,,
\eeq
where \gls{gstars} denotes the effective number of relativistic degrees of freedom for entropy, we can substitute into (\ref{eq:fnonthermal}) to obtain
\beq \label{eq:finfdec}
f_{\chi}(p,t_{\rm reh}) \;\simeq\; \frac{4\pi^4 {\rm Br}_{\chi} g_{*s}^{\rm reh}}{5 g_{\chi}} \left(\frac{T_{\rm reh}}{m_{\phi}}\right)^4 \left(\frac{m_{\phi}}{2p}\right)^{3/2} e^{1-(2p/m_{\phi})^{3/2}} \theta(m_{\phi}/2-p)\,.
\eeq
{Naively, this distribution would evolve at later times simply in accordance to (\ref{eq:freestream}). However,} the production of entropy from inflaton decay does not suddenly stop at $t_{\rm reh}$, but continues for some time into the radiation domination era. The continuous transition $w=0\rightarrow 1/3$ makes the analytical estimation of $f_{\chi}$ beyond $t_{\rm reh}$ complicated, although not impossible (see e.g.~\cite{EGNOP,Ellis:2015pla}). Nevertheless, Eqs.~(\ref{eq:fnonthermal}) and (\ref{eq:nphi1}) make an estimate of the shape of the tail of the distribution straightforward. During radiation domination $a\propto t^{1/2}$, implying that for momenta which satisfy the relation $t_{\rm reh} \ll \hat t=(2p/m_{\phi})^2t$, the time-dependence of $n_{\phi}$ yields
\begin{flalign} \label{eq:ftail}
& \text{($t_{\rm reh} \ll (2p/m_{\phi})^2t$)} & \Cen{3}{ f_{\chi}(p,t) \;\propto\; \exp\left[\left(\frac{2p}{m_{\phi}}\right)^2 \frac{t}{t_{\rm reh}} \right] \theta(m_{\phi}/2-p)\,, }      &&  
\end{flalign}
i.e.~a Gaussian tail.\par\bigskip

\begin{figure}[!t]
\centering
    \includegraphics[width=0.7\textwidth]{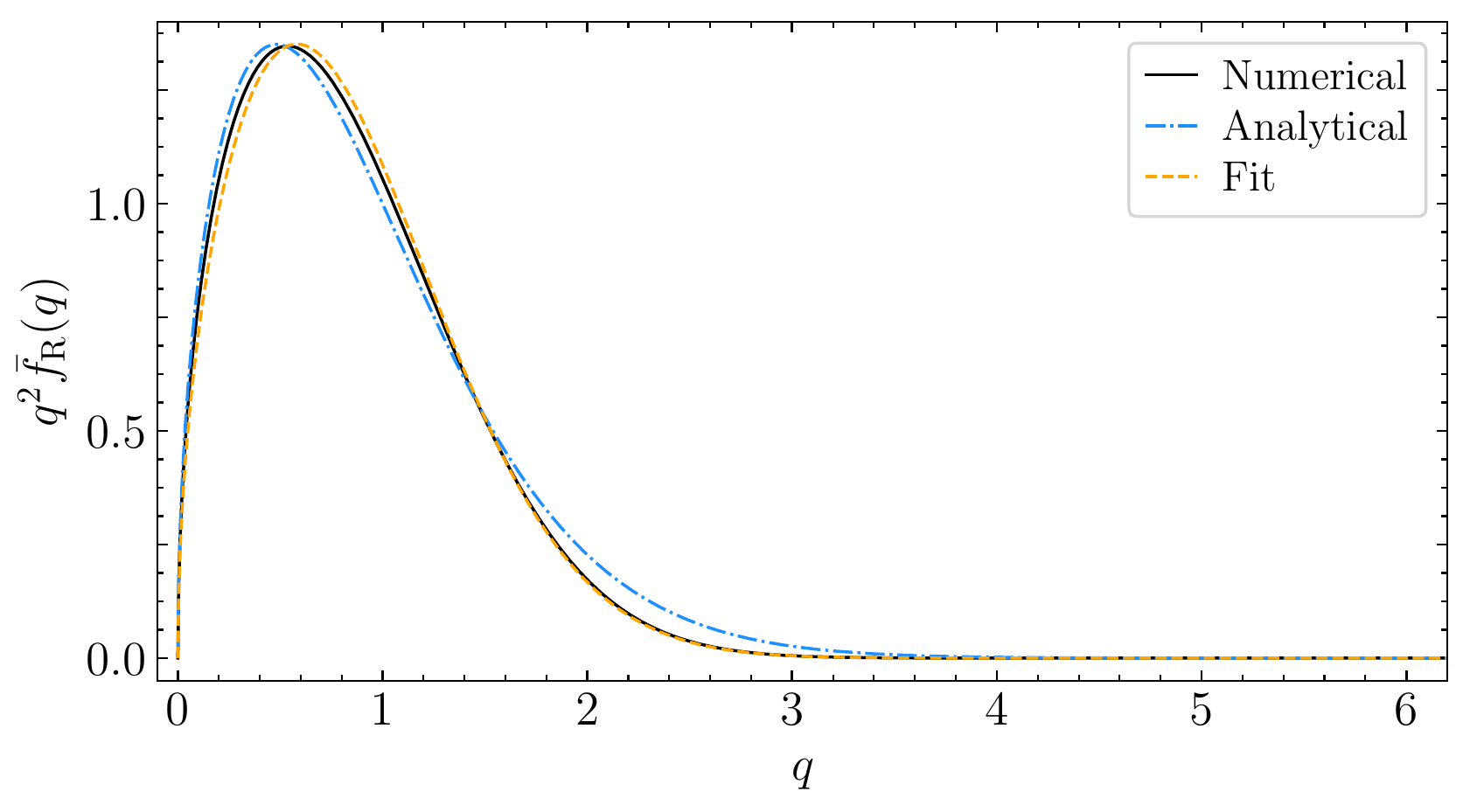}
    \caption{The rescaled distribution function $\bar{f}_{{\rm R}}$, defined in (\ref{eq:resdistinf}), as a function of the rescaled momentum $q$, for DM produced from inflaton decay. Solid, black: the numerically computed result. Dashed-dotted, blue: the analytical result (\ref{eq:finfdec}) without the Heaviside function. Dashed, orange: the phenomenological fit (\ref{eq:infphenfit}). The part of the distribution for which $q<1$ is populated during $t<t_{\rm reh}$. The part of the distribution for which $q>1$ is populated during $t>t_{\rm reh}$.}
    \label{fig:distinf}
\end{figure}

A better approximation for $f_{\chi}(p,t)$ beyond the end of reheating can be constructed by solving numerically the Friedmann-Boltzmann system (\ref{eq:FB1})-(\ref{eq:FB3}) together with (\ref{eq:boltzmanneq}) with collision term (\ref{eq:Cinfdec}). This solution is shown as the continuous black curve in Fig.~\ref{fig:distinf}, in the form of the rescaled distribution $\bar{f}_{{\rm R}}$, defined through the relation
\beq\label{eq:resdistinf}
f_{\chi}(p,t)\, \diff ^3 \bp \;=\; \frac{4\pi^4 {\rm Br}_{\chi} g_{*s}^{\rm reh}}{5 g_{\chi}} \left(\frac{T_{\rm reh}}{m_{\phi}}\right)^4  \, \left(\frac{a_0}{a(t)}\right)^3 \tncdm^3\, \bar{f}_{\rm R}(q)\, \diff ^3 \boldsymbol{q}\,.
\eeq
Here $q$ is defined as in (\ref{eq:qdef}), and in this scenario
\beq\label{eq:TncdmR}
\tncdm \;=\; \frac{m_{\phi}}{2}\frac{a_{\rm reh}}{a_0} \;=\; \left(\frac{g_{*s}^{0}}{g_{*s}^{\rm reh}}\right)^{1/3}\frac{m_{\phi}}{2T_{\rm reh}}\, T_0\,.
\eeq
The numerical solution was computed at $t=50t_{\rm reh}$, well beyond the matter-radiation equality that signals the end of reheating. At this time the universe is dominated by radiation, and the production of entropy from inflaton decay has ceased. The particle population that was produced during $t<t_{\rm reh}$ occupies the distribution at $q<1$, while the population created during $t>t_{\rm reh}$ corresponds to the $q>1$ tail. Shown in Fig.~\ref{fig:distinf} is also the analytical solution (\ref{eq:finfdec}), ignoring the Heaviside cutoff at $q=1$. As expected, this expression accurately describes the distribution at small momenta, $\bar{f}_{\rm R}\propto q^{-3/2}$, but the tail is not matched. Given that we expect the large momentum regime to be described by (\ref{eq:ftail}), we also show in the figure, as an orange dashed curve, a fitting function that mimics the low- and high-energy behavior of the distribution, 
\beq\label{eq:infphenfit}
 \bar{f}_{{\rm R}}(q) \;\simeq\; 2.28\, q^{-3/2} e^{-0.74q^2}\,.
\eeq
This approximation is of the form (\ref{eq:allfit}), and provides an excellent fit to the exact form of $\bar{f}_{\rm R}$. Note the seeming mismatch between the ratio $(t/t_{\rm reh})^{1/2}$ and the ratio $a(t)/a_{\rm reh}$ through which $q$ is defined, quantified by the factor 0.74 in the exponent. This is due to the relatively complicated dependence of the scale factor on time in the matter-radiation transition at the end of reheating, affecting the high-energy tail of the distribution. 

\subsubsection{Power spectrum and Ly-$\alpha$ constraints}\label{sec:lyainfdec}

With the phase space distribution for DM produced from direct inflaton decay, we can now make use of Eq.~(\ref{eq:mncdm_fromeos}) to map the WDM Ly-$\alpha$ constraints on the DM mass for this scenario. Straightforward calculation gives the following rescaling of the bound on the DM mass,
\begin{align}\notag
   \hatmncdm \; \gtrsim \;  \left( \dfrac{m_\text{WDM}}{3~\text{keV}} \right)^{4/3} &\left( \dfrac{106.75}{g_{*s}^{\rm reh}} \right)^{1/3} \\
    & \times \left( \dfrac{m_\phi}{3\times 10^{13}~\text{GeV}} \right)\left( \dfrac{10^{10}~\text{GeV}}{T_{\text{reh}}} \right) \begin{cases}
3.78~\text{MeV}\,, ~ & {\rm Numerical}\,,\\
4.11~\text{MeV}\,, ~& {\rm Analytical}\,,\\
3.79~\text{MeV}\,, ~& {\rm Fit}\,.
\end{cases}
\label{eq:mDM_bound_ID}
\end{align}
The numerical, analytical and fit approximations correspond to the numerically computed distribution shown in Fig.~\ref{fig:distinf}, to (\ref{eq:finfdec}), and to (\ref{eq:infphenfit}), respectively. For low reheating temperatures the bound on the NCDM mass becomes significantly larger than that for WDM. This can be understood by fixing the inflaton mass and decreasing progressively the reheating temperature. The bulk of DM is produced around the reheating temperature with typical momentum $p\sim m_\phi/2$, regardless of the radiation temperature. Reducing the reheating temperature therefore prevents the momentum of the DM particle from redshifting too much, resulting in a hotter spectrum at the present time than that expected for large reheating temperatures. 
\begin{figure}[!t]
\centering
    \includegraphics[width=0.75\textwidth]{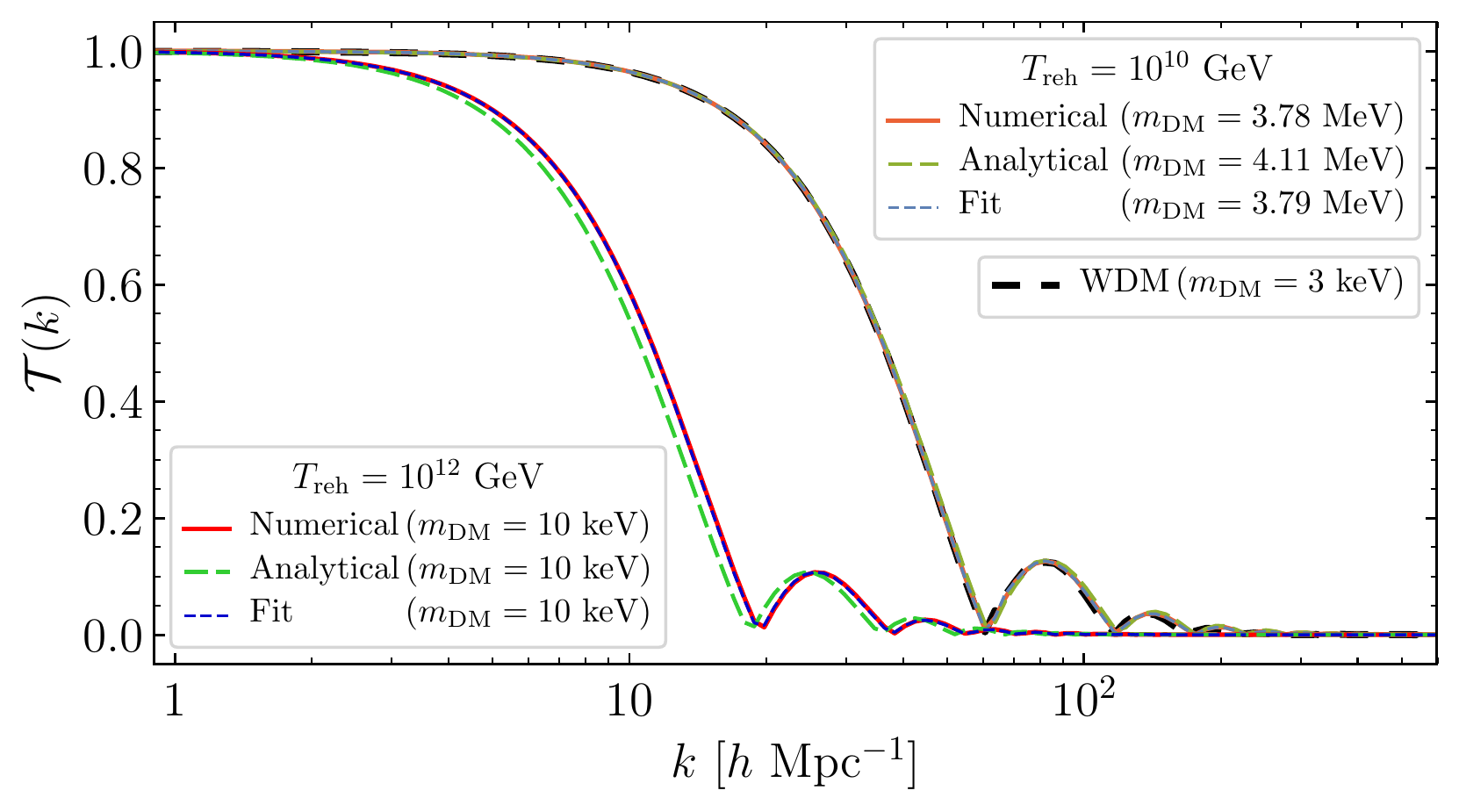}
    \caption{Linear transfer function for the scenario in which DM is produced by inflaton decay, assuming the
numerical, analytical or fitted phase space distributions described in Section~\ref{sec:perturbative_inflaton_decay}, by taking the mass estimated in (\ref{eq:mDM_bound_ID}) and identical reheating temperatures, $T_{\rm reh}=10^{10}\,{\rm GeV}$. The transfer function for the WDM case is shown for comparison with a dashed black line. We depicted as well the transfer function for numerical,
analytical or fitted phase space distributions with identical masses and reheating temperature, $T_{\rm reh}=10^{12}\,{\rm GeV}$.}
    \label{fig:TF_ID}
\end{figure}\par\bigskip

Fig.~\ref{fig:TF_ID} shows the form of the transfer function for the matter power spectrum, as computed with \texttt{CLASS} \cite{Blas:2011rf,Lesgourgues:2011rh}. Depicted are the results for the numerical, analytical and fit approximations to $f_{\chi}$. The rightmost set of curves shows the form of each $\mathcal{T}(k)$ for a reheating temperature $T_{\rm reh}=10^{10}\,{\rm GeV}$, and masses given by Eq.~(\ref{eq:mDM_bound_ID}). The overlap of all three curves with each other, and with the reference WDM transfer function, demonstrates the validity of our method for this DM production mechanism. At $k_{1/2}^{\rm WDM}$ the relative difference between WDM and the numerical result is in particular smaller than $10^{-3}$, c.f.~Fig.~\ref{fig:transfer_function_residual}. The leftmost cluster of curves shows the form of $\mathcal{T}(k)$ for the three approximations for a larger reheating temperature, $T_{\rm reh}=10^{12}\,{\rm GeV}$, assuming a mass of $10\,{\rm keV}$. These curves do not overlap with the WDM bound, and they differ slightly between each other, albeit the agreement between the numerical and fit cases is still excellent.\par\bigskip

\subsubsection{Relic density and phenomenology}

We now discuss the phenomenological implications of a lower bound on a light DM particle produced from inflaton decay. Given a reheating temperature and a DM mass, the normalization of the distribution function is determined by the value of the present DM 
fraction $\Omega_{\chi}=\rho_{\chi}/\rho_c$, where $\rho_c \simeq 1.05 \times 10^{-5} h^2 {\rm GeV\, cm}^{-3}$ is the present critical density of the Universe~\cite{Tanabashi:2018oca}. Integration of (\ref{eq:resdistinf}) at $t\gg t_{\rm reh}$ gives
\beq
n_{\chi}(t) \;\simeq\; 0.70\pi^2 {\rm Br}_{\chi} g_{*s}^{\rm reh} \left(\frac{T_{\rm reh}}{m_{\phi}}\right)^4  \, \left(\frac{a_0}{a(t)}\right)^3 \tncdm^3\,,
\eeq
which in turn yields
\beq\label{eq:omegainfdec}
\Omega_{\chi}h^{2} \;\simeq\; 0.1 \left(\frac{{\rm Br}_{\chi}}{5.5\times 10^{-4}}\right) \left(\frac{\mncdm}{1\,{\rm MeV}}\right)\left(\frac{T_{\rm reh}}{10^{10}\,{\rm GeV}}\right)\left(\frac{3\times 10^{13}\,{\rm GeV}}{m_{\phi}}\right)\,.
\eeq
Combining the bounds on the DM mass (\ref{eq:mDM_bound_ID}) and on the relic abundance (\ref{eq:omegainfdec}), the following constraint can be derived for the branching ratio of the decay of the inflaton into dark matter,
\beq\label{eq:brlim}
{\rm Br}_{\chi} \;\lesssim\; 1.5\times 10^{-4}\, \left(\frac{g_{*s}^{\rm reh}}{106.5}\right)^{1/3} \left(\frac{3\,{\rm keV}}{m_{\rm WDM}}\right)^{4/3}\,.
\eeq
Note the universality of this bound: it is independent of the inflaton mass and the reheating temperature. As mentioned earlier, such a limit will apply even in the absence of tree-level couplings between the inflaton and DM. Assuming a dominant fermionic decay channel of the inflaton, with these decay products in turn coupled to DM through an effective interaction of the following form,
\beq\label{eq:effLdecphi}
\mathcal{L} \;=\; y\phi\bar{f}f + \frac{1}{\Lambda^2} \bar{f}f\bar{\chi}\chi\,,
\eeq
(which could arise from the exchange of a massive field with mass $\sim\Lambda$), a non-vanishing decay rate for the $\phi\rightarrow\bar{\chi}\chi$ process is induced at 1-loop~\cite{Kaneta:2019zgw},
\beq
\Gamma_{\phi\rightarrow\bar{\chi}\chi} \;\simeq\; \frac{y^2}{128\pi^5}\left(1+\frac{\pi^2}{4}\right)\frac{m_{\phi}^5}{\Lambda^4}\,,
\eeq
corresponding to ${\rm Br}_{\chi}=\frac{1}{16\pi^4}(1+\frac{\pi^2}{4})(\frac{m_{\phi}}{\Lambda})^4$. Substitution into (\ref{eq:brlim}) reveals that
\beq
\Lambda \;\gtrsim\; 2\, m_{\phi} \left(\frac{106.5}{g_{*s}^{\rm reh}}\right)^{1/12} \left(\frac{m_{\rm WDM}}{3\,{\rm keV}}\right)^{1/3}\,,
\eeq
a condition consistent with the form of the effective action (\ref{eq:effLdecphi}), assumed to be valid at all times during reheating.\par\medskip

We finish this section by emphasizing that the bounds (\ref{eq:mDM_bound_ID}) and (\ref{eq:brlim}) apply for the perturbative decay of the inflaton $\phi$ while it oscillates about a quadratic minimum. A different production mechanism, e.g.~through perturbative decay in a non-quadratic potential~\cite{Shtanov:1994ce,Ichikawa:2008ne,Garcia:2020eof}, or via non-adiabatic particle production~\cite{Kofman:1997yn,Greene:1997fu,Felder:1998vq,Amin:2014eta,Lozanov:2016hid,Lozanov:2017hjm}, will lead to a different constraint on ${\rm Br}_{\chi}$.


\subsection{Moduli decays}
\label{sec:moduli_decay}
The inflaton is not necessarily the only scalar condensate that can decay in the early Universe. In many BSM constructions, notably supersymmetric and string SM extensions, a plethora of weakly-interacting unstable scalar fields, collectively known as {\em moduli}, arise~\cite{Coughlan:1983ci,Ellis:1986zt,Banks:1993en,deCarlos:1993wie,Banks:1995dp,Banks:1995dt}.
During inflation, these moduli can be excited away from the minima of their potential, resulting in a posterior roll towards these minima. Depending on the initial misalignment, and the masses of the moduli, the subsequent oscillations about the minima may eventually dominate the energy density of the Universe. The decay of these fields would then reheat the Universe at temperatures below the inflationary reheating temperature, diluting any relics produced earlier (such as DM) and the baryon asymmetry. This process would also lead to deviations from the standard Big Bang Nucleosynthesis (BBN), which is strongly constrained by the data, unless $T_{\rm reh}\gtrsim 1\,{\rm MeV}$~\cite{Fields:2019pfx,Hasegawa:2019jsa}. 

\begin{figure}[!t]
\centering
    \includegraphics[width=0.56\textwidth]{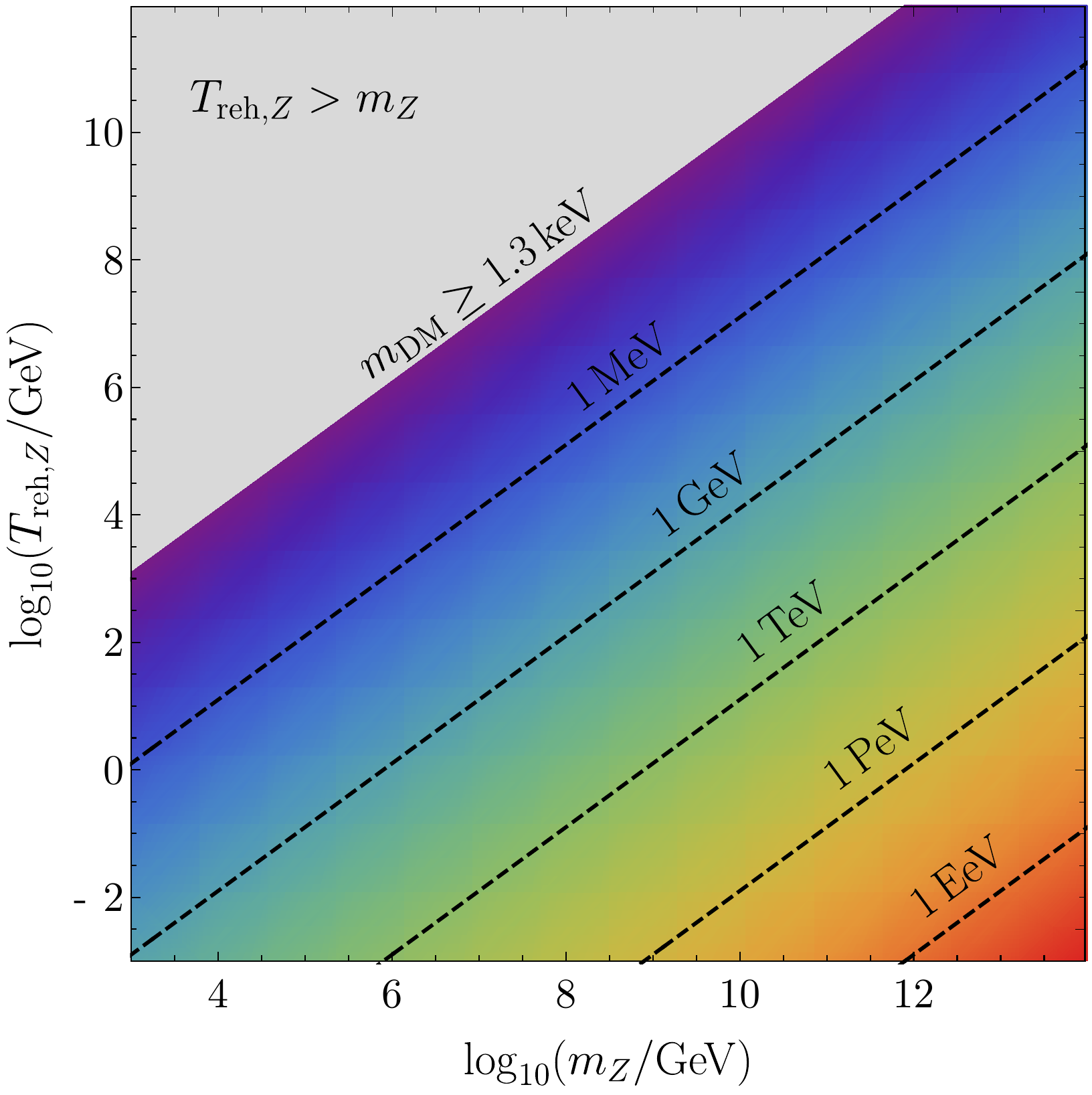}
    \caption{Ly-$\alpha$ constraint on the DM mass, as a function of the modulus mass $m_Z$ and reheating temperature, in the case in which the oscillation of $Z$ dominates the energy density of the Universe, leading to entropy production upon its decay. The gray region corresponds to $T_{{\rm reh},Z}>m_Z$, where in-medium and/or non-perturbative effects may determine the decay of $Z$. At $m_Z=3\times 10^{13}\,{\rm GeV}$ the inflaton decay scenario is recovered.}
    \label{fig:modmass}
\end{figure}

 If a modulus $Z$ has a non-vanishing branching ratio to DM, the Ly-$\alpha$ bounds derived in the previous section can  be mapped to its decay (provided that $m_Z\gg m_{\rm DM}$) simply by replacing the inflaton mass and reheating temperatures with their corresponding modulus values. In particular, for the mass bound, we can write
\beq\label{eq:mdmmodulifirst}
m_{\rm DM} \;\gtrsim\; 12.6\,{\rm GeV}\left(\frac{m_{\rm WDM}}{3\,{\rm keV}}\right)^{4/3} \left(\frac{106.75}{g_{*s}^{{\rm reh},Z}}\right)^{1/3} \left(\frac{m_Z}{10\,{\rm TeV}}\right)\left(\frac{1\,{\rm MeV}}{T_{{\rm reh},Z}}\right)\,,
\eeq
while (\ref{eq:brlim}) remains unchanged, except for the replacement $g_{*s}^{\rm reh} \rightarrow g_{*s}^{{\rm reh},Z}$. Note that for moduli with masses $m_Z\gtrsim 100\,{\rm TeV}$, the lower bound on the DM mass is $\gtrsim 100\,{\rm GeV}$, on the range of electroweak-scale DM candidates such as the lightest neutralino. Moreover, the late decay of $Z$ would ensure that the non-thermal phase space distribution (\ref{eq:infphenfit}) remains imprinted into this relic. This is due to the fact that most of the DM is produced around $T_{{\rm reh},Z}$, well below the corresponding thermal decoupling (freeze-out) temperature. Fig.~\ref{fig:modmass} shows the limit (\ref{eq:mdmmodulifirst}) in the mass vs.~reheating temperature plane, excluding the model-dependent $T_{{\rm reh},Z}>m_Z$ region. Note the wide range of values for $m_{\rm DM}$.
We emphasize that {this kind of constraint} must be accounted for in any discussion regarding DM production in non-standard thermal histories, with an intermediate matter-dominated epoch between the end of reheating and BBN~\cite{Watson:2009hw,Kane:2015jia,Aparicio:2016qqb,Waldstein:2016blt,Acharya:2017szw,Drees:2017iod,Allahverdi:2018aux}. For a sufficiently large branching ratio of $Z$ to $\chi$, this non-thermal production can dominate over DM freeze-out, which would have occurred during the modified expansion history. We finally mention that it is typical of the decay of a modulus into dark matter to occur in two stages, $Z\rightarrow A \rightarrow \chi$, where $A$ is an intermediate unstable particle, such as the gravitino. This scenario is studied in detail in Section~\ref{sec:nonthermaldecay}. As we show there, although the phase space distribution of $A$ and $\chi$ differ noticeably in their shape, the rescaled bound on $m_{\rm DM}$ is only corrected by an $\mathcal{O}(1)$ factor in some regimes. \par\medskip

Our main focus in this section is instead stabilized moduli: scalar condensates that oscillate and subsequently decay in the early Universe, while never dominating the energy budget of the Universe~\cite{Grana:2005jc,Douglas:2006es,Dine:2006ii,Kitano:2006wz,Kallosh:2006dv,Abe:2007yb,Fan:2011ua,Dudas:2006gr,Abe:2006xp,Linde:2011ja,Dudas:2012wi,Evans:2013lpa,Garcia:2013bha,Evans:2013nka,Ellis:2015kqa,Dudas:2017kfz,Ellis:2019dtx,Kaneta:2019yjn,Ellis:2020xmk}. Without modifying inflation~\cite{Randall:1994fr}, this is typically achieved in model-building by introducing additional interactions that rise the mass of the modulus, increasing its decay rate, and by decreasing the amount of initial misalignment. %

It is important to realize that for a subdominant decaying scalar, the post-inflationary background dynamics will be determined by either the oscillating inflaton, or by its redshifting relativistic decay products. Therefore, it is necessary to distinguish between three different scenarios: (a) the modulus begins oscillating and decays during reheating, (b) the modulus begins oscillating during reheating, but decays during radiation domination, or (c) the modulus oscillates and decays during radiation domination. We now proceed to determine the phase space distribution in all three cases, to subsequently determine the Ly-$\alpha$ bounds and the corresponding phenomenologies. As we discuss below, the observed DM abundance can be obtained from the decay of a stabilized modulus when its energy density is much smaller than that of radiation.

\subsubsection{DM phase space distribution}
\label{sec:moduli_PSD}

\noindent
{\bf Case a: Oscillation and decay during reheating ($m_Z>\Gamma_Z>\Gamma_{\phi}$)}\par\smallskip

We begin by studying the scenario in which the field $Z$ begins its oscillations during the matter-dominated reheating, and fully decays before the end of reheating. Given that we follow the decay of a classical condensate, its distribution function will be of the form $f_Z(k,t)=(2\pi)^3n_Z(t) \delta^{(3)}(\bk)$, where $n_Z$ is the modulus number density (see Eq.~(\ref{eq:nfrel})), and hence the DM distribution will be given by (\ref{eq:fnonthermal}), upon replacing $\phi\rightarrow Z$. The solution of Eq.~(\ref{eq:thateq}), necessary to determine the cosmic-time dependence of $f_{\chi}$, can be found in a straightforward way, and is given by $\hat{t}=(2p/m_Z)^{3/2}t$. Moreover, the number density of the decaying $Z$ is found by integration of (\ref{eq:FB1}), again replacing $\phi\rightarrow Z$,
\beq
n_Z(t) \;=\; \frac{\rho_{\rm osc}}{m_Z}\left(\frac{a(t)}{a_{\rm osc}}\right)^{-3}e^{-\Gamma_Z(t-t_{\rm osc})}\,.
\eeq
Here the subindex `osc' refers to the beginning of the oscillation of $Z$, which occurs at $t_{\rm osc}\simeq \frac{3}{2}H_{\rm osc} \simeq m_Z$. Assuming, as we did for the inflaton, a quadratic minimum for the potential of $Z$, we can write $\rho_{\rm osc}\simeq \frac{1}{2}m_Z^2 Z_0^2$, where $Z_0$ denotes the value of $Z$ at the initial misalignment. Straightforward substitution gives then
\beq
f_{\chi}(p,t) \;\simeq\; \frac{12\pi^2{\rm Br}_{\chi}}{g_{\chi}(\Gamma_Z t)} \left(\frac{Z_0\Gamma_Z}{m_Z^2}\right)^2 \left(\frac{m_Z}{2p}\right)^{3/2} e^{-(\Gamma_Zt)(2p/m_Z)^{3/2}+(\Gamma_Z/m_Z)} \theta(m_Z/2-p)\,.
\eeq
In analogy to the inflaton case, we estimate the decoupling time to be $t=\Gamma_Z^{-1}$. The effect of any subsequent production is to populate the exponential tail of the distribution. Hence, in what follows we evaluate the distribution at this decoupling time, and disregard the effect of the Heaviside function. Moreover, we will always work in the limit when $\Gamma_Z \ll m_Z$, as is the case even for stabilized moduli. \par\medskip

To evolve the distribution at later times we make use of the decoupled-regime solution (\ref{eq:freestream}). Note that in order to apply it we need to account for the redshift that occurs from the decay of $Z$ to the end of reheating, and the subsequent redshift from the end of reheating to present times. Since
\beq
\frac{a(t)}{a_{\rm dec}} \;=\; \frac{a(t)/a_0}{a_{\rm dec}/a_0} \;=\; \frac{a(t)/a_0}{(a_{\rm dec}/a_{\rm reh})(a_{\rm reh}/a_0)} \;\simeq \; \frac{a(t)}{a_0} \left(\frac{g_{*s}^{\rm reh}}{g_{*s}^0}\right)^{1/3} \left(\frac{T_{\rm reh}}{T_0}\right) \left(\frac{\Gamma_Z}{\Gamma_{\phi}}\right)^{2/3}\,,
\eeq
we can finally write, at late times,
\beq\label{eq:modulifa}
f_{\chi}(p,t)\,\diff^3\bp \;\simeq\; \frac{16\pi^2 {\rm Br}_{\chi}}{g_{\chi}} \left(\frac{Z_0}{m_Z}\right)^2  \left(\frac{\Gamma_Z}{m_Z}\right)^2 \left(\frac{a(t)}{a_0}\right)^3 T_{\star,a}^3 \bar{f}_{{\rm M},a}(q) \, \diff^3\boldsymbol{q}\,,
\eeq
where
\beq
T_{\star,a} \;=\;  \frac{m_Z}{2T_{\rm reh}} \left(\frac{g_{*s}^0}{g_{*s}^{\rm reh}}\right)^{1/3} \left(\frac{\Gamma_{\phi}}{\Gamma_{Z}}\right)^{2/3} T_0\,,
\eeq
and
\beq\label{eq:fmaq}
\bar{f}_{{\rm M},a}(q) \;=\; \frac{3}{4}q^{-3/2} e^{-q^{3/2}}\,.
\eeq
\begin{figure}[!t]
\centering
    \includegraphics[width=0.7\textwidth]{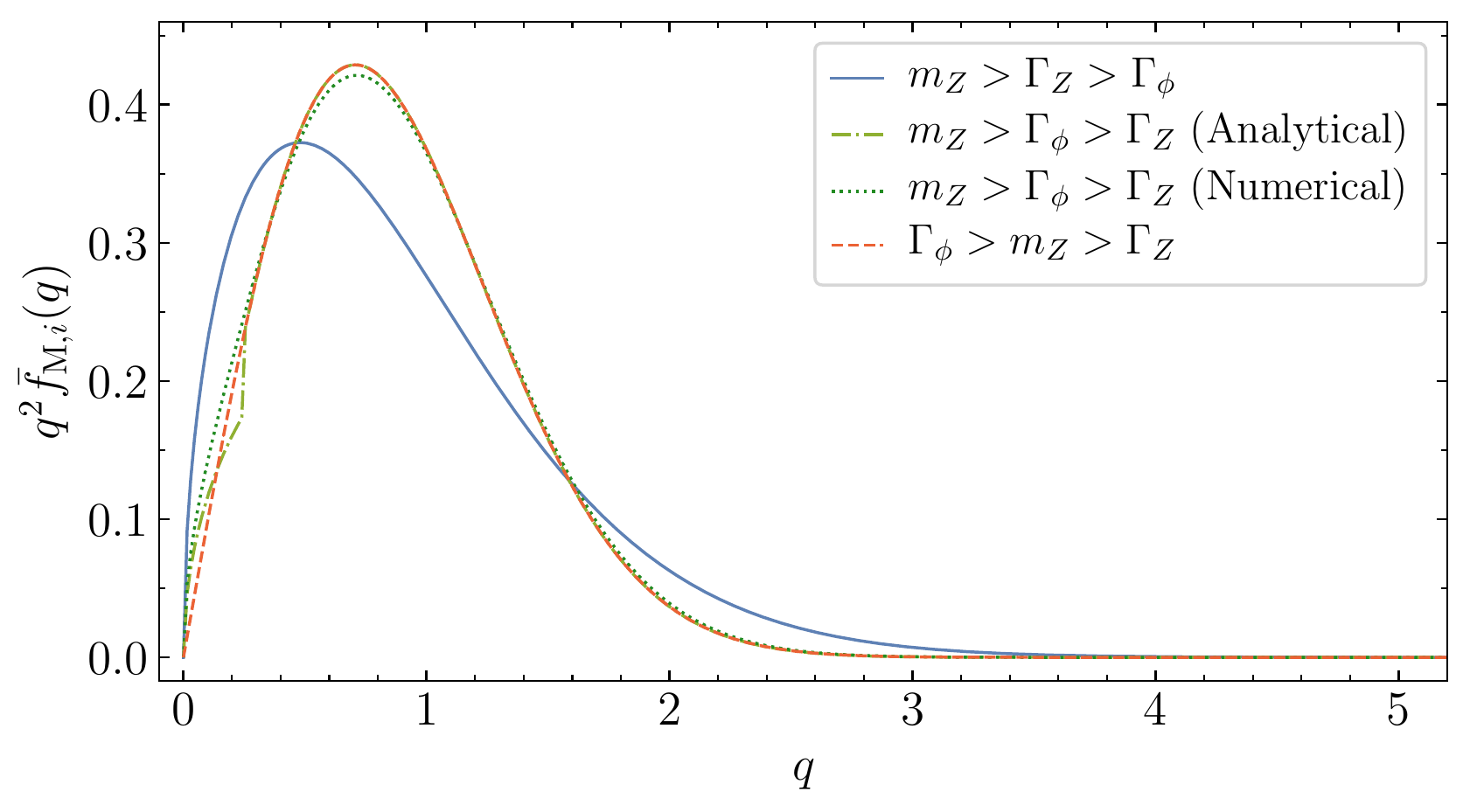}
    \caption{The rescaled distribution functions $\bar{f}_{{\rm M},i}$, $i=\{a,b,c\}$, defined in (\ref{eq:fmaq}), (\ref{eq:fmbq}) and (\ref{eq:fmcq}), as a functions of the rescaled momentum $q$, for DM produced from modulus decay. For the case $m_Z>\Gamma_{\phi}>\Gamma_Z$, the reheating-radiation domination transition scale $q_Z$ has been chosen here to be $q_Z=1/4$, and both the analytical approximation and a numerical solution are shown.}
    \label{fig:distmod}
\end{figure}
Fig.~\ref{fig:distmod} shows the form of this rescaled distribution (blue, solid curve). The low momentum power-law dependence and the exponential tail are evident. Clearly, this distribution is of the form (\ref{eq:allfit}) with $\alpha=-\gamma=3/2$ and $\beta=1$. \par\bigskip

\noindent
{\bf Case b: Oscillation during reheating, decay after reheating  ($m_Z>\Gamma_{\phi}>\Gamma_Z$)}\par\smallskip

Let us now consider the case for which $Z$ starts oscillating during reheating, and its decay is not completed until the subsequent radiation domination. It is crucial to notice that when this occurs there are two possible solutions for Eq.~(\ref{eq:thateq}),
\beq
\hat{t} \;\simeq\; \begin{cases}
t\left(\dfrac{2p}{m_Z}\right)^2\,, & p>p_{\rm reh}\\[10pt]
t_{\rm reh}\left(\dfrac{t}{t_{\rm reh}}\right)^{3/4}\left(\dfrac{2p}{m_Z}\right)^{3/2}\,,& p< p_{\rm reh}\,,
\end{cases}\,,
\qquad p_{\rm reh} \equiv \frac{m_Z}{2}\left(\frac{t_{\rm reh}}{t}\right)^{1/2}\,.
\eeq
Here we have assumed for simplicity a sharp transition from matter to radiation domination at $t_{\rm reh}$, with $a\propto t^{2/3}$ in the former case and $a\propto t^{1/2}$ in the later case. This approximation necessarily leads to a discontinuity in the Hubble parameter, which will translate into a discontinuity in the distribution function $f_{\chi}$. This is nothing but an artifact of our approximations, and it has minimal phenomenological consequences as we will show below.

For $p>p_{\rm reh}$ we have $\hat{t}>t_{\rm reh}$. In this case we write the number density of $Z$ as follows,
\begin{align}\notag
n_Z(\hat{t}) \;&\simeq\; \frac{\rho_{\rm osc}}{m_Z} \left(\frac{a_{\rm reh}}{a_{\rm osc}}\right)^{-3} \left(\frac{a(\hat{t})}{a_{\rm reh}}\right)^{-3}e^{-\Gamma_Z(\hat{t}-t_{\rm osc})}\\
&\simeq\; \frac{1}{2}m_Z Z_0^2 \left(\Gamma_{\phi} t \right)^{-3/2} \left(\frac{\Gamma_{\phi}}{m_Z}\right)^2  \left(\frac{m_Z}{2p}\right)^3 e^{-(\Gamma_Z t)(2p/m_Z)^2}\,,
\end{align}
and
\beq
H(\hat{t}) \;\simeq\; \frac{1}{2t} \left(\frac{m_Z}{2p}\right)^2\,.
\eeq
On the other hand, if $p<p_{\rm reh}$, $\hat{t}<t_{\rm reh}$. Therefore, 
\begin{align}\notag
n_Z(\hat{t}) \;&\simeq\; \frac{\rho_{\rm osc}}{m_Z} \left(\frac{a(t)}{a_{\rm osc}}\right)^{-3} e^{-\Gamma_Z(\hat{t}-t_{\rm osc})}\\
&\simeq\; \frac{1}{2}m_Z Z_0^2 \left( \Gamma_{\phi}t\right)^{-3/2} \left(\frac{\Gamma_{\phi}}{m_Z}\right)^2 \left(\frac{m_Z}{2p}\right)^3 \exp\left[-\left(\Gamma_{\phi}t\right)^{3/4}\left(\frac{\Gamma_Z}{\Gamma_{\phi}}\right)\left(\frac{2p}{m_Z}\right)^{3/2}   \right]\,,
\end{align}
and
\beq
H(\hat{t}) \;\simeq\; \frac{2}{3}\Gamma_{\phi}\left(\Gamma_{\phi} t\right)^{-3/4} \left(\frac{m_Z}{2p}\right)^{3/2}\,.
\eeq
By substituting into (\ref{eq:fnonthermal}) and evaluating at $t_{\rm dec}=\Gamma_Z^{-1}$ we obtain the distribution at decoupling. Moreover, noting that in this case the redshift occurs in the absence of intermediate entropy production, we can finally write the form of the distribution at late times in the following simplified way,
\beq\label{eq:modulifb}
f_{\chi}(p,t)\,\diff^3\bp \;\simeq\; \frac{16\pi^2 {\rm Br}_{\chi}}{g_{\chi}} \left(\frac{Z_0}{m_Z}\right)^2\left(\frac{\Gamma_{\phi}}{m_Z}\right)^2 \left(\frac{\Gamma_Z}{\Gamma_{\phi}}\right)^{3/2} \left(\frac{a(t)}{a_0}\right)^3 T_{\star,b}^3 \bar{f}_{{\rm M},b}(q) \, \diff^3\boldsymbol{q}\,,
\eeq
with
\beq
T_{\star,b} \;=\; \frac{m_Z}{2T_{\rm dec}} \left(\frac{g_{*s}^0}{g_{*s}^{\rm reh}}\right)^{1/3}  T_0\,,
\eeq
where $T_{\rm dec}=(45/(2\pi^2 g_{*s}^{\rm dec}))^{1/4}(\Gamma_Z M_P)^{1/2}$ denotes the background temperature at the moment of decay, and
\beq\label{eq:fmbq}
\bar{f}_{{\rm M},b}(q) \;=\; \begin{cases}
q^{-1} e^{-q^2}\,, & q>q_Z\\[10pt]
\dfrac{3}{4}q_Z^{1/2} q^{-3/2}e^{-q_Z^{1/2}q^{3/2}}\,, & q<q_Z
\end{cases}\,, \qquad q_Z\;\equiv\; \left(\frac{\Gamma_Z}{\Gamma_{\phi}}\right)^{1/2}\,.
\eeq

This rescaled distribution is shown in Fig.~\ref{fig:distmod} for $q_Z=1/4$. The analytical expression (\ref{eq:fmbq}) is shown as the light green dot-dashed curve. It shows the different scaling with $q$ for $q>q_Z$ and $q<q_Z$, with a jump at $q=q_Z$. As we mention above, this discontinuity is an artifact of our approximations, demonstrated by the dark green, dotted curve in this same figure, which shows the fully numerical solution, which interpolates smoothly between the two regimes. Note that for $q_Z\sim 1$ the fitting function (\ref{eq:allfit}) fails to accurately describe the distribution. Nevertheless, for $q_Z\ll 1$, it accurately describes the DM phase space distribution for any $q\sim \mathcal{O}(1)$, with $\alpha=-\beta=-1$ and $\gamma=2$. \par\bigskip

\par\bigskip

\noindent
{\bf Case c: Oscillation and decay during radiation domination  ($\Gamma_{\phi}>m_Z>\Gamma_Z$)}\par\smallskip

For the last case we assume that the beginning of the oscillation of $Z$ is delayed beyond the end of reheating, due to a rapidly decaying inflaton, a relatively light $Z$, or a combination of both. The absence of a matter-radiation crossover during oscillations, and of an intermediate entropy production regime, make this analysis straightforward. The solution of (\ref{eq:thateq}) is simply given by $\hat{t} = t(2p/m_Z)$, and from it we obtain the following expressions for the number density in $Z$,
\beq
n_Z(\hat{t}) \;\simeq\; \frac{1}{2}m_Z Z_0^2 (m_Zt)^{3/2} \left(\frac{m_Z}{2p}\right)^2 e^{-(\Gamma_Z t)(2p/m_Z)^2}\,,
\eeq
and the Hubble parameter,
\beq
H(\hat{t}) \;=\; \frac{1}{2t}\left(\frac{m_Z}{2p}\right)^2\,.
\eeq
Substitution into (\ref{eq:fnonthermal}) and (\ref{eq:freestream})
\beq\label{eq:modulifc}
f_{\chi}(p,t)\,\diff^3\bp \;\simeq\; \frac{16\pi^2 {\rm Br}_{\chi}}{g_{\chi}} \left(\frac{Z_0}{m_Z}\right)^2\left(\frac{\Gamma_{Z}}{m_Z}\right)^{3/2}   \left(\frac{a(t)}{a_0}\right)^3 T_{\star,c}^3 \bar{f}_{{\rm M},c}(q) \, \diff^3\boldsymbol{q}\,,
\eeq
with $T_{\star,c}=T_{\star,b}$ and
\beq\label{eq:fmcq}
\bar{f}_{{\rm M},c}(q) \;=\; q^{-1}e^{-q^2}\,.
\eeq
The resulting distribution is trivially of the form (\ref{eq:allfit}), and is shown in Fig.~\ref{fig:distmod} as the red, dashed curve. 

\subsubsection{Power spectrum and Ly-$\alpha$ constraints}

The analytical determination of the phase space distributions in all cases allows us to map the WDM Ly-$\alpha$ constraints to the production of DM from moduli decay. The main hurdle consists in the evaluation of the second moment of the distribution in the case when the oscillation and the decay of $Z$ occur in different epochs,
\beq
\langle q^2\rangle \;=\; \begin{cases}
\Gamma(7/3)\,, & m_{Z}>\Gamma_Z>\Gamma_{\phi}\\[10pt]
e^{-q_Z^2}(1+q_Z^2) - q_Z^4E_{-4/3}(q_Z^2) + \dfrac{\Gamma(7/3)}{q_Z^{2/3}}\,, & m_Z>\Gamma_{\phi}>\Gamma_Z\,,\\[10pt]
1\,, & \Gamma_{\phi}>m_Z>\Gamma_Z\,.
\end{cases}
\eeq
Here $E_n(x)$ denotes the exponential integral function. Nevertheless, we find the following to be a good approximation,
\begin{equation}
    m_{\rm DM} \;\gtrsim\; 3.78\,{\rm keV}\,\left(\frac{m_{\rm WDM}}{3\,{\rm keV}}\right)^{4/3} \left(\frac{g_{*s}^0}{g_{*s}^{\rm reh}}\right)^{1/3}
\frac{m_Z}{T_{\rm dec}} \times \begin{cases}
\sqrt{\Gamma(7/3)}q_Z^{-4/3}\,, & \Gamma_Z>\Gamma_{\phi}\,,\\[10pt]
1\,, & \Gamma_{\phi}>\Gamma_Z\,.
\end{cases}
\label{eq:mDM_bound_MD}
\end{equation}
As expected, the limit on the DM mass is weakened if $Z$ decays during reheating, relative to $Z$ decay during radiation domination. In this case, DM is cooled down in two stages: from the redshift from $t_{\rm dec}$ to $t_{\rm reh}$ and from the subsequent redsift from the end of inflation to the present epoch. 

\begin{figure}[!t]
\centering
    \includegraphics[width=0.75\textwidth]{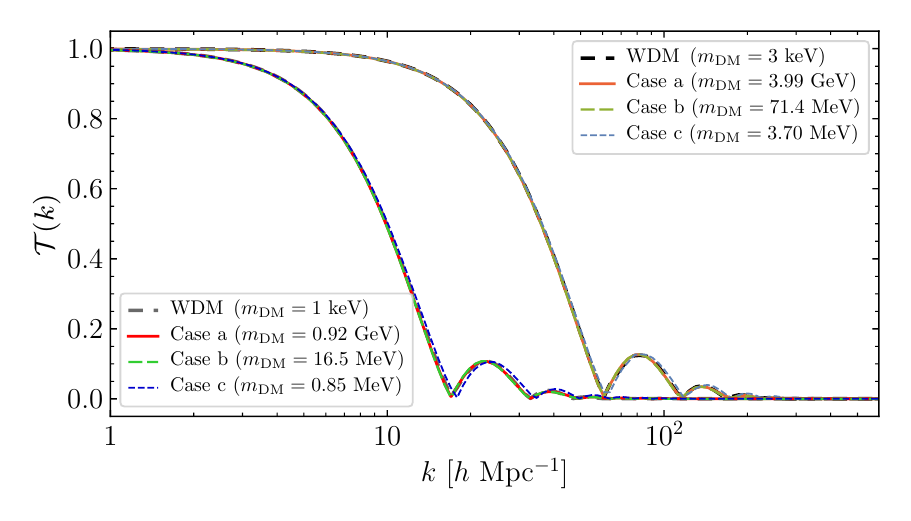}
    \caption{Linear transfer function for the scenario where DM is produced by moduli decay, for the cases (a), (b) and (c) described in Section~\ref{sec:moduli_PSD}. The transfer function for the Warm Dark Matter case is shown in gray and black dashed lines, with $m_\text{DM}=1,\,3$ keV, respectively, for comparison. The numerical values chosen for $m_{\rm DM}$ are estimated from Eq.~(\ref{eq:mDM_bound_MD}), with $m_Z$ and $T_{\rm dec}$ given by the values that saturate the Ly-$\alpha$ and abundance constraints for the strongly stabilized Polonyi scenario discussed in Section~\ref{sec:modulipheno}, shown as stars in Fig.~\ref{fig:moduliplot} (see text for details). }
    \label{fig:TF_MD}
\end{figure}

Fig.~\ref{fig:TF_MD} shows the transfer function for stabilized modulus decay compared to WDM with $m_{\rm WDM}=1$ and $3\,{\rm keV}$, for the three cases discussed in this section. The overlap between NCDM and WDM is good for all shown scales, albeit a slight shift can be observed for $m_{\rm WDM}=1\,{\rm keV}$. As Fig.~\ref{fig:transfer_function_residual} shows, the relative difference is in all cases $\lesssim 1\%$. The DM masses are taken from (\ref{eq:mDM_bound_MD}), where the modulus mass and decay temperature are in turn chosen to be $m_Z\simeq 3\times 10^6\,{\rm GeV}$ and $T_{\rm dec}\simeq 1\,{\rm GeV}$ for case (a), $m_Z\simeq 5\times 10^7\,{\rm GeV}$ and $T_{\rm dec}\simeq 800\,{\rm GeV}$ for case (b), and $m_Z\simeq 3\times 10^8\,{\rm GeV}$ and $T_{\rm dec}\simeq 10^5\,{\rm GeV}$ for case (c). These values are motivated by our discussion of the phenomenology of a strongly stabilized Polonyi-like modulus, in Sec.~\ref{sec:modulipheno}. For these choices of the $Z$ mass and decay temperature both the Lyman-$\alpha$ bound and the closure fraction bound $\Omega_{\chi}h^2\simeq 0.1$ are saturated (see Fig.~\ref{fig:moduliplot}).

\subsubsection{Relic density and phenomenology}\label{sec:modulipheno}

We now consider the possible phenomenological consequences of the Ly-$\alpha$ bound on $m_{\rm DM}$ found above. We first determine the DM relic abundance from stabilized moduli decays. Integration of Eqs.~(\ref{eq:modulifa}), (\ref{eq:modulifb}) and (\ref{eq:modulifc}) provides the following expression for the late-time DM number density,
\beq
n_{\chi}(t_0) \;=\; 4 {\rm Br}_{\chi} \left(\frac{Z_0}{m_Z}\right)^2 \times \begin{cases}
\left(\dfrac{\Gamma_Z}{m_Z}\right)^2 T_{\star,a}^3 \,, & m_{Z}>\Gamma_Z>\Gamma_{\phi}\\[10pt]
\left(\dfrac{\Gamma_{\phi}}{m_Z}\right)^2 \left(\dfrac{\Gamma_Z}{\Gamma_{\phi}}\right)^{3/2} T_{\star,b}^3\,, & m_Z>\Gamma_{\phi}>\Gamma_Z\,,\\[10pt]
\left(\dfrac{\Gamma_Z}{m_Z}\right)^{3/2} T_{\star,c}^3\,, & \Gamma_{\phi}>m_Z>\Gamma_Z\,.
\end{cases}
\eeq
As mentioned above, the discontinuity in the phase space distribution for $\chi$ is not inherited by the number density, justifying our approximations. 
We emphasize that our results are valid only if the field  $Z$ does not dominate the energy budget of the Universe at any time. For the first scenario, decay before reheating, this is ensured for $Z_0\ll M_P$, since if $\rho_{\rm osc}< \rho_{\phi}(t_{\rm osc})$ then it will continue being so until the decay of $Z$. For the other two cases we must ensure that the energy density in radiation, $\rho_r$, is always greater than $\rho_Z$. Since the oscillating modulus redshifts more slowly than the background radiation, it is sufficient to enforce this condition at $Z$-decay. With $\rho_Z(t_{\rm dec}) \simeq \frac{1}{2}m_Z^2Z_0^2 (a_{\rm osc}/a_{\rm dec})^3$ and $\rho_r(t_{\rm dec})\simeq \Gamma_{\phi}^2 M_P^2 (a_{\rm reh}/a_{\rm dec})^4$, we can evaluate the scale factors explicitly to obtain that
\beq
\frac{\rho_Z(t_{\rm dec})}{\rho_r(t_{\rm dec})} \;\simeq \frac{1}{2}\left(\frac{Z_0}{M_P}\right)^2 \times \begin{cases}
\left(\dfrac{\Gamma_{\phi}}{\Gamma_Z}\right)^{1/2}\,, & m_Z > \Gamma_{\phi}>\Gamma_Z\\[10pt]
\left(\dfrac{m_Z}{\Gamma_Z}\right)^{1/2}\,, & \Gamma_{\phi}>m_Z>\Gamma_Z\,.
\end{cases}
\eeq
This ratio must be $<1$ if our present analysis is to be valid. Otherwise, a $Z$-dominated epoch occurs, and the bound (\ref{eq:mdmmodulifirst}) applies. Note that for a branching ratio ${\rm Br}_{\chi}=1$, the condition $\rho_Z\ll \rho_r$ is necessary to obtain the observed DM abundance. For example, saturating the Ly-$\alpha$ bound (\ref{eq:mDM_bound_MD}) one obtains $\Omega_{\chi}h^2 \sim 270 (\rho_Z/\rho_r)$ for cases b and c. \par\medskip

We now consider as a proof-of-concept example a particular realization of modulus stabilization, corresponding to a strongly stabilized Polonyi field\footnote{The Polonyi field, if left unstabilized, is an example of a problematic modulus for BBN that can arise in $\mathcal{N}=1$ supergravity~\cite{Polonyi:1977pj,Goncharov:1984qm,Kawasaki:1995cy,Endo:2006zj,Nakamura:2006uc}. This field, responsible for the breaking of supersymmetry, communicates with the SM through Planck-suppressed interactions. It is also relatively light: its mass of the order of the gravitino mass, which in turn is parametrically related to the scale at which supersymmetry is broken. Moreover, typically its initial misalignment is $\mathcal{O}(M_P)$.} in $\mathcal{N}=1$ supergravity, stabilized by the non-minimal addition to the K{\"a}hler potential $\Delta K = - (Z\bar{Z})^2/\Lambda_Z^2$~\cite{Dine:2006ii,Kitano:2006wz,Kallosh:2006dv,Abe:2007yb,Fan:2011ua,Evans:2013nka}. For our purposes it is sufficient to note the following values of the Polonyi modulus mass, its misalignment, and its decay rate
\beq\label{eq:stabpolonyi}
m_Z \;=\; \sqrt{12}\,m_{3/2} \left(\frac{M_P}{\Lambda_Z}\right)\,, \qquad Z_0 \;=\; \frac{\Lambda_Z^2}{\sqrt{6}M_P}\,,\qquad \Gamma_Z\;=\; \frac{3\sqrt{3}m_{3/2}^3M_P^3}{\pi \Lambda_Z^5}\,.
\eeq
Here $m_{3/2}\gtrsim \mathcal{O}(10\,{\rm TeV})$ is the gravitino mass for this particular case of gravity-mediated supersymmetry breaking, and $\Lambda_Z\ll M_P$. Hence, the entropy production problem is averted by simultaneously increasing the $Z$ mass well above the electroweak scale, by reducing the misalignment to deep sub-Planckian values, and by enhancing the decay rate. The dominant decay channel of $Z$ is to two gravitinos, which then subsequently decay into the lightest neutralino. Although in this example this decay chain implies that the (rescaled) DM distribution will not be exactly given by the $\bar{f}_{{\rm M},i}(q)$, the scaling of the Ly-$\alpha$ constraint will be maintained up to $\mathcal{O}(1)$ corrections (see Section~\ref{sec:nonthermaldecay}).

\begin{figure}[!t]
\centering
    \includegraphics[width=0.75\textwidth]{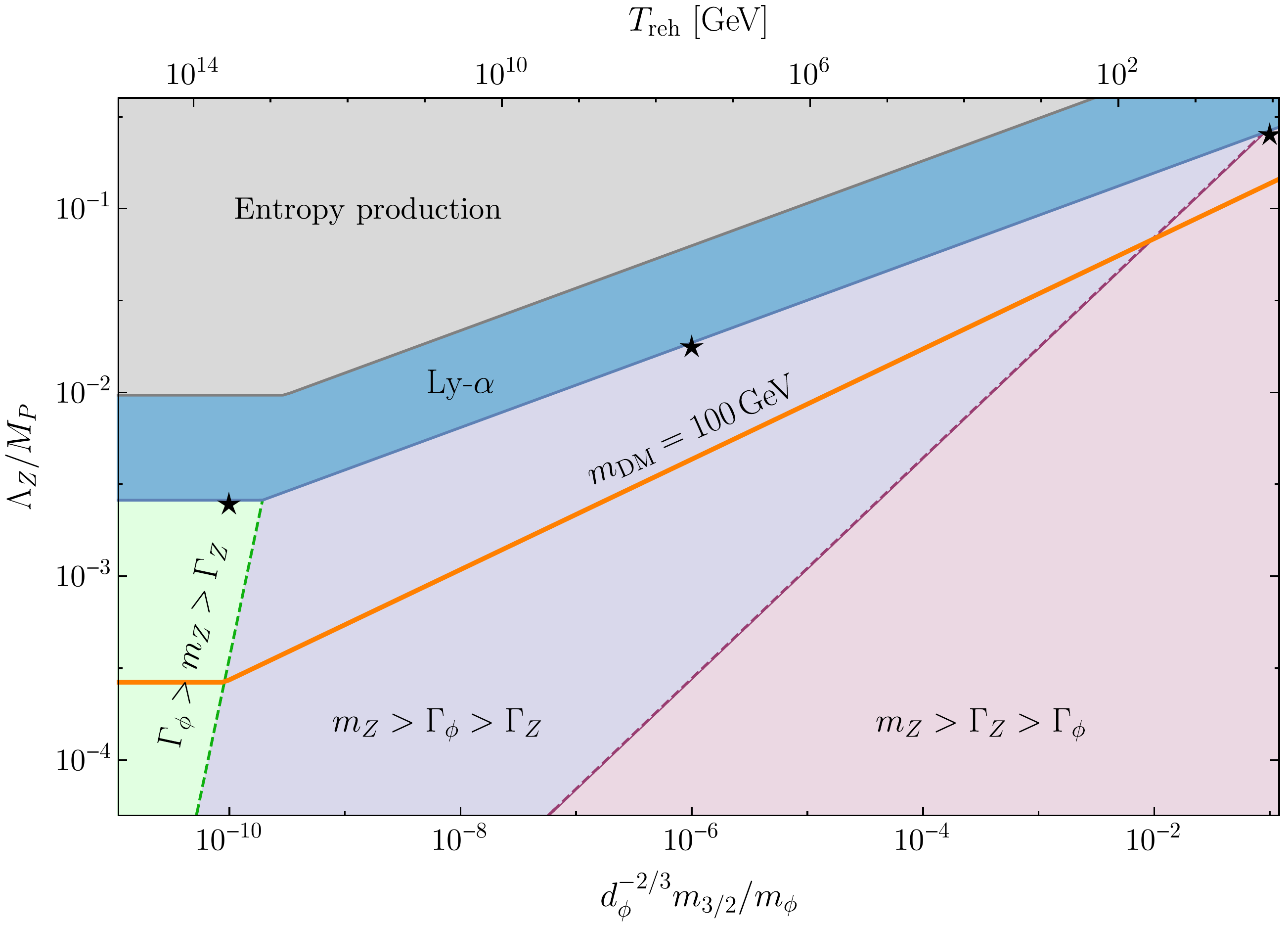}
    \caption{Allowed range for $\Lambda_Z$ as a function of $d_{\phi}^{-2/3}m_{3/2}/m_{\phi}$, where $\Gamma_{\phi}=d_{\phi}^2m_{\phi}^3/M_P^2$, for the stabilized modulus defined by (\ref{eq:stabpolonyi}). Shown are the regions excluded by $Z$-domination (entropy production) and by the Ly-$\alpha$ constraint, assuming ${\rm Br}_{\chi}=1$. The allowed parameter space is divided into the regions where $Z$ oscillates and decays after reheating (left), where it begins oscillations during reheating, and decays after reheating (middle), and where it oscillates and decays during reheating (right). The orange curve corresponds to $\Omega_{\chi}h^2=0.1$ for $m_{\rm DM}=100\,{\rm GeV}$. Above it, DM is overproduced. The stars correspond to the points selected to construct the transfer functions shown in Fig.~\ref{fig:TF_MD}. Where necessary, the gravitino mass is chosen $m_{3/2}=10^{-13}M_P$. See~\cite{Evans:2013nka} for further details.}
    \label{fig:moduliplot}
\end{figure}
Fig.~\ref{fig:moduliplot} shows the allowed parameter space for $\Lambda_Z$ as a function of the quantity $d_{\phi}^{-2/3}m_{3/2}/m_{\phi}$, where $\Gamma_{\phi}=d_{\phi}^2m_{\phi}^3/M_P^2$. Here $d_{\phi} \lesssim \mathcal{O}(10^{-1})$ includes the inflaton-matter (or radiation) couplings and the phase space factors of the width. This parametrization is chosen to coincide with that of~\cite{Evans:2013nka}, and is inspired by the Planck-suppressed decays which are a generic feature of supersymmetric reheating (see e.g.~\cite{Endo:2006xg,Ellis:2015kqa}). As it can be seen, when the stabilization scale is close to the Planck scale, the modulus ceases to be strongly stabilized, and it dominates the energy budget of the Universe after inflation. This is averted for \beq
\Lambda_Z \;\lesssim\; M_P\times \begin{cases}
1.5\left(\dfrac{m_{3/2}^3}{M_P^2\Gamma_{\phi}}\right)^{1/13}\,, & m_Z>\Gamma_{\phi}>\Gamma_Z\,,\\[10pt]
1.4\left(\dfrac{m_{3/2}}{M_P}\right)^{1/6}\,, & \Gamma_{\phi}>m_Z>\Gamma_Z\,.
\end{cases}
\eeq
In this figure we have also shown the domain restricted by Ly-$\alpha$ observations. We observe that it extends 
the disallowed region (due to entropy production) by about an order of magnitude in $\Lambda_Z$. Its boundary, and the orange line for which the observed DM abundance is obtained for $m_{\rm DM}=100\,{\rm GeV}$, are determined through the following expression,
\begin{align}\notag
\Omega_{\chi}h^2 \;\simeq\; 0.1 &\left(\frac{106.75}{g_{*s}^{\rm reh}}\right)^{1/4} \left(\frac{m_{\chi}}{100\,{\rm GeV}}\right) \\
& \times \begin{cases}
\left(\dfrac{\Lambda_Z}{6.2\times 10^{14}\,{\rm GeV}}\right)^{9/2} \left(\dfrac{10^{-13}M_P}{m_{3/2}}\right)^{-1/2} \,, &\Gamma_{\phi}>m_Z\,,  \\[10pt]
d_{\phi}  \left(\dfrac{\Lambda_Z}{2.4\times 10^{15}\,{\rm GeV}}\right)^5 \left(\dfrac{m_{\phi}}{3\times10^{13}\,{\rm GeV}}\right)^{3/2}\left(\dfrac{10^{-13}M_P}{m_{3/2}}\right)\,, & m_Z>\Gamma_{\phi}\,.
\end{cases}
\end{align}
In the parameter range shown in the figure, the Ly-$\alpha$ and DM abundance constraints are simultaneously saturated for 
\beq
m_{\rm DM} \;\simeq\; 3.5\,{\rm MeV}\,\left(\frac{m_{\rm WDM}}{3\,{\rm keV}}\right)^{4/3} \times \begin{cases}
\left(\dfrac{1.3\times 10^{-5}}{d_{\phi}}\right)^{3/13} \left(\dfrac{10^{-13}M_P}{m_{3/2}}\right)^{2/13}\,, & m_Z>\Gamma_{\phi}>\Gamma_Z\,,\\[10pt]
1 \,, & \Gamma_{\phi}>m_Z>\Gamma_Z\,,
\end{cases}
\eeq
assuming ${\rm Br}_{\chi}=1$. For $d_{\phi}^{-2/3}m_{3/2}/m_{\phi}=\{10^{-1},10^{-6},10^{-10}\}$, $m_{\rm DM}\simeq \{3.99\,{\rm GeV}, 71.3\,{\rm MeV}, \allowbreak 3.70\,{\rm MeV}\}$, c.f.~Fig.~\ref{fig:TF_MD}.  We finish by noting that for this particular stabilization scenario, the Ly-$\alpha$ constraint is irrelevant compared to the requirement that $\Omega_{\chi}h^2\simeq 0.1$ assuming electroweak-scale LSP masses. Nevertheless, the power spectrum bound may be relevant for alternative constructions in which the modulus mass and the DM mass are independent.
%


\section{Freeze-in via decay}\label{sec:freezeindecay}

In the previous section we considered the production of DM from the decay of the spatially homogeneous 
condensate. We now extend our discussion to decays of particles with distributions populated above the zero-momentum mode. Specifically, we will determine the phase space distribution and the mass lower bound for DM produced from the decay of a thermalized relic, and from the decay of a non-thermalized inflaton decay product. As in all cases, we will assume that DM interactions are sufficiently suppressed to prevent it from reaching kinetic and/or chemical equilibrium. For this reason we dub this scenario freeze-in through decays~\cite{Hall:2009bx}.

\subsection{Thermal decay}\label{sec:thermaldecay}

\subsubsection{DM phase space distribution}
\label{sec:thermal_decay_PSD}

Let us first consider the decay of a population of particles in thermal equilibrium, which decays during radiation domination totally or partially into DM. For definiteness we will assume again that the unstable particle, denoted here by $A$, decays to DM, $\chi$, via a two-body channel, $A \rightarrow \chi+\psi$.

The integration of the corresponding collision term can be performed in complete analogy to the inflaton decay scenario (see Eq.~(\ref{eq:Cdecay})). Noting in particular that, for a two-body decay, the unpolarized amplitude squared is determined solely by the masses of the initial and final state particles, we can write
\begin{align}\notag
\mathcal{C}[f_{\chi}(p,t)] \;&=\; \frac{ |\mathcal{M}|^2_{A\rightarrow \chi\psi}}{2p_0} \int \frac{\diff ^3 {\boldsymbol{k}}}{(2\pi)^3 2k_0} \frac{g_{\psi} \diff^3 {\bp}_{\psi}}{(2\pi)^3 2p_{\psi}^0} (2\pi)^4 \delta^{(4)}(k-p-p_{\psi}) f_{A}(k_0)\\ \label{eq:Cfordecay}
&=\; \frac{{\rm Br}_{\chi}\Gamma_{A}m_{A}}{p_0\sqrt{p_0^2 - \mncdm^2}} \int_{k_{-}}^{k_{+}} \diff k_0\, f_{A}(k_0)\,,
\end{align}
where
\begin{align}\notag
2 \mncdm^2k_{\pm} \;=\; p_0 &(m_{A }^2 +\mncdm^2-m_{\psi }^2)\\
& \pm \sqrt{ (p_0^2-\mncdm^2 )  (m_{A }^4+\mncdm^4+m_{\psi }^4-2 \mncdm^2 m_{\psi }^2-2 \mncdm^2 m_{A }^2-2 m_{\psi }^2 m_{A }^2 )}\,.
\end{align}
Note that up to this point no assumptions have been made regarding the form of $f_A$. For our exploration of the decay of a thermalized relic $A$ into 
DM, we can assume that $m_{A}\gg \mncdm,m_{\psi}$, and substitute a thermal Bose-Einstein (BE) of Fermi-Dirac (FD) form for $f_A$,
\beq
f_A(k_0) \;=\; \frac{1}{e^{k_0/T}\pm 1}\,.
\eeq
Substitution into (\ref{eq:Cfordecay}) yields the following collision term,
\begin{align} \notag
\mathcal{C}[f_{\chi}(p,t)] \;&\simeq\; \frac{{\rm Br}_{\chi}\Gamma_{A}m_{A}}{p^2} \int_{p+\frac{m_{A}^2}{4p}}^{\infty} \frac{\diff k_0}{e^{k_0/T} \pm 1}\\ \label{eq:Cthdec}
&= \; (\pm) \frac{{\rm Br}_{\chi}\Gamma_{A}m_{A}T}{p^2} \ln\left[1 \pm \exp\left( -\frac{p}{T} - \frac{m_{A}^2}{4pT}\right) \right]\,.
\end{align}
Disregarding the inverse decay process, and recalling the relation between time and temperature during radiation domination, 
\beq\label{eq:HRD}
H \;=\; \left(\frac{\pi^2 g_{*\rho}(T)}{90}\right)^{1/2} \frac{T^2}{M_P} \;\simeq\; \frac{1}{2t}\,,
\eeq
the solution of the transport equation (\ref{eq:boltzmanneq}) is a straightfoward application of the freeze-in solution (\ref{eq:Cgensol}). After some algebraic manipulation, the DM phase space distribution can be cast in the following form~\cite{Shaposhnikov:2006xi,Petraki:2007gq}
\begin{align}
 f_\chi \left(p,T\right) \,=\, & (\pm) {\rm Br}_{\chi} \frac{\Gamma_{A}T^2 M_{\rm Pl}}{ p^2 m_{A}^2}  \left(\dfrac{90}{ \pi^2} \right)^{1/2} g_{*s}^{2/3}(T)\int_0^{m_A/T} \diff x \,  x^2 \, g_{*s}^{-2/3}(m_A/x) \, g_{*\rho}^{-1/2} (m_A/x)  \nonumber   \\&  \times   \left(1-\dfrac{1}{3}\dfrac{\diff \log g_{*s}}{\diff \log x} \right) \ln\left[1 \pm \exp\left( -\dfrac{p}{T} \left( \dfrac{g_{*s}(m_A/x)}{g_{*s}(T)}\right)^{\frac{1}{3}} - \frac{x^2T }{4p} \left( \dfrac{g_{*s}(T)}{g_{*s}(m_A/x)}\right)^{\frac{1}{3}}\right) \right]\,.
 \label{eq:TDexact}
\end{align}
Such expression is valid up to the decoupling temperature $T>T_{\rm dec}\sim m_A$ below which the dark matter production from the thermal bath is negligible.

\begin{figure}[!t]
\centering
    \includegraphics[width=0.7\textwidth]{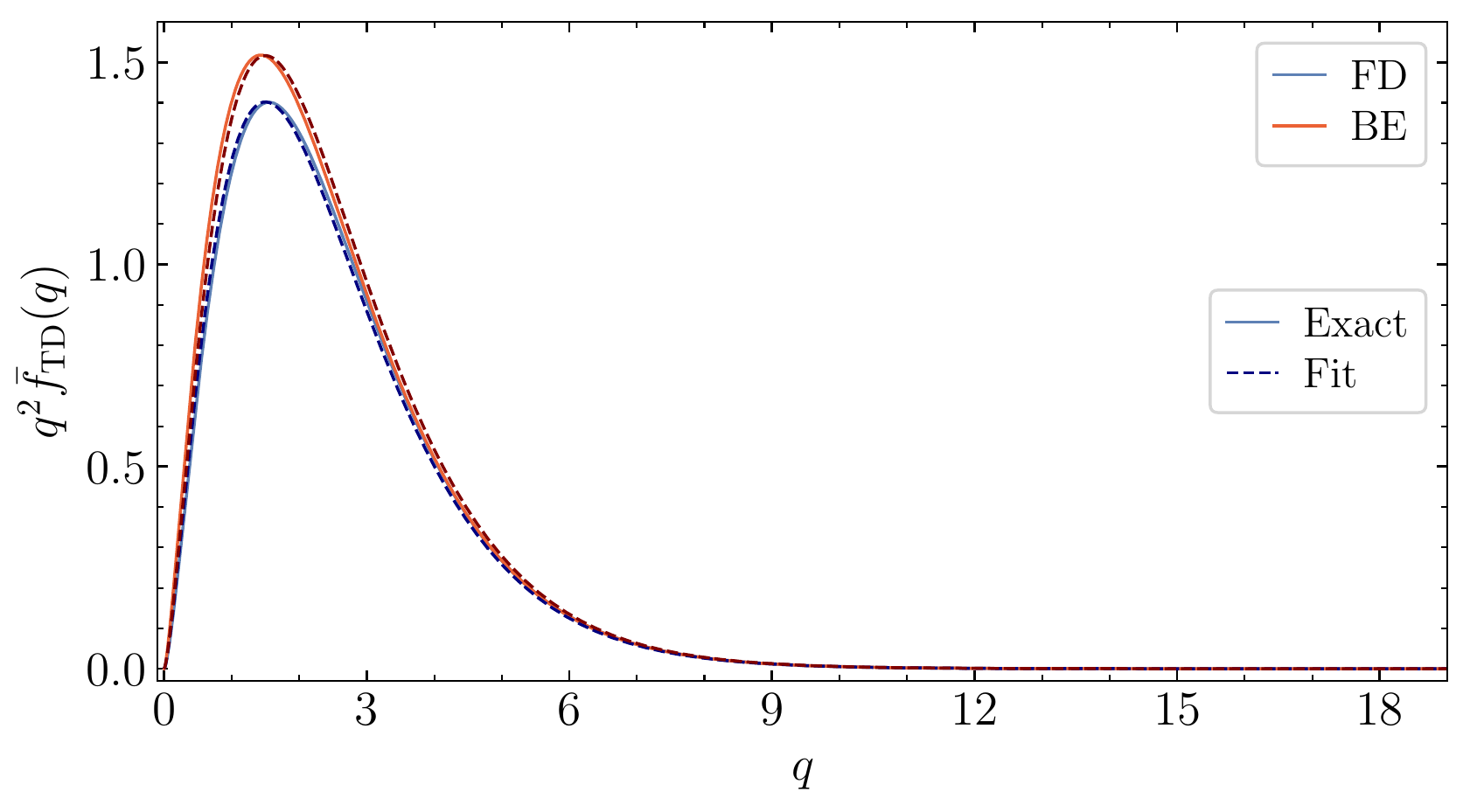}
    \caption{The rescaled distribution function $\bar{f}_{{\rm TD}}$, defined in (\ref{eq:resdistTD}), as a function of the rescaled momentum $q=p/T$, assuming $T\ll m_{A} \ll T_0$. Solid: the numerically computed phase space distributions for a fermionic (blue) or bosonic (red) decaying thermalized particle. Dashed: the phenomenological fits (\ref{eq:TDphenfit}).}
    \label{fig:distTD}
\end{figure}
A closed form for $f_{\chi}$ for either bosonic or fermionic $A$ is not available, and (\ref{eq:TDexact}) must be integrated numerically. These distributions are presented in Fig.~\ref{fig:distTD} in the limit when $T\ll m_{A} \ll T_{\rm reh}$,  by neglecting the temperature evolution of the effective degrees of freedom during production, in terms of the rescaled distribution
\beq  \label{eq:resdistTD}
\bar{f}_{{\rm TD}}(q) \;\equiv\; \sqrt{\frac{g_{*s}^{\rm dec}}{90}}\frac{\pi m_{A}^2}{{\rm Br}_{\chi}\Gamma_{A} M_P} f_{\chi}(q)\,.
\eeq
Here $q=p/T$, noting that (\ref{eq:HRD}) can be extended up to recombination, where $g_{*s}\simeq g_{*s}^{0}$. The continuous red (blue) curve corresponds to a decaying fermion (boson) $A$. It is worth noting that the difference between the two curves is relatively small, which suggests that a phenomenological Maxwell-Boltzmann-like fit could describe these distributions. Indeed, Fig.~\ref{fig:distTD} also shows two dashed curves which correspond to the following fitting functions,
\beq\label{eq:TDphenfit}
\bar{f}_{{\rm TD}}(q) \;\simeq\; q^{-1/2} e^{-q} \times \begin{cases}
3.38\,, & \text{FD}\,,\\
3.77\,, & \text{BE}\,.
\end{cases}
\eeq
Save for the fitting factors, the functional form for this expression may trivially be obtained from (\ref{eq:TDexact}) in the Maxwell-Boltzmann limit, for which $\ln\left[1 \pm \exp\left(-\frac{p}{T} - \frac{x^2T}{4p}\right)\right] \rightarrow\pm \exp\left(-\frac{p}{T} - \frac{x^2T}{4p}\right)$ \cite{Heeck:2017xbu,Boulebnane:2017fxw,Kamada:2019kpe}. 
Worth noting is the mapping of the exponential tail from the thermalized progenitor $A$ to the daughter particles. Nevertheless, the low-momentum behavior is different, manifesting the lack of thermal equilibrium in the $\chi$ sector. This distribution is of the form (\ref{eq:allfit}), with $\gamma=1$.

\subsubsection{Power spectrum and Ly-$\alpha$ constraints}

\begin{figure}[!t]
\centering
    \includegraphics[width=0.75\textwidth]{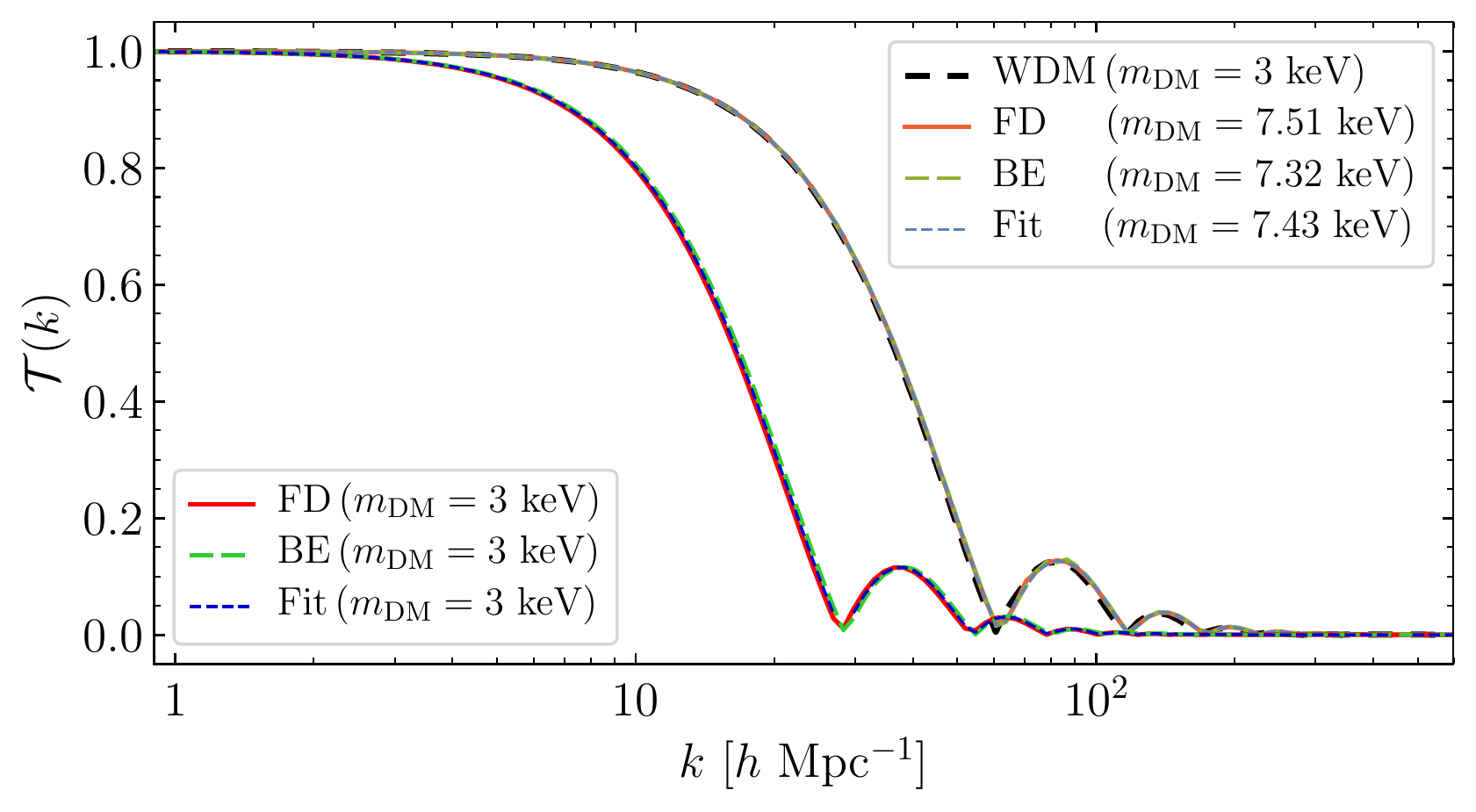}
    \caption{Linear transfer function for the scenario where DM is produced by decay of a thermalized particle (denoted by $A$ in the main text), assuming a Fermi-Dirac (FD), Bose-Einstein (BE) or a fitting phase space distribution as described in Section~\ref{sec:perturbative_inflaton_decay}, by taking the mass estimated in Eq.~(\ref{eq:mDM_bound_TD}). The transfer function for the WDM case is shown for comparison in a black dashed line. Also depicted here are the transfer functions for FD, BE and fitted phase space distribution (\ref{eq:TDphenfit}) with identical masses $m_\text{DM}=~3\ \text{keV}$.}
    \label{fig:TF_TD}
\end{figure}

The fact that the phase space distribution of $\chi$ is quasi-thermal suggests that the power spectrum should match the one of WDM. Fig.~\ref{fig:TF_TD} attests the reliability of this matching. The leftmost set of curves shows the transfer functions for the thermal decay cases with BE or FD initial states with masses determined by Eq.~(\ref{eq:mncdm_fromeos}), which in this case corresponds to the following rescaled bound,
\beq
    \hatmncdm \, \gtrsim \,  \,  \left( \dfrac{m_\text{WDM}}{3~\text{keV}} \right)^{4/3} \left( \dfrac{106.75}{g_{* s}(m_A)} \right)^{1/3} \times \begin{cases}
7.51~\text{keV}\,, \quad & {\rm FD}\,,\\
7.32~\text{keV}\,, \quad & {\rm BE}\,,\\
7.43~\text{keV}\,, \quad & {\rm Fit}\,.
\end{cases}
\label{eq:mDM_bound_TD}
\eeq
Here `fit' stands for both the FD and BE approximations (\ref{eq:TDphenfit}), which differ only by a $q$-independent numerical factor. The overlap of these transfer functions with the WDM result is evident in the whole range of scales shown in the figure, the relative deviation being $\simeq 1\%$ at $k_{1/2}^{\rm WDM}$ (see Fig.~\ref{fig:transfer_function_residual}). In Fig.~\ref{fig:TF_TD} we also show the form of $\mathcal{T}(k)$ if we consider a smaller DM mass, and ignore the difference in statistics. In this case, all three curves  shift to the left, as expected, but the difference between them remains small. As mentioned earlier, this is the result of the relatively minimal dependence of $f_{\chi}$ on the spin of the decaying particle $A$. 

\subsubsection{Relic density and phenomenology}

In addition to the power spectrum constraint on the mass discussed above, one must address the limit from the DM abundance
which determines the normalization of the $\chi$ distribution function. Integration of $f_{\chi}$ gives the following expression for the DM number density at late times, $T\ll T_{\rm reh}$,
\beq
n_{\chi}(T) \;\simeq\; \sqrt{90} \frac{g_{\chi} {\rm Br}_{\chi} \Gamma_A M_P T^3}{2\pi^3 m_A^2}  g_{*s}(T)\left(  \dfrac{1}{g_{*s}^{\rm dec}}\right)^{3/2} \times \begin{cases}
4.58\,, \quad & {\rm FD}\,,\\
4.89\,, \quad & {\rm BE}\,.
\end{cases}
\eeq
Correspondingly, 
\begin{equation}
\Omega_{\chi}h^2 \;\simeq\;0.12 \,{\rm Br}_{\chi} \,\left( \dfrac{g_{\chi}}{2} \right) \left(\frac{106.5}{g_{*s}^{\rm dec}}\right)^{3/2} \left(\frac{m_{\rm DM} }{6\,{\rm keV}}\right) \left(\frac{\Gamma_A}{10^{-14}\,{\rm GeV}}\right) \left(\frac{1\,{\rm TeV}}{ m_A}\right)^2 \times \begin{cases}
1.17\,, \quad & {\rm FD}\,,\\
1.02\,, \quad & {\rm BE}\,.
\end{cases}
\end{equation}
Except for the number of degrees of freedom, which we consistently normalize to the SM value, the normalizations chosen in the previous equation are inspired by the decay of thermalized supersymmetric particles into light DM candidates, such as the Higgsino $\rightarrow$ axino + Higgs production process in $R$-parity violating DFSZ models~\cite{Bae:2011iw,Bae:2017dpt}, for which
\beq
\Gamma(\tilde{H}\rightarrow \tilde{a}+H) \;=\; \frac{1}{8\pi} \left(\frac{\mu}{f_a}\right)^2\mu\,,
\eeq
with the $\mu$-term parameter $\mu\sim 500\,{\rm GeV}$, and the Peccei-Quinn scale $f_a\sim 10^{10}\,{\rm GeV}$.
Similarly to the inflaton decay case, a mass-independent constraint on the branching ratio to DM from the decay of the thermalized $A$ could be derived. Nevertheless, this bound would not be universal, as the mass and width of $A$ are model dependent, as opposed to the inflaton decay case (see Eq.\ \eq{eq:brlim}).

\subsection{Non-thermal decay}\label{sec:nonthermaldecay}

\subsubsection{DM phase space distribution}
\label{sec:non-thermal_decay_PSD}

Let us now assume that the {particle $A$ whose decay produces the DM} interacts very weakly with the {SM}
and was produced via inflaton decay, but does not reach thermal equilibrium. Unlike in the previously studied thermal case, this particle cannot be assumed to be {produced abundantly}
in the thermal plasma 
{during the decay of the latter,}
Therefore, in principle the imprint that its decay leaves on its phase space distribution must be taken into account.

Disregarding the effect of Bose enhancement/Pauli blocking, and the inverse decay process, the Boltzmann equation satisfied by this non-thermal unstable relic is given by~\cite{Kawasaki:1992kg}
\beq\label{eq:fadecaya}
\frac{\partial f_A}{\partial t} - H p \frac{\partial f_A}{\partial p} \;=\; -\frac{m_A\Gamma_A}{\sqrt{m_A^2+p^2}}f_A\,.
\eeq
This equation can be exactly solved in the relativistic and non-relativistic regimes. In both cases the decay of $A$ proceeds exponentially in time. For this reason we will be content to approximate the evolution of $f_A$ as that of a free-streaming particle until its sudden decay, which occurs at
\beq \label{eq:RorNR}
t_{\rm dec} \;\simeq\; \begin{cases}
\Gamma_A^{-1}\,, & \dfrac{\Gamma_A}{H_A} \ll 1\,,\\[10pt]
\left(\dfrac{m_{\phi}\langle q_A\rangle}{2m_A\Gamma_A\Gamma_{\phi}^{1/2}}\right)^{2/3}\,, & \dfrac{\Gamma_A}{H_A} \gg 1\,.
\end{cases}
\eeq
Here $H_A$ denotes the Hubble parameter at the time when $A$ becomes non-relativistic. We have estimated the effective lifetime as the inverse of the mean $f_A$ prefactor in the right-hand side of (\ref{eq:fadecaya})~\cite{Chacko:2019nej}.

With the previous arguments in mind, for $t<t_{\rm dec}$ we write the collision term for $\chi$ (\ref{eq:Cfordecay}) as 
\begin{align} \notag
\mathcal{C}[f_{\chi}(p,t)] \;=\; \frac{4\pi^4 g_{*s}^{\rm reh} {\rm Br}_{\chi}{\rm Br}_{A}  \Gamma_{A} m_{A}}{5 g_{A}p^2} & \left(\frac{T_{\rm reh}}{m_{\phi}}\right)^4 \left(\frac{m_{\phi}}{2}\right)\left(\frac{a_{\rm reh}}{a(t)}\right)\\
&\times \int_{\left|\frac{2p}{m_{\phi}}\frac{a(t)}{a_{\rm reh}}-\frac{m_{A}^2}{2pm_{\phi}}\frac{a(t)}{a_{\rm reh}}\right|}^{\infty} \frac{z\,\diff z}{\sqrt{z^2 + \left(\frac{2m_A a(t)}{m_{\phi}a_{\rm reh}}\right)^2}}\, \bar{f}_{\rm R}\left(z\right)\,,
\end{align}
where the distribution for inflaton decay products $\bar{f}_\text{R}$, given in terms of the 3D momentum magnitude, was defined in (\ref{eq:resdistinf}). In this expression ${\rm Br}_A$ stands for the branching ratio of the decay from inflaton to $A$. Substitution into the general freeze-in solution (\ref{eq:Cgensol}) gives
\begin{align}\notag
f_{\chi}(p,t_{\rm dec}) \;=\;\; &\frac{8\pi^4 g_{*s}^{\rm reh}{\rm Br}_{\chi} {\rm Br}_{A} \Gamma_{A} m_A}{5g_{A} m_{\phi}} \left(\frac{T_{\rm reh}}{m_{\phi}}\right)^4 q_{\rm dec}^{-2} \\
\label{eq:fattdec}
& \times \int_{t_{\rm reh}}^{t_{\rm dec}} \diff t'\, \frac{a(t')}{a_{\rm reh}} \int_{ \left|q_{\rm dec}-\frac{1}{q_{\rm dec}}\left( \frac{m_A}{m_{\phi}}\frac{a(t')}{a_{\rm reh}}\right)^2\right|}^{\infty} \frac{z\,\diff z}{\sqrt{z^2 + \left(\frac{2m_A a(t')}{m_{\phi}a_{\rm reh}}\right)^2}}\, \bar{f}_{\rm R}(z)\,,
\end{align}
where $q_{\rm dec}=(2p/m_{\phi})(a_{\rm dec}/a_{\rm reh})$. The ratio
\begin{align}
\frac{m_A a(t)}{m_{\phi} a_{\rm reh}} \;\propto\; \frac{m_A}{\langle p\rangle}\,,
\end{align} 
quantifies how relativistic the distribution for $A$ is at a given moment of time. In particular, we define
\beq
\Ydec \;\equiv\; \frac{m_A a_{\rm dec}}{m_{\phi} a_{\rm reh}} \;=\; \left(\frac{g_{*s}^{\rm reh}}{g_{*s}^{\rm dec}}\right)^{1/3}\frac{m_A T_{\rm reh}}{m_{\phi} T_{\rm dec}} \;=\; \begin{cases}
\left(\dfrac{2H_A}{\Gamma_A}\right)^{1/2}\gg 1 \quad  &\text{for}\quad  \dfrac{\Gamma_A}{H_A} \ll 1\,,\\[10pt]
\left(\langle q_A\rangle \dfrac{3 H_A}{2\Gamma_A}\right)^{1/3}\ll 1 \quad  &\text{for}\quad \dfrac{\Gamma_A}{H_A} \gg 1\,.
\end{cases}\,.
\eeq
Extending the solution past $t_{\rm dec}$ we can write
\begin{align} \label{eq:fchind}
f_{\chi}(p,t)\, \diff^3\bp \;=\; \frac{24\pi^3 \sqrt{10 g_{*s}^{\rm reh}} {\rm Br}_{\chi} {\rm Br}_{A} \Gamma_{A} M_P}{5g_A m_A^2} \left(\frac{T_{\rm reh}}{m_{\phi}}\right)^2 \mathcal{F}(q,\Ydec) \left(\frac{a_0}{a(t)}\right)^3 \tncdm^3 \diff^3\boldsymbol{q}\,,
\end{align}
where
\beq\label{eq:anotherf}
\mathcal{F}(q,\Ydec) \;=\; q^{-2} \int_0^{\Ydec} \diff y\, y^2 \int_{\left|q-\frac{y^2}{q}\right|}^{\infty} \frac{z\,\diff z}{\sqrt{q^2 + 4y^2}}\, \bar{f}_{\rm R}(z) \;\simeq\; \begin{cases}
\bar{f}_{\rm D,NR}(q)\,,& \Ydec \gg 1\,,\\[10pt]
\dfrac{\Ydec^3}{3} \bar{f}_{\rm D,R}(q)\,, & \Ydec \ll 1\,.
\end{cases}
\eeq
Here $q$ and $\tncdm$ are the same as in (\ref{eq:resdistinf}) and (\ref{eq:TncdmR}). The rescaled distributions $\bar{f}_{\rm D,NR}$ and $\bar{f}_{\rm D,R}$ can be computed by making use of the fit approximation (\ref{eq:infphenfit}) for $\bar{f}_{\rm R}$. We obtain 
\begin{align} \label{eq:resdistND0}
\bar{f}_{\rm D,NR}(q) \;&\simeq\; 0.36\, q^{-1} \left[0.43 \,q\, \Gamma \left(\frac{1}{4},0.19\, q^2\right)-\Gamma \left(\frac{3}{4},0.19\, q^2\right)+2\, \Gamma \left(\frac{3}{4}\right)\right]\theta \big(\Ydec-q \big) \,, \\ \label{eq:resdistND}
\bar{f}_{\rm D,R}(q) \;&=\; q^{-2} \int_q^{\infty} \diff z\, \bar{f}_{\rm R}(z) \;\simeq\; 1.06\, q^{-2}\,\Gamma\left(-\frac{1}{4},0.74\, q^2\right)\,,
\end{align}
where $\Gamma(a,x)$ denotes the upper incomplete gamma function.\par\medskip
\begin{figure}[!t]
\centering
    \includegraphics[width=0.7\textwidth]{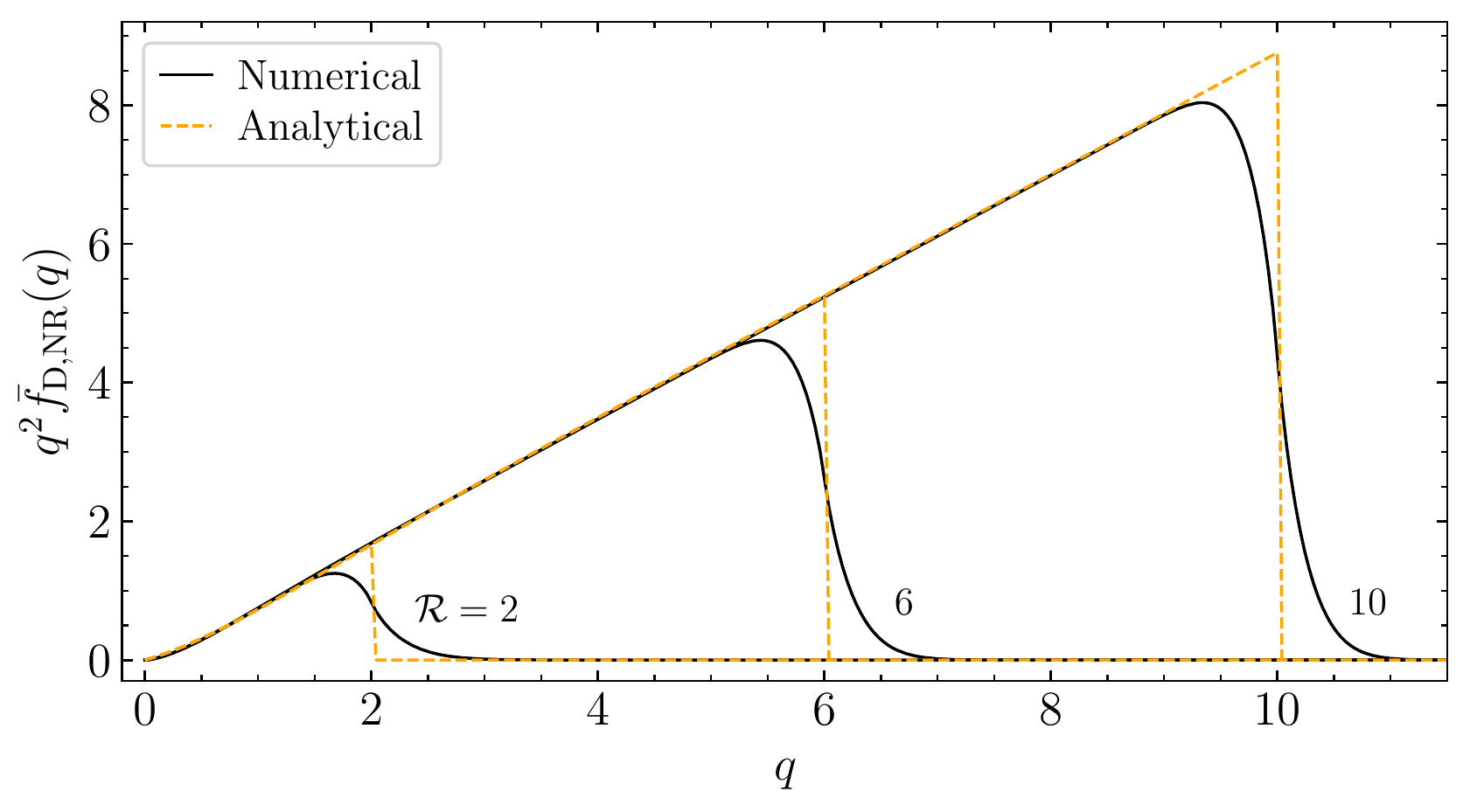}
    \caption{The rescaled distribution function $\bar{f}_{{\rm D,NR}}$, defined in (\ref{eq:resdistND0}), as a function of the rescaled momentum $q$ and the order parameter $\Ydec=(m_A/m_{\phi})(\Gamma_{\phi}/\Gamma_{A})^{1/2}$. Solid, black: numerically computed phase space distribution. Dashed, orange: the fit approximation (\ref{eq:resdistND0}).}
    \label{fig:distNTDNR}
\end{figure}

The DM phase space distribution corresponding to the decay of a non-relativistic particle $A$ is shown in Fig.~\ref{fig:distNTDNR} as a function of $q$ and  $\Ydec> 1$. The solid black line shows the result of the numerical integration of (\ref{eq:anotherf}). The distribution grows with an almost linear universal envelope, independent of the decay rate of $A$, until $q\sim \Ydec$, at which point the distribution sharply decreases. This non-universality of the cutoff prevents us from constructing a reasonable fit approximation of the form (\ref{eq:allfit}) for generic values of $\Ydec$. 
In the same figure, the orange dashed lines show the analytical approximation (\ref{eq:resdistND0}), which as can be seen is equivalent to imposing a hard cutoff at $q=\Ydec$ on the universal envelope.\footnote{The numerical distribution can be well fitted by substituting the $\theta$ function in Eq.~(\ref{eq:resdistND0}) by a logistic function.}

\begin{figure}[!t]
\centering
    \includegraphics[width=0.7\textwidth]{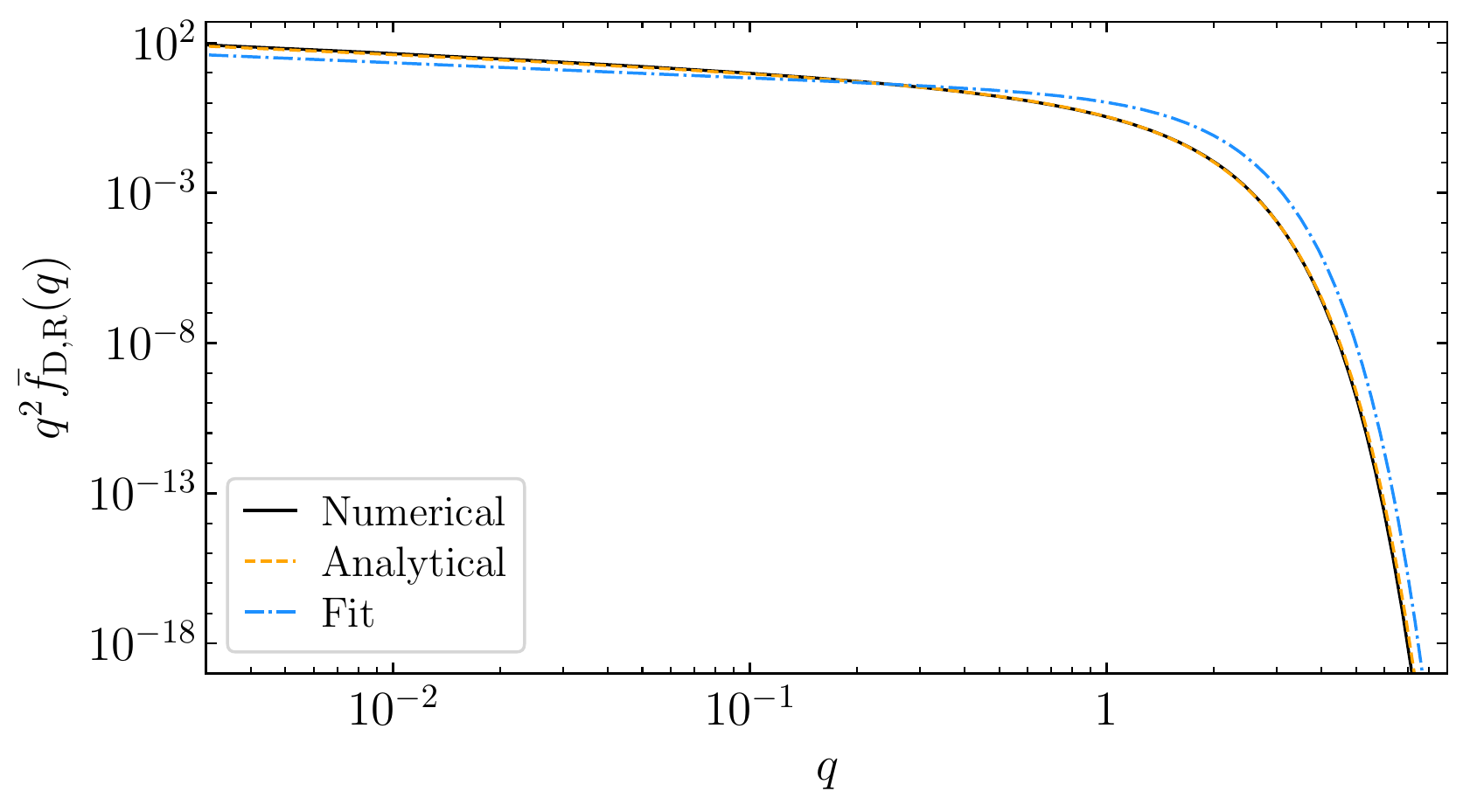}
    \caption{The rescaled distribution function $\bar{f}_{{\rm D,R}}$, defined in (\ref{eq:resdistND}), as a function of the rescaled momentum $q$. Solid, black: numerically computed phase space distribution. Dashed, orange: the analytical approximation (\ref{eq:resdistND}). Dashed-dotted, blue: the fit approximation (\ref{eq:appND2}).}
    \label{fig:distNTD}
\end{figure}

Fig.~\ref{fig:distNTD} shows the numerically computed relativistic distribution $\bar{f}_{\rm D,R}$ as the solid black curve, and the analytical approximation given by (\ref{eq:resdistND}) as the orange, dashed curve. In the same figure a `fit' approximation of the form (\ref{eq:allfit}) is also shown. This approximation is obtained by mimicking the asymptotic behavior of the gamma function at large and small $q$, while preserving the normalization, and is given by
\begin{align}\label{eq:appND2}
\bar{f}_{\rm ND}(q) \;&\approx\; 2.19 q^{-5/2} e^{-0.74 q^2}\,.
\end{align}
It is worth noting that in this case the Gaussian tail is of the same form as that of the parent unstable particle. It is important to emphasize that this distribution is obtained in the limit $\Ydec\rightarrow 0$, as we discuss below. 

\begin{figure}[!t]
\centering
    \includegraphics[width=0.73\textwidth]{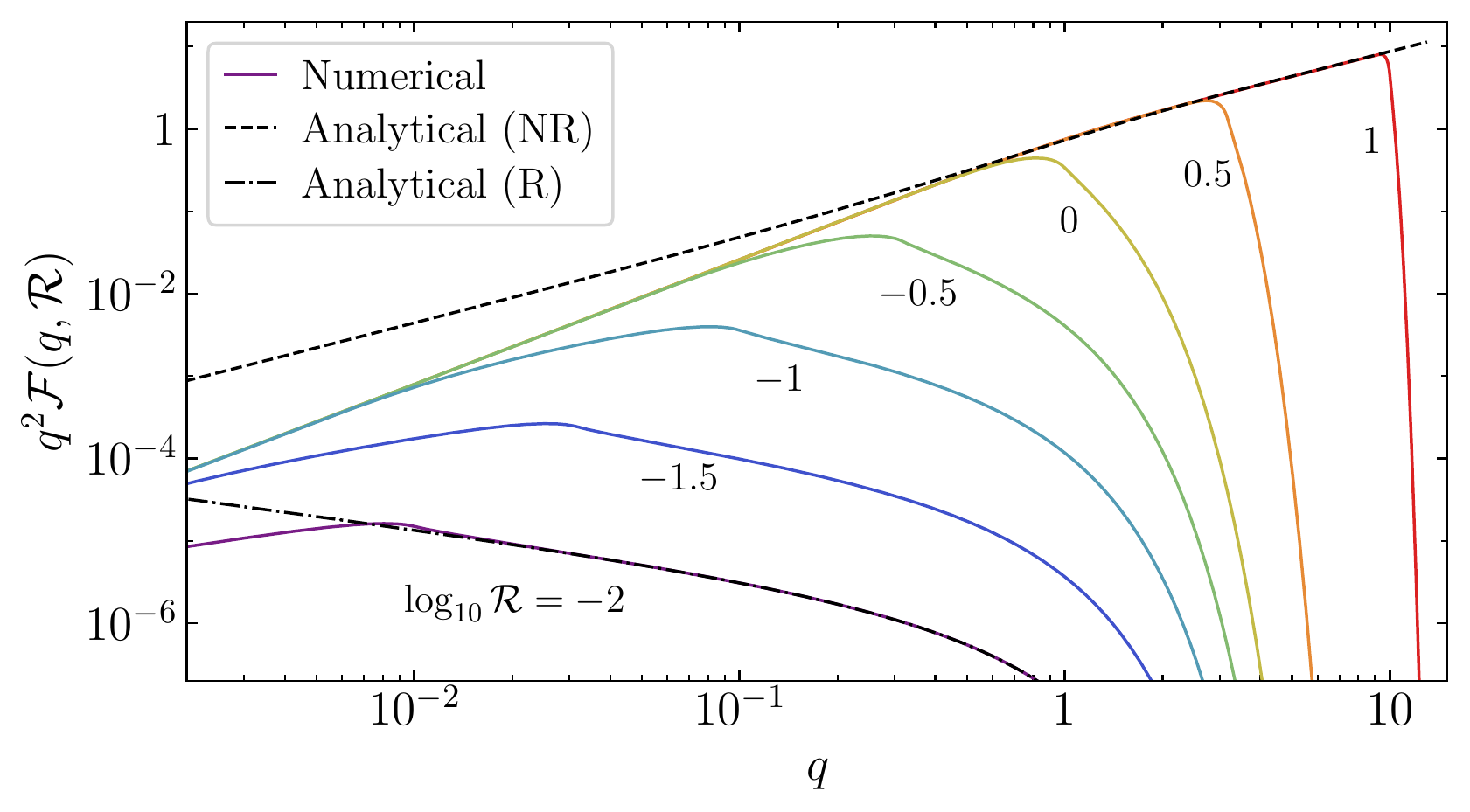}
    \caption{The function $\mathcal{F}(q,\Ydec)$, defined in (\ref{eq:anotherf}), as a function of the rescaled momentum $q$ and the decay parameter $\Ydec$. Solid: numerically computed distributions. Dashed: the non-relativistic analytical approximation (\ref{eq:resdistND0}) {for $\Ydec\gg 1$.}
Dashed-dotted: the relativistic analytical approximation (\ref{eq:resdistND}).}
    \label{fig:distNTDAll}
\end{figure}

Fig.~\ref{fig:distNTDAll} shows the form of the function $\mathcal{F}(q,\Ydec)$, defined in Eq.~(\ref{eq:anotherf}), for several values of $\Ydec$, ranging from $10^{-2}$ to 10. Here we can appreciate the transition between the relativistic and non-relativistic decay cases. In all cases
the phase space distribution peaks at $q\simeq \Ydec$, with a positive skew for a relativistic $A$, and a negative skew for non-relativistic $A$. For $\Ydec< 1$ the analytical approximation (\ref{eq:resdistND}) describes well the exact distribution for $q\gtrsim \Ydec$. For $\Ydec>1$, the non-relativistic approximation (\ref{eq:resdistND0}) is in turn a good fit for the exact distribution for $q\gtrsim 1/2$. 

\subsubsection{Power spectrum and Ly-$\alpha$ constraints}

For the distributions that we have derived, we can make use of (\ref{eq:mncdm_fromeos}) to determine the rescaling of the bound on the DM mass. For the case of a relativistic (R) decay we find that
\begin{align} \notag
   \hatmncdm \, \gtrsim \,  \, & \left( \dfrac{m_\text{WDM}}{3~\text{keV}} \right)^{4/3} \left( \dfrac{106.75}{g_{*s}^{\rm reh}} \right)^{1/3}\\ 
    &\times \left( \dfrac{m_\phi}{3\times 10^{13}~\text{GeV}} \right) \left( \dfrac{10^{10}~\text{GeV}}{T_{\text{reh}}} \right) \times \begin{cases}
1.23~\text{MeV}\,, ~ & {\rm Numerical}~{\rm (R)}\,,\\
1.26~\text{MeV}\,, ~& {\rm Analytical}~{\rm (R)}\,,\\
2.19~\text{MeV}\,, ~& {\rm Fit}~{\rm (R)}\,,\\
\end{cases}
\label{eq:mDM_bound_NTD}
\end{align}
while for the non-relativistic (NR) case,
\begin{align} \notag
   \hatmncdm \, \gtrsim \,  \, & \left( \dfrac{m_\text{WDM}}{3~\text{keV}} \right)^{4/3} \left( \dfrac{106.75}{g_{*s}^{\rm reh}} \right)^{1/3}\\ 
    &\times \left( \dfrac{m_\phi}{3\times 10^{13}~\text{GeV}} \right) \left( \dfrac{10^{10}~\text{GeV}}{T_{\text{reh}}} \right)\left( \dfrac{\Ydec}{6} \right)  \times \begin{cases}
16.1~\text{MeV}\,, ~ & {\rm Analytical}~{\rm (NR)}\,,\\
16.2~\text{MeV}\,, ~ & {\rm Numerical}~{\rm (NR)}\,.
\end{cases}
\label{eq:mDM_bound_NTD_NR}
\end{align}
The $\hatmncdm \propto \Ydec$ behavior is only correct for large $\Ydec \gg 1 $ but remains a reasonable approximation for $\Ydec \sim \mathcal{O}(1-10)$. We note here that the lower bound on the NCDM mass can be many orders of magnitude larger than the corresponding WDM bound, and it increases as the reheating temperature is decreased. This is expected, as in this case the parent particle is produced from inflaton decay (see Sec.~\ref{sec:lyainfdec}). The difference between the numerical and analytical results is minimal, consistent with the agreement between both curves in Fig.~\ref{fig:distNTDNR} and Fig.~\ref{fig:distNTD}. However, the fit approximation for the relativistic case provides a relatively poor approximation to the bound, overestimating it by a factor of $\sim1.8$.\par\bigskip
\begin{figure}[!t]
\centering
    \includegraphics[width=0.75\textwidth]{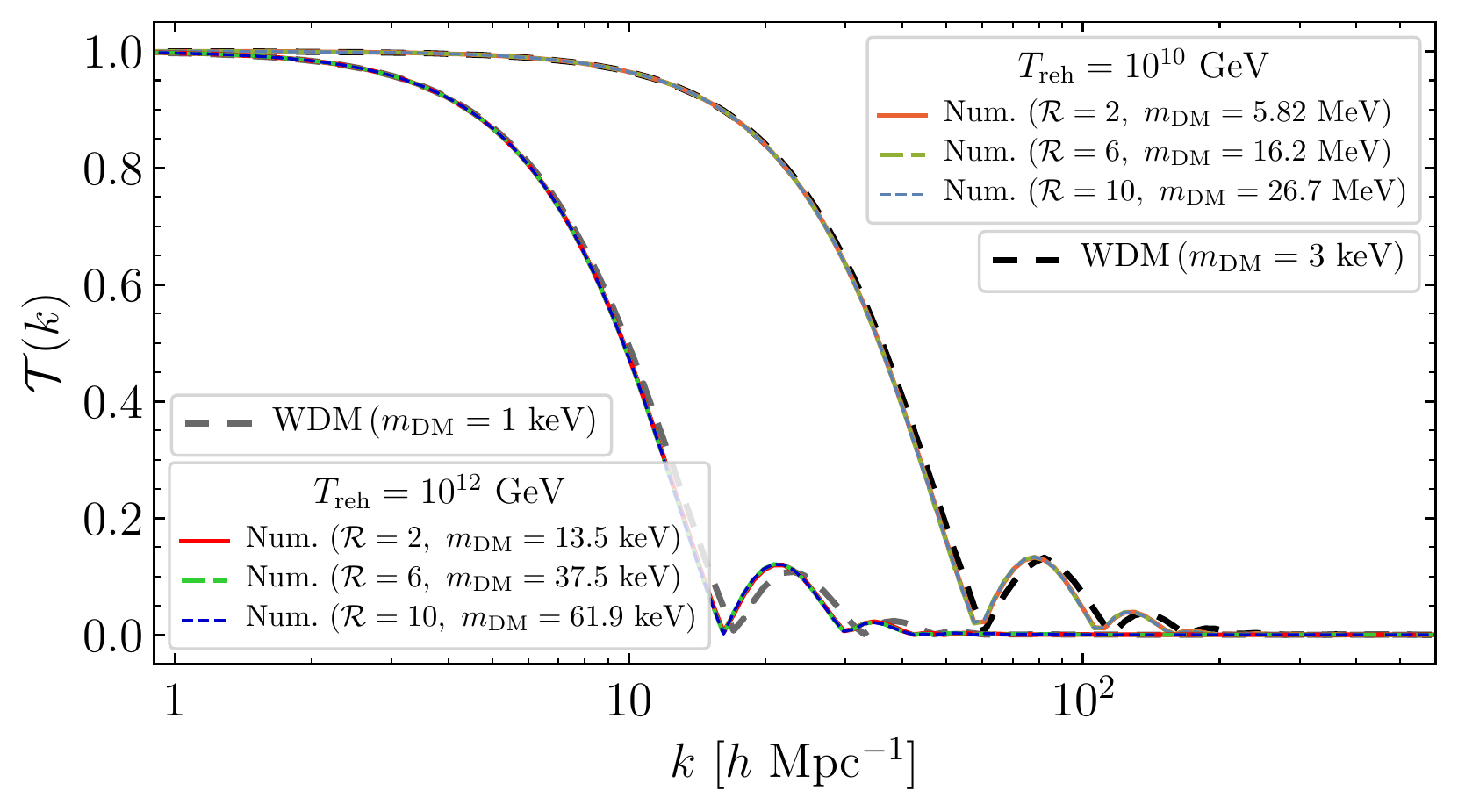}
    \caption{Linear transfer function for DM produced by the decay of a non-thermalized non-relativistic particle. We show here the numerical results for two sets of cosmological parameters: $T_{\rm reh}=10^{12}\,{\rm GeV}$ and $m_{\rm WDM}=1\,{\rm keV}$, and $T_{\rm reh}=10^{10}\,{\rm GeV}$ and $m_{\rm WDM}=3\,{\rm keV}$, making use of the rescaled bound  (\ref{eq:mDM_bound_NTD_NR}). For comparison we also show the transfer function for the corresponding WDM cases.}
    \label{fig:TF_NTDNR}
\end{figure}
\begin{figure}[!t]
\centering
    \includegraphics[width=0.75\textwidth]{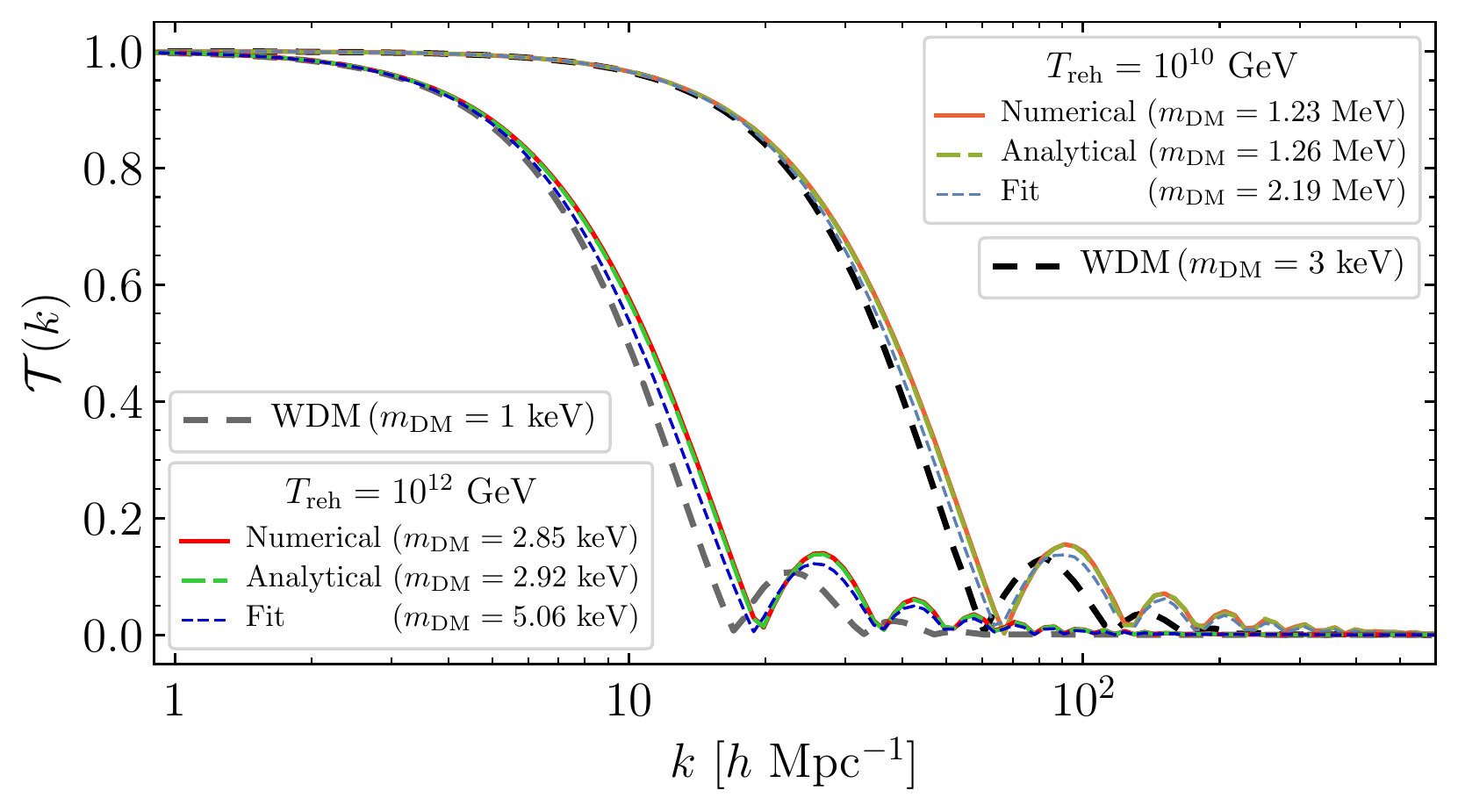}
    \caption{Linear transfer function for DM produced by the decay of a relativistic non-thermalized particle. We show here the numerical, analytical and fit approximations discussed in the text, for two sets of cosmological parameters: $T_{\rm reh}=10^{12}\,{\rm GeV}$ and $m_{\rm WDM}=1\,{\rm keV}$, and $T_{\rm reh}=10^{10}\,{\rm GeV}$ and $m_{\rm WDM}=3\,{\rm keV}$, making use of the rescaled bound  (\ref{eq:mDM_bound_NTD}). For comparison we also show the transfer function for the corresponding WDM cases.}
    \label{fig:TF_NTD}
\end{figure}
Figs.~\ref{fig:TF_NTDNR} and \ref{fig:TF_NTD} show the results of the numerical evaluation of the transfer functions with \texttt{CLASS}~\cite{Blas:2011rf,Lesgourgues:2011rh}, and their comparison with the WDM case.\footnote{For the relativistic case, the small disagreement between the numerical transfer function with values from Eq.~(\ref{eq:mDM_bound_NTD}) and the corresponding WDM spectrum is also attributed to the sharp drop of the phase space distribution for $q<\Ydec\ll 1$, akin to a low-momentum cutoff. Such a cutoff results in a loss of numerical precision if reasonable computation times are required.} {For the two sets of curves shown in each figure, we use the rescaled Ly-$\alpha$ bound (\ref{eq:mDM_bound_NTD}) or (\ref{eq:mDM_bound_NTD_NR}).}
For the leftmost set we take $T_{\rm reh}=10^{12}\,{\rm GeV}$ and $m_{\rm WDM}=1\,{\rm keV}$, while for the rightmost set we consider $T_{\rm reh}=10^{10}\,{\rm GeV}$ and $m_{\rm WDM}=3\,{\rm keV}$. {For the decay of a non-relativistic particle, a comparison is made between the three different choices for $\Ydec=2,6$ and $10$. Note the overlap between the three curves, with a relative difference of $\sim 1\%$ (see Fig.~\ref{fig:transfer_function_residual}, where the relative difference is plotted as a function of $k$ for $\mathcal{R}=6$). For the decay of a relativistic particle, the agreement between the numerical and analytical results can be immediately appreciated, as well as the difference between these and the result of using the fit approximation~(\ref{eq:appND2}) for the DM distribution.}\footnote{The analytical expression of Eq.~(\ref{eq:resdistND0}) is not represented in Fig.~\ref{fig:TF_NTDNR} as the sharp $\theta$-function cannot be handled properly with \texttt{CLASS} as it requires the distribution function to smoothly decrease at large $q$.}
 Even more evident though is the difference of the NCDM transfer functions with respect to the one for WDM, of around $10\%$ at $k_{1/2}^{\rm WDM}$, c.f.~Fig.~\ref{fig:transfer_function_residual}. For the relativistic case, the distribution $f_{\chi}$ has a very non thermal shape, monotonically decreasing with $p$, resulting in a power spectrum that, although not too dissimilar from the WDM case, exhibits in the figure an appreciable difference from it.

\subsubsection{Relic density and phenomenology}

The present relic abundance of DM is obtained from integration of (\ref{eq:fchind}). To do this we make use of the (numerical) result
\beq
\int_0^{\infty} \diff q \, q^2\bar{f}_{\rm D,NR}(q) \;\simeq\; 0.4\,\Ydec^2\,.
\eeq
At $t\gg t_{\rm dec}$ the number density has the form
\beq
n_{\chi}(t) \;\simeq\; g_{*s}^0 {\rm Br}_{\chi} {\rm Br}_A \left(\frac{g_{\chi}}{g_A}\right) \left(\frac{T_{\rm reh}}{m_{\phi}}\right)  \left(\frac{a_0}{a(t)}\right)^3 T_0^3 \times \begin{cases}
\left(\dfrac{g_{*s}^{\rm reh}}{g_{*s}^{\rm dec}}\right)^{1/6} \,, & \Ydec \gg 1\,,\\[10pt]
 \left(\dfrac{g_{*s}^{\rm reh}}{g_{*s}^{\rm dec}}\right)^{1/4} \,,& \Ydec \ll 1\,.
\end{cases}
\eeq
Note that both expressions agree up to a different power of the number of relativistic degrees of freedom. This agreement is to be expected, as the total number of the decaying particle $A$ and its decay product must be a constant in a comoving volume. Considering for definiteness the case of a relativistic decaying particle, we determine that the present abundance is given by
\begin{align} \label{eq:omegachi_NTD}
\Omega_{\chi} h^2 \;\simeq\; 0.1\, {\rm Br}_{\chi} &\left(\frac{{\rm Br}_{A}}{5.5\times10^{-4}}\right)   \left(\frac{g_{\chi}}{g_A}\right) \left(\dfrac{g_{*s}^{\rm reh}}{g_{*s}^{\rm dec}}\right)^{1/4} \left(\frac{m_{\rm DM}}{1\,{\rm MeV}}\right) \left(\frac{T_{\rm reh}}{10^{10}\,{\rm GeV}}\right)\left(\frac{3\times 10^{13}\,{\rm GeV}}{m_{\phi}}\right) \,.
\end{align}
As expected, $\Omega_{\chi}$ is independent of the properties of $A$, and corresponds simply to a re-scaling by degrees of freedom of the inflaton decay result (\ref{eq:omegainfdec}). \par\medskip

\begin{figure}[!t]
\centering
    \includegraphics[width=0.73\textwidth]{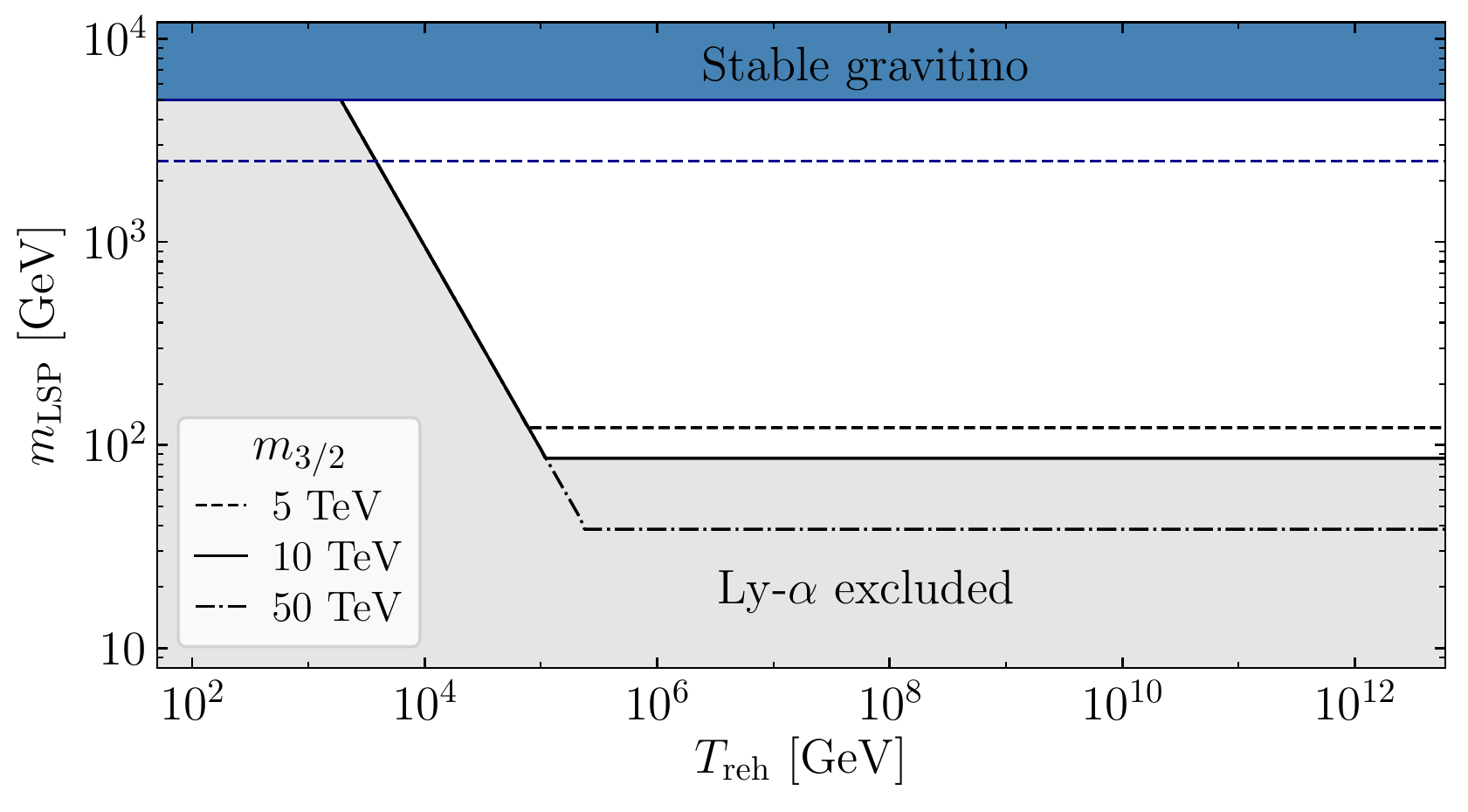}
    \caption{Ly-$\alpha$ constraint on the LSP mass, as a function of temperature, in the case of production through the decay chain inflaton $\rightarrow$ gravitino $\rightarrow$ LSP. For $T_{\rm reh}>10^5\,{\rm GeV}(m_{3/2}/10\,{\rm TeV})^{1/2}$ the decay of the gravitino occurs when it is non-relativistic. For $T_{\rm reh}<10^5\,{\rm GeV}(m_{3/2}/10\,{\rm TeV})^{1/2}$, the gravitino is relativistic at the moment of decay.}
    \label{fig:LSPlim}
\end{figure}

Given this result, a universal lower bound on ${\rm Br}_{\chi}$ can be obtained, in full analogy with the inflaton decay scenario. Let us discuss it in the context of a specific model. Consider the decay chain inflaton $\rightarrow$ gravitino $\rightarrow$ LSP (lightest supersymmetric particle), which is generically present in supersymetric models of inflation with supersymmetry breaking mediated gravitationally~\cite{Kallosh:1999jj,Giudice:1999yt,Nilles:2001ry,Nilles:2001my,Ellis:2015jpg}.\footnote{In typical gauge-mediation scenarios, the gravitino can be very light, $m_{3/2}\sim {\rm keV}$ and is produced through thermal freeze-out, thus being an example of WDM~\cite{Bond:1982uy,Pierpaoli:1997im,Baltz:2001rq,Fujii:2002fv,Viel:2005qj}.} Assuming a minimal supersymmetric extension of the Standard Model (MSSM), the decay rate of the spin-3/2 gravitino is~\cite{Moroi:1995fs} 
\beq
\Gamma_{3/2} \;=\; \frac{193}{384\pi}\frac{m_{3/2}^3}{M_P^2}\,.
\eeq
Generically, ${\rm Br}_{\rm LSP}=\mathcal{O}(1)$. Substitution into (\ref{eq:mDM_bound_NTD}) and (\ref{eq:omegachi_NTD}) leads to the following absolute constraints on the branching ratio of the decay of the inflaton into gravitinos, independent of the DM mass: For non-relativistic decaying particles, $T_{\rm reh}\gg 10^5\,{\rm GeV}(m_{3/2}/10\,{\rm TeV})^{1/2}$ and
\beq\label{eq:br32a}
{\rm Br}_{3/2} \;\lesssim\; 
1.3\times 10^{-8}\left(\dfrac{3\,{\rm keV}}{m_{\rm WDM}}\right)^{4/3} \left(\frac{m_{\phi}}{3\times 10^{13}\,{\rm GeV}}\right)\left(\frac{10^{10}\,{\rm GeV}}{T_{\rm reh}}\right) \left(\frac{m_{3/2}}{10\,{\rm TeV}}\right)^{1/2}\,,
\eeq
while for relativistic decaying ones, $T_{\rm reh}\ll 10^5\,{\rm GeV}(m_{3/2}/10\,{\rm TeV})^{1/2}$ and
\beq\label{eq:br32b}
{\rm Br}_{3/2} \;\lesssim\; 1.2\times 10^{-3}\left(\dfrac{3\,{\rm keV}}{m_{\rm WDM}}\right)^{4/3}\,.
\eeq
In this (MSSM) scenario, the excluded DM masses span a phenomenologically interesting
 region in the parameter space of the model, as shown in Fig.~\ref{fig:LSPlim}. The exclusion region corresponds to 
\beq
m_{\rm LSP} \;\lesssim\; \begin{cases}
86\,{\rm GeV} \left(\dfrac{m_{\rm WDM}}{3\,{\rm keV}}\right)^{4/3}\left(\dfrac{10\,{\rm TeV}}{m_{3/2}}\right)^{1/2}\,, & T_{\rm reh}\gg10^{5}\,{\rm GeV} \left(\dfrac{m_{3/2}}{10\,{\rm TeV}}\right)^{1/2}\,,\\[10pt]
95\,{\rm GeV}\left(\dfrac{m_{\rm WDM}}{3\,{\rm keV}}\right)^{4/3}\left(\dfrac{10^5\,{\rm GeV}}{T_{\rm reh}}\right) \,, & T_{\rm reh}\ll10^{5}\,{\rm GeV} \left(\dfrac{m_{3/2}}{10\,{\rm TeV}}\right)^{1/2}\,.
\end{cases}
\eeq
These bounds are or the order of the electroweak scale, and are comparable to collider and direct detection limits~\cite{Barman:2020vzm,Barman:2020zpz}. Note that for a model-fixed LSP mass, the Lyman-$\alpha$ constraint puts a bound on the inflaton-matter couplings. For $m_{\rm LSP}\gtrsim 100\,{\rm GeV}$, $T_{\rm reh}\lesssim 100\,{\rm TeV}$ are excluded.

A straightforward computation shows that independently of the mean momentum of the decaying gravitino, the decay occurs at temperatures at which the LSP can be safely assumed to be decoupled from the thermal plasma, and hence preserves its non-equilibrium phase space distribution. 

\section{Ultraviolet freeze-in via scatterings}\label{sec:UVfreezein}

In this section we consider the production of light DM from scatterings in the primordial plasma. We will restrict ourselves to $2\rightarrow 2$ processes, for which the integrated effective cross section is assumed to be of the form
\beq\label{eq:sigmas}
\sigma(s) \;=\; \frac{s^{\frac{n}{2}}}{\gls{Lambda}^{n+2}}\,,
\eeq
where $n$ is an integer and $s$ is the Mandelstam variable, related at high energies with the center of mass energy $E$ by $\sqrt{s}=E$. Although for non-negative $n$ this cross section naively violates unitarity at high energies, we assume that it merely corresponds to the low-energy effective description of a UV-complete model. The energy scale $\Lambda$ can be thought to be parametrically related to the mass of a heavy mediator.
The suppression by $\Lambda$ guarantees that the primordial abundance is determined by forward processes (plasma $\rightarrow$ DM) rather than by annihilations. Therefore, Pauli-blocking/Bose-enhancement for $\chi$ can be safely disregarded, and in the absence of other interaction channels, $\chi$ never reaches thermal equilibrium with the plasma. Thus, freeze-in is realized~\cite{Hall:2009bx,McDonald:2001vt}.

Assuming no post-reheating entropy production (that is, a standard thermal history), particle production is dominated by
temperatures $T\geq T_{\rm reh}$ if $n>-1$. This is referred to as ultraviolet (UV) freeze-in~\cite{Elahi:2014fsa,Garcia:2017tuj,Garcia:2018wtq}. Moreover, for $n>-1$, and for a sufficiently large reheating temperature, we can safely assume that both the parent scatterers and the produced DM are ultrarelativistic at the time of production,\footnote{This justifies disregarding any dependence on thresholds. For $n\leq-1$, the masses of the scatterers play an important role to determine the lower bound on $m_{\rm DM}$~\cite{Bae:2019sby,Kamada:2019kpe}.}  if the former are in thermal equilibrium. If the parent scatterers are not in equilibrium at production time, the condition that their masses are $\ll m_{\phi}$ suffices. Here we will consider both scenarios.

In order to evaluate the necessary collision terms for thermal and non-thermal production, we need to make assumptions regarding the form of the scattering amplitude. Its dependence on the angles (or Mandelstam variables $s,t,u$) involved in the scattering varies between different microscopic descriptions of the process. We will assume that for the scattering process $A(k)+B(\tilde{k}) \rightarrow \chi(p) + \psi(\tilde{p})$, the mean, unpolarized squared scattering amplitude can be parametrized in the following way,
\beq\label{eq:M2sn}
|\mathcal{M}|^2 \;=\; 16\pi \frac{s^{\frac{n}{2}+1}}{\Lambda^{n+2}}\,.
\eeq
Integration with respect to the two-particle phase space recovers (\ref{eq:sigmas}). For a different combination of $s,t,u$, our results will generically only differ by numerical factors, which can be absorbed into {the value of $\Lambda$}.\footnote{Exceptions include those cases in which finite-temperature in-medium effects are necessary to regulate infrared divergences, which arise from the exchange of massless mediators. Thermal axion production and gravitino production in low-scale supersymmetry are included in these cases~\cite{Braaten:1991dd,Bolz:2000fu,Pradler:2006qh,Rychkov:2007uq,Eberl:2020fml,Salvio:2013iaa}. } 

Under the freeze-in assumption, and with the square amplitude given by (\ref{eq:M2sn}), the collision term for the production of $\chi$ can be written as follows,
\beq\label{eq:CgenFI}
\mathcal{C}[f_{\chi}] \;=\; \frac{16\pi g_A g_B g_{\psi}}{\Lambda^{n+2}2 p_0} \int \frac{\diff^3\tilde{\bp}}{2(2\pi)^3 \tilde{p}_0} \frac{\diff^3 \bk}{2(2\pi)^3 k_0} \frac{\diff^3 \tilde{\bk} }{2(2\pi)^3 \tilde{k}_0}  (2\pi)^4 \delta^{(4)}(p+\tilde{p}-k-\tilde{k}) s^{\frac{n}{2}+1} f_A (k_0) f_B(\tilde{k}_0)\,.
\eeq
The integration of this collision term for arbitrary $f_{A,B}$ can be easily done following the steps of~\cite{Bolz:2000fu,GarciaGarcia:2016nhj}. We detail these steps in Appendix~\ref{app:figen}. As a result, we obtain Eq.~(\ref{eq:cfgen}), which will be the starting point of our discussion of thermal and non-thermal UV freeze-in.

\subsection{Thermal freeze-in}\label{sec:thermalfreezein}

We begin by applying the general solution (\ref{eq:cfgen}) to the production of DM from thermalized scatterers, i.e.~with Fermi-Dirac or Bose-Einstein distributions. As stated earlier in this section, we focus on UV freeze-in, for which the bulk of the DM relic abundance is produced during reheating. Thermal production during reheating is the dominant production channel in the absence of significant direct inflaton $\rightarrow$ DM decays for $-1<n\leq 2$ in (\ref{eq:sigmas}). Moreover, for higher $n$, thermal production can dominate over non-thermal effects if the parent scatterers that couple to the dark sector are not directly produced from inflaton decay (see e.g.~\cite{Garcia:2020hyo}). Given the need to compute the integrals in (\ref{eq:cfgen}) in a case-by-case basis for scatterers outside of the Maxwell-Boltzmann limit, we will focus on the lowest even values of $n$, namely $n=\{0,2,4,6\}$. In doing so, we will recover the results of~\cite{Garcia:2017tuj}, showing that for $n\geq 6$, thermal production is dominated by the highest temperature during reheating, \gls{Tmax}.\footnote{More precisely, $T_{\rm max}$ denotes the maximum temperature of the Universe after the thermalization of the primordial plasma during reheating. In the regime where non-perturbative particle production is the subdominant decay channel for the inflaton, this $T_{\rm max}\gg T_{\rm reh}$, although it is smaller than the value that it would naively have assuming instantaneous thermalization~\cite{Harigaya:2014waa,Harigaya:2013vwa,Mukaida:2015ria}. }

\subsubsection{DM phase space distribution}
\label{sec:therma_FI_PSD}

\begin{figure}[!t]
\centering
    \includegraphics[width=\textwidth]{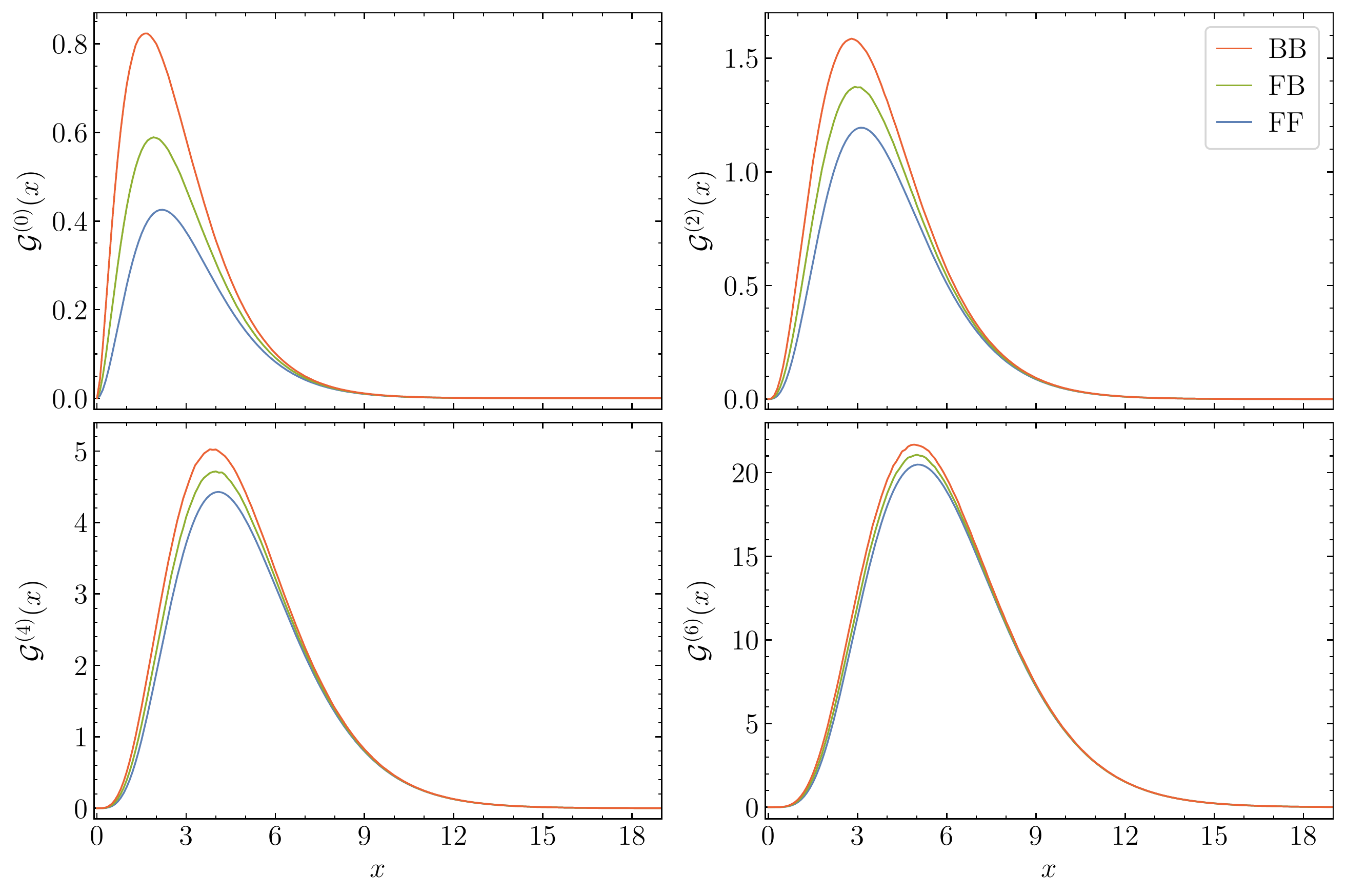}
    \caption{The collision term function $\mathcal{G}^{(n)}$ for thermal freeze-in defined in (\ref{eq:CnFIG}), for $n=\{0,2,4,6\}$. Shown in each panel are the corresponding forms for this function in the case of a fermion-fermion scattering (FF, blue, lowest curve), for fermion-boson scattering (FB, green, middle curve) and boson-boson scattering (BB, red, highest curve).}
    \label{fig:Gn}
\end{figure}
For any $n$, the computation of the innermost integrals in (\ref{eq:cfgen}) can be performed analytically {(in terms of polylogarithmic functions) if the initial states have} the form fermion+fermion (FF), fermion+boson (FB) or boson+boson (BB). The outermost integral is however more challenging, and we compute it numerically. The result can be written as
\beq\label{eq:CnFIG}
\mathcal{C}[f_{\chi}]^{(n)} \;=\; \frac{ g_A g_B g_{\psi} 2^{n+2} \Gamma(\frac{n+4}{2}) T^{n+5}}{(2\pi)^2\Lambda^{n+2} p^2}\, \mathcal{G}^{(n)}(p/T)\,,
\eeq
where we have simplified the notation assuming that the DM particles are ultrarelativistic at production, and where the functions $\mathcal{G}^{(n)}(x)$ are shown in Fig.~\ref{fig:Gn} for the four values of $n$ that we consider. In the Maxwell-Boltzmann limit, $\mathcal{G}^{(n)}(x)=x^{\frac{n+4}{2}}e^{-x}$. This collision term can now be substituted into the general solution (\ref{eq:Cgensol}) of the transport equation (\ref{eq:boltzmanneq}). In order to translate the time integral into a temperature integral, we make use of an approximate solution for the Friedmann-Boltzmann system (\ref{eq:FB1})-(\ref{eq:FB3}), which for $t_{\rm end}\ll t\ll t_{\rm reh}$ gives
\begin{align}\label{eq:atoandapp}
\frac{a(t)}{a_{\rm end}} \;&\simeq\; \frac{9 \rho_{\rm end} t^2}{4M_P^2}\,,
\end{align}
and~\cite{Garcia:2017tuj}
\begin{align} \label{eq:Tvstreh}
T \;&\simeq\; \left(\frac{24}{\pi^2 g_{*s}^{\rm reh}}\right)^{1/4}(\Gamma_{\phi}M_P)^{1/2}\left(\Gamma_{\phi}t\right)^{-1/4}
\equiv\; \gls{Treh} (\Gamma_{\phi} t/{ \gls{b}})^{-1/4}\,,
\end{align}
We have defined $T_{\rm reh}$ here as a function of $\Gamma_{\phi}$ up to an $\mathcal{O}(1)$ numerical factor that we denote by ${ b}$.\footnote{The factor ${ b}$ depends on how the transition between matter and radiation domination at the end of reheating is described, which complicates the analytical determination of $T_{\rm reh}$.  Extrapolating Eq.~(\ref{eq:Tvstreh}) to $\Gamma_{\phi} t=1$ yields ${ b}=1$, while substitution of $H\simeq 2\Gamma_{\phi}/3$ at $\rho_{\phi}=\rho_r$ in Eq.~(\ref{eq:FB3}) gives ${ b}=6/5$. Numerical solution of (\ref{eq:FB1})-(\ref{eq:FB3}) reveals that ${ b}\simeq 1.6$ at (inflaton) matter-radiation equality. For convenience we keep ${ b}$ unspecified.} 
Straightforward algebraic manipulation reveal the following form for the DM phase space distribution at the end of reheating, 
\begin{align}\notag
f_{\chi}(p,T_{\rm reh}) \;\simeq\; \left(\frac{6{ b}}{g_{*s}^{\rm reh}}\right)^{1/2} &\frac{3\cdot 2^{n+6}\Gamma(\frac{n+4}{2}) g_A g_B g_{\psi} M_P T_{\rm reh}^{n+1}}{5(2\pi)^3  \Lambda^{n+2} } \\ \label{eq:fgenTFI}
& \times \left(\frac{T_{\rm reh}}{p}\right)^{\frac{3(n-1)}{5}} \int_{p/T_{\rm reh}}^{pT_{\rm max}^{5/3}/T_{\rm reh}^{8/3}} \diff x\, x^{\frac{3(n-6)}{5}} \mathcal{G}^{(n)}(x)\,.
\end{align}
Disregarding the residual production of DM for $T\lesssim T_{\rm reh}$, we can extend this solution by means of the free streaming expression (\ref{eq:freestream}) with $a_{\rm dec}\simeq a_{\rm reh}$. Hence, at late times we can finally write 
\begin{align}
f_{\chi}(p,t)\, \diff^3 \bp \;\simeq\; \left(\frac{6{ b}}{g_{*s}^{\rm reh}}\right)^{1/2} &\frac{3\cdot 2^{n+6}\Gamma(\frac{n+4}{2}) g_A g_B g_{\psi} M_P T_{\rm reh}^{n+1}}{5(2\pi)^3  \Lambda^{n+2} } \left(\frac{a_0}{a(t)}\right)^3 \tncdm^3\, \bar{f}_{\rm TF}^{(n)}(q)\, \diff^3\boldsymbol{q}\,,
\end{align}
where
\begin{align}\label{eq:fTFless}
\bar{f}_{\rm TF}^{(n)}(q) \;&\equiv\; q^{\frac{3(1-n)}{5}} \int_q^{q(T_{\rm max}/T_{\rm reh})^{5/3}} \diff x\, x^{\frac{3(n-6)}{5}} \mathcal{G}^{(n)}(x)\,,\\
\tncdm \;&=\;  \left(\frac{g_{*s}^{0}}{g_{*s}^{\rm reh}}\right)^{1/3} T_0\,.
\end{align} 

\begin{figure}[!t]
\centering
    \includegraphics[width=0.7\textwidth]{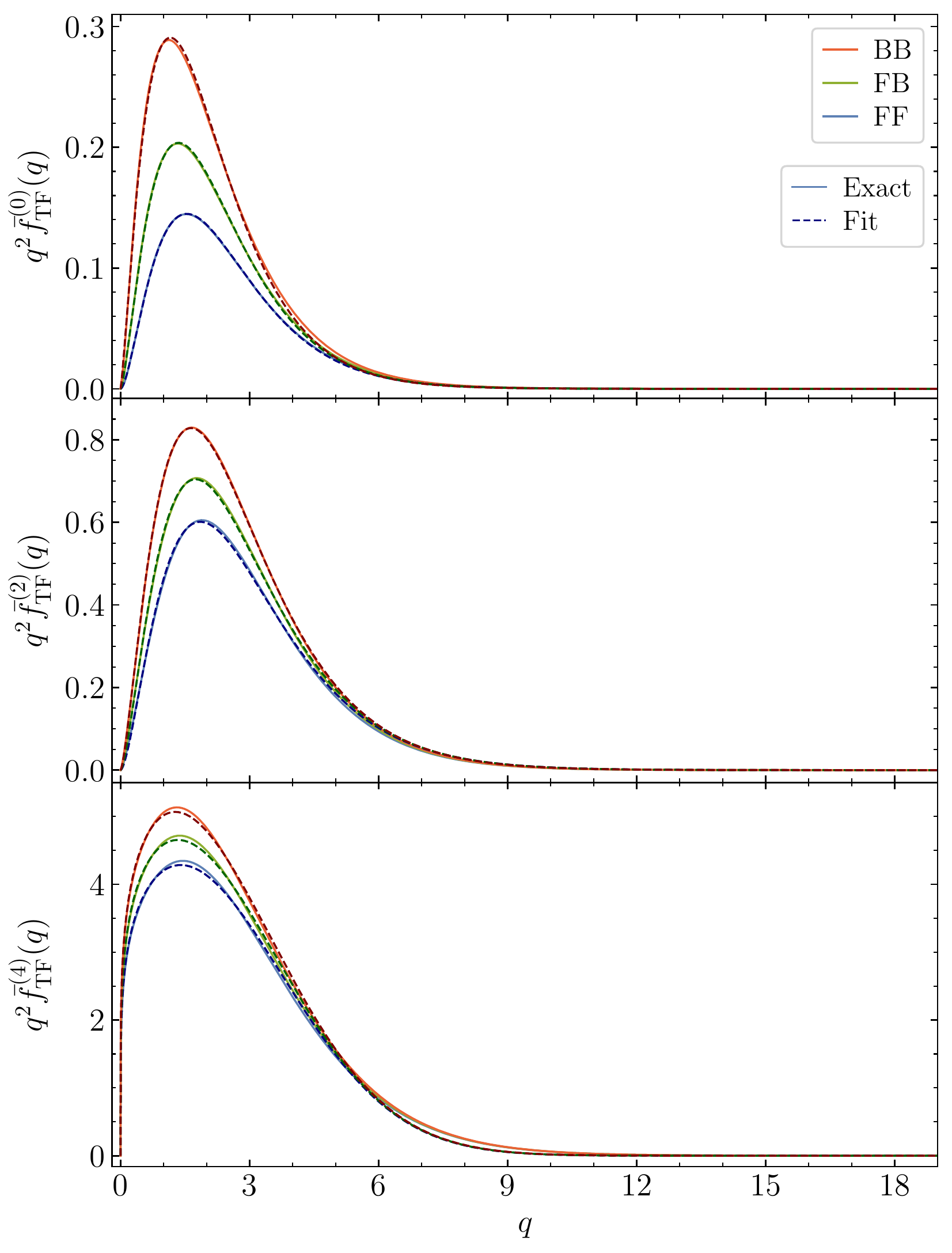}
    \caption{The rescaled distribution function $\bar{f}_{\rm TF}^{(n)}$ defined in (\ref{eq:fTFless}), as a function of the rescaled momentum $q$, for $n=\{0,2,4\}$. Solid: numerically computed phase space distribution. Dashed: the fit (\ref{eq:allfit}) with the parameters shown in Table~\ref{tab:TFIfits}. As in Fig.~\ref{fig:Gn}, each panel shows the form of the distribution in the case of fermion-fermion, fermion-boson or boson-boson scatterings.}
    \label{fig:low_n}
\end{figure}

\begin{table}[t]
\centering
\begin{tabular}{|c|c|c|c|c|c|}
\hline \multicolumn{2}{|c|}{Scenario} & Prefactor & $\alpha$ & $\beta$ & $\gamma$ \\
    \hline \hline \multirow{3}{*}{$n=0$} & BB & 0.88 & -0.70 & 1.13 & 1.00\\
\cline{2-6} & FB & 0.58 &-0.51 & 1.10 & 1.00\\
\cline{2-6} & FF & 0.38 & -0.29 & 1.11 & 1.00\\
 \hline \multirow{3}{*}{$n=2$} & BB & 1.76 & -0.51 & 0.91 & 1.00\\
\cline{2-6} & FB & 1.42 & -0.42 & 0.90 & 1.00\\
\cline{2-6} & FF & 1.14 & -0.33 & 0.90 & 1.00\\
 \hline \multirow{3}{*}{$n=4$} & BB & 5.35 & -1.79 & 0.06 & 1.98\\
\cline{2-6} & FB & 4.85 & -1.79 & 0.06 & 2.04\\
\cline{2-6} & FF & 4.41 &  -1.79 & 0.05 & 2.10\\
\hline
\end{tabular}
\caption{
Fit parameters of Eq.~(\ref{eq:allfit}) for the thermal freeze-in distributions $\bar{f}_{\rm TF}^{(n)}(q)$, with $n=\{0,2,4\}$. Here $\gamma=1$ is fixed for $n=0,2$, while $\gamma$ is left as free parameter for the $n=4$ case.
}
\label{tab:TFIfits}
\end{table}
The functions $\bar{f}_{\rm TF}^{(n)}(q)$ for $n=\{0,2,4\}$ are shown in Fig.~\ref{fig:low_n}. In these three cases the integral in (\ref{eq:fTFless}) is dominated by the lower limit: production is peaked at $T_{\rm reh}$~\cite{Garcia:2017tuj}. Therefore the approximation $q(T_{\rm max}/T_{\rm reh})^{5/3}\rightarrow \infty$ can be taken, implying a loss of dependence on the maximum temperature of the relic density. In all three cases, and for all three different scatterer configurations, the low-$q$ part of the distribution grows as a power-law, while the large-$q$ part retains the exponential tail of the thermalized parent particles. Hence, phenomenological fits of the form (\ref{eq:allfit}) can be constructed in all cases, with fit parameters as shown in Table~\ref{tab:TFIfits}. To construct these fits we impose $\gamma=1$ for $n=0,2$ as the fitted function is already matching well the numerical distributions. However, we left $\gamma$ as free parameter for the $n=4$ case in order for the fitting function to accurately describe the distribution. Note that in none of these cases the resulting distribution function fully inherits the FD or BE distribution of the parent scatterers, as it is sometimes assumed~\cite{Baur:2015jsy}.\footnote{For $n=0$, a thermal distribution provides an adequate fit, although worse than a fit of the form~(\ref{eq:allfit})~\cite{McDonald:2015ljz}.} It is also worth noting that, although we have not attempted to obtain a closed form expression for the exact distribution, we can compute its integral analytically. Namely, integration of the Boltzmann equation (\ref{eq:boltzmanneq}) with  the collision term (\ref{eq:CgenFI}) leads to the following evolution equation for the DM number density,
\beq\label{eq:boltznchi}
\dfrac{\diff n_{\chi}}{\diff t} + 3H n_{\chi} \;=\; 2 g_A g_B g_{\psi} g_{\chi} \int \frac{\diff^3\bp_1}{(2\pi)^3 2p_1^0} \frac{\diff ^3\bp_2}{(2\pi)^3 2p_2^0}\, s\sigma(s)\, f_A(p_1) f_B(p_2)\,.
\eeq
Upon integration and evaluation at $T_{\rm reh}$, the following result is obtained, valid for $n<6$,
\beq
\int_0^{\infty} \diff q\, q^2 \bar{f}^{(n)}_{\rm TF}(q) \;=\; \frac{5\, \Gamma(\frac{n}{2}+3)\zeta(\frac{n}{2}+3)^2 \mathcal{S}(n)}{3(6-n)}\,.
\eeq
Here $\mathcal{S}$ is a statistics-dependent function of $n$,
\beq
\mathcal{S}(n) \;=\;  \begin{cases}
1\,, \quad & \text{BB}\,,\\ 
\left(1-2^{-(\frac{n}{2}+2)}\right)\,, \quad & \text{FB}\,,\\
\left(1-2^{-(\frac{n}{2}+2)}\right)^2 \,, \quad & \text{FF}\,.
\end{cases}
\eeq 

\begin{figure}[!t]
\centering
    \includegraphics[width=0.7\textwidth]{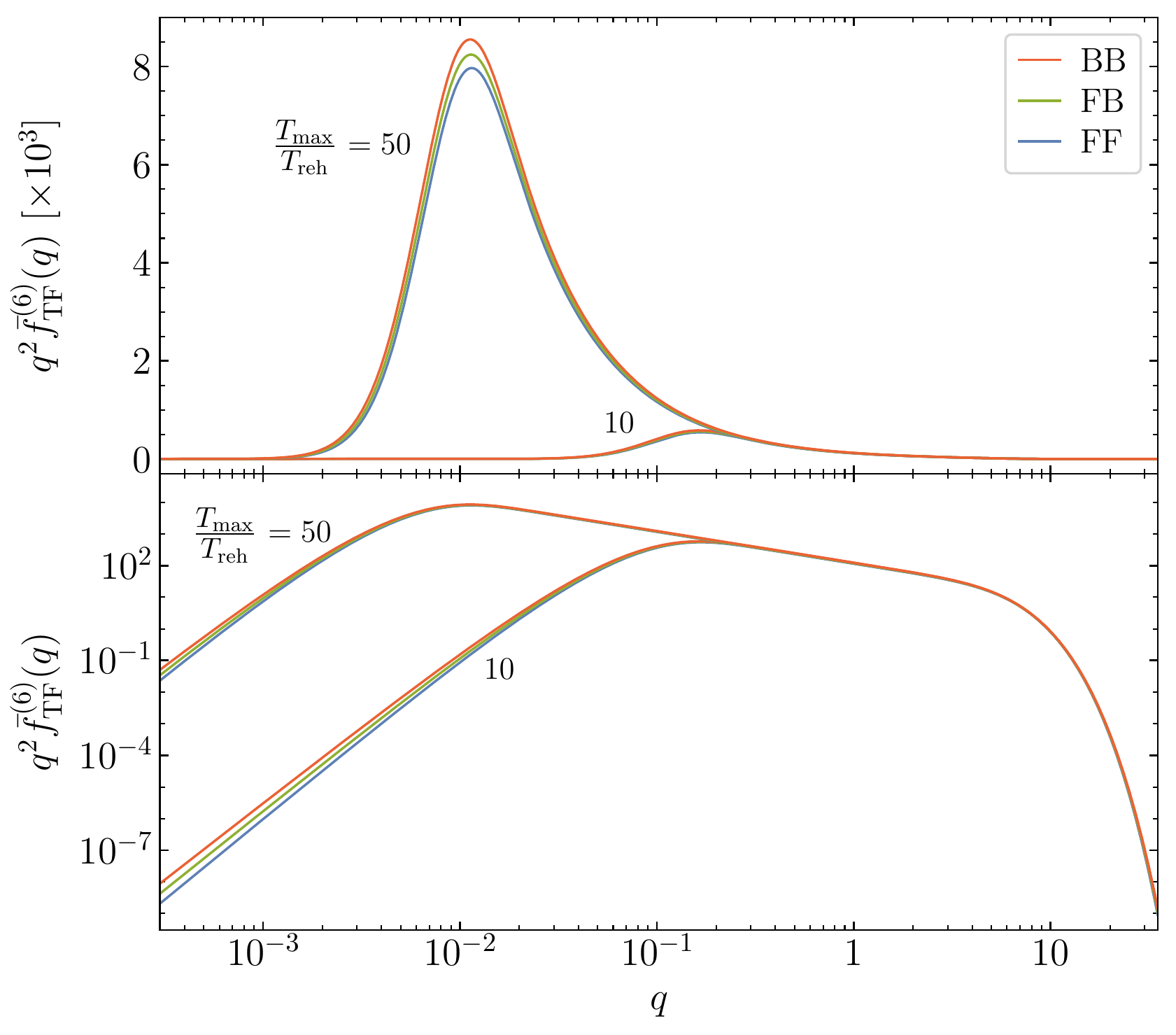}
    \caption{The rescaled distribution function $\bar{f}_{\rm TF}^{(6)}$ defined in (\ref{eq:fTFless}), as a function of the rescaled momentum $q$. The top panel shows the form of the distribution for $T_{\rm max}/T_{\rm reh}=10$ and 50 in a log-linear scale. The bottom panel displays the same distributions in a log-log scale.}
    \label{fig:neq6}
\end{figure}

The case $n=6$ is special and must be treated separately. In this case, both limits of integration in (\ref{eq:fTFless}) must be kept to obtain a finite result. This results in a dependence on $T_{\rm max}$ of the relic density: production is in this case peaked at the maximum temperature after thermalization. This feature manifests itself noticeably in the shape of the resulting phase space distribution. Fig.~\ref{fig:neq6} shows this distribution for two cases, $T_{\rm max}/T_{\rm reh}=10$ and 50.\footnote{These relatively low values of the temperature ratio are chosen to simplify numerical integrations and the reading of the resulting plots. Additionally, low $T_{\rm max}$ is required to match the observed DM abundance in the case of light DM.} The top panel shows the distributions in a log-linear scale, in order to showcase the significant difference that the value of the temperature ratio makes in the location and amplitude of the peak of $\bar{f}_{\rm TF}^{(6)}(q)$. The lower panel of this same figure, in turn, shows the same distributions in a log-log scale, to demonstrate the three different regimes in $q$. For $q$ below the peak, the distribution increases as a power law, $q^2\bar{f}_{\rm TF}^{(6)}(q)\sim q^5$, while $q^2\bar{f}_{\rm TF}^{(6)}(q)\sim q^{-1}$ below the peak for $q\lesssim 10$. For $q\gtrsim 10$, the distribution has the expected exponential tail due to the thermal nature of the scatterers. For these reasons we have not presented a fit of the form (\ref{eq:allfit}) for this ($n=6$) case, as it would inevitably fail to mimic at least one of the three scalings of the distribution. As mentioned in the Introduction, this is one of the cases for which the phenomenological fit is not applicable. Nevertheless, an analytical closed-form solution for the integrated distribution function is available,
\beq
\int_0^{\infty} \diff q\, q^2 \bar{f}^{(6)}_{\rm TF}(q) \;=\; \frac{8\pi^{12} \mathcal{S}(6)}{35721}\,\ln\left(\frac{T_{\rm max}}{T_{\rm reh}}\right)\,,
\eeq
where for this case,
\beq
\mathcal{S}(6) \;=\;  \begin{cases}
1\,, \quad & \text{boson-boson}\,,\\[10pt]
\dfrac{31}{32}\,, \quad & \text{fermion-boson}\,,\\[10pt]
\dfrac{961}{1024} \,, \quad & \text{fermion-fermion}\,.
\end{cases}
\eeq\par\bigskip

We finish this section with a word on the relevance of quantum statistics. Despite the fact that the use of Maxwell-Boltzmann statistics for the parent scatterers will necessarily lead to errors in the DM and relic abundance, it is instructive to show how much our previous computations are simplified in this limit. As mentioned above, $\mathcal{G}^{(n)}(x)=x^{\frac{n+4}{2}}e^{-x}$ in this case, and therefore the corresponding integration of the collision term proceeds in a straightforward manner to give
\beq\label{eq:ftfapp}
\bar{f}(q)_{\rm TF} \;=\; q^{\frac{3}{5}(1-n)} \left[ \Gamma\left(\frac{11}{10}n-\frac{3}{5},q\right) - \Gamma\left(\frac{11}{10}n-\frac{3}{5},q\left(\frac{T_{\rm max}}{T_{\rm reh}}\right)^{5/3}\right) \right]\,.
\eeq
As $n$ is increased, this approximation becomes a better fit for the distributions with the correct statistics, and in fact converges to the FB case. It can also be shown that this convergence is exponential in the case of the relic abundance. 

\subsubsection{Power spectrum and Ly-$\alpha$ constraints}

We now proceed to discuss the phenomenological implications of the Ly-$\alpha$ rescaled constraints on UV freeze-in. For the low-$n$ cases, the WDM rescaling relation (\ref{eq:mncdm_fromeos}) leads to the following constraints on the DM mass, based on the numerical and fit distributions:\par\bigskip

\begin{figure}[!t]
\centering
    \includegraphics[width=0.75\textwidth]{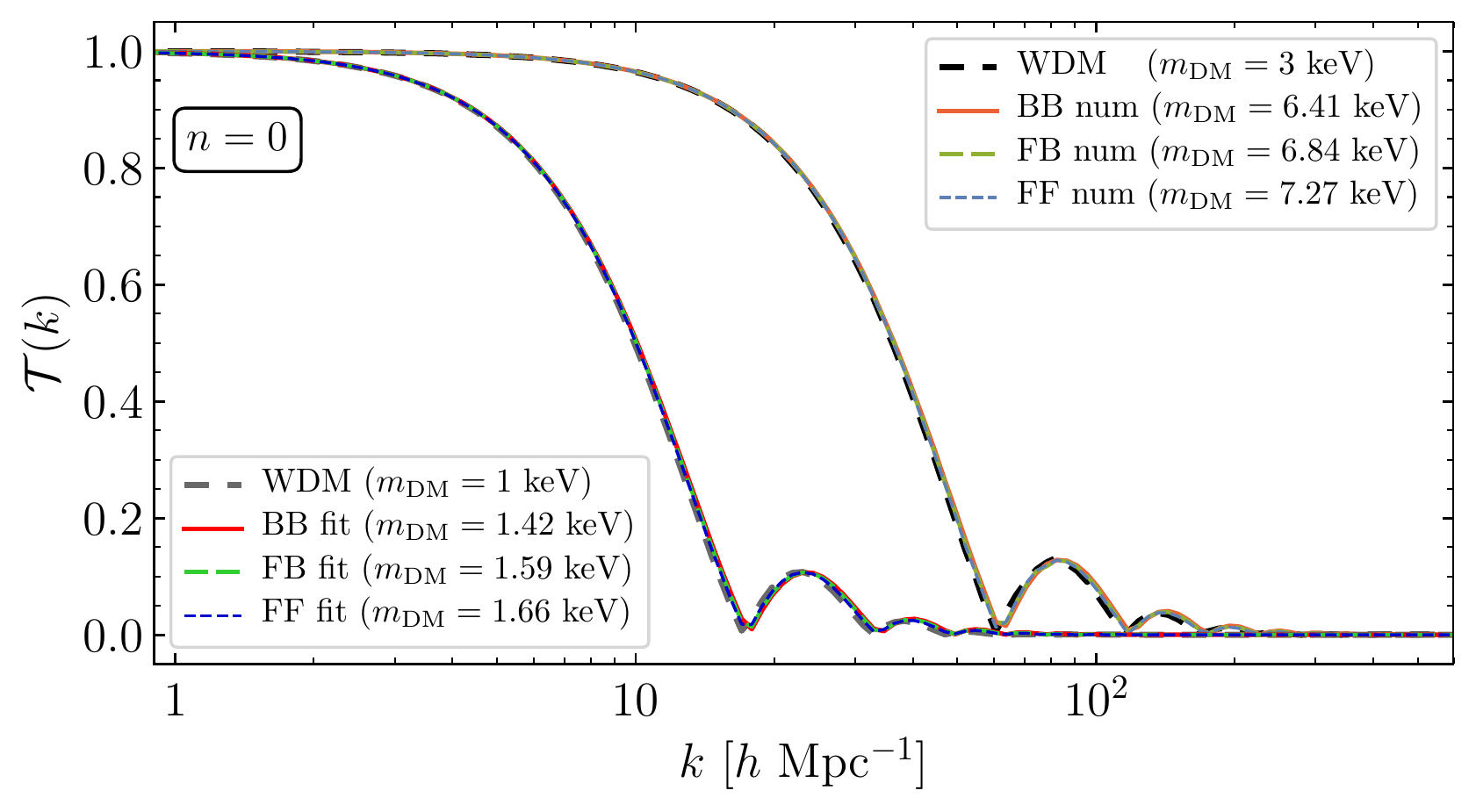}\vspace{-3pt}
     \includegraphics[width=0.75\textwidth]{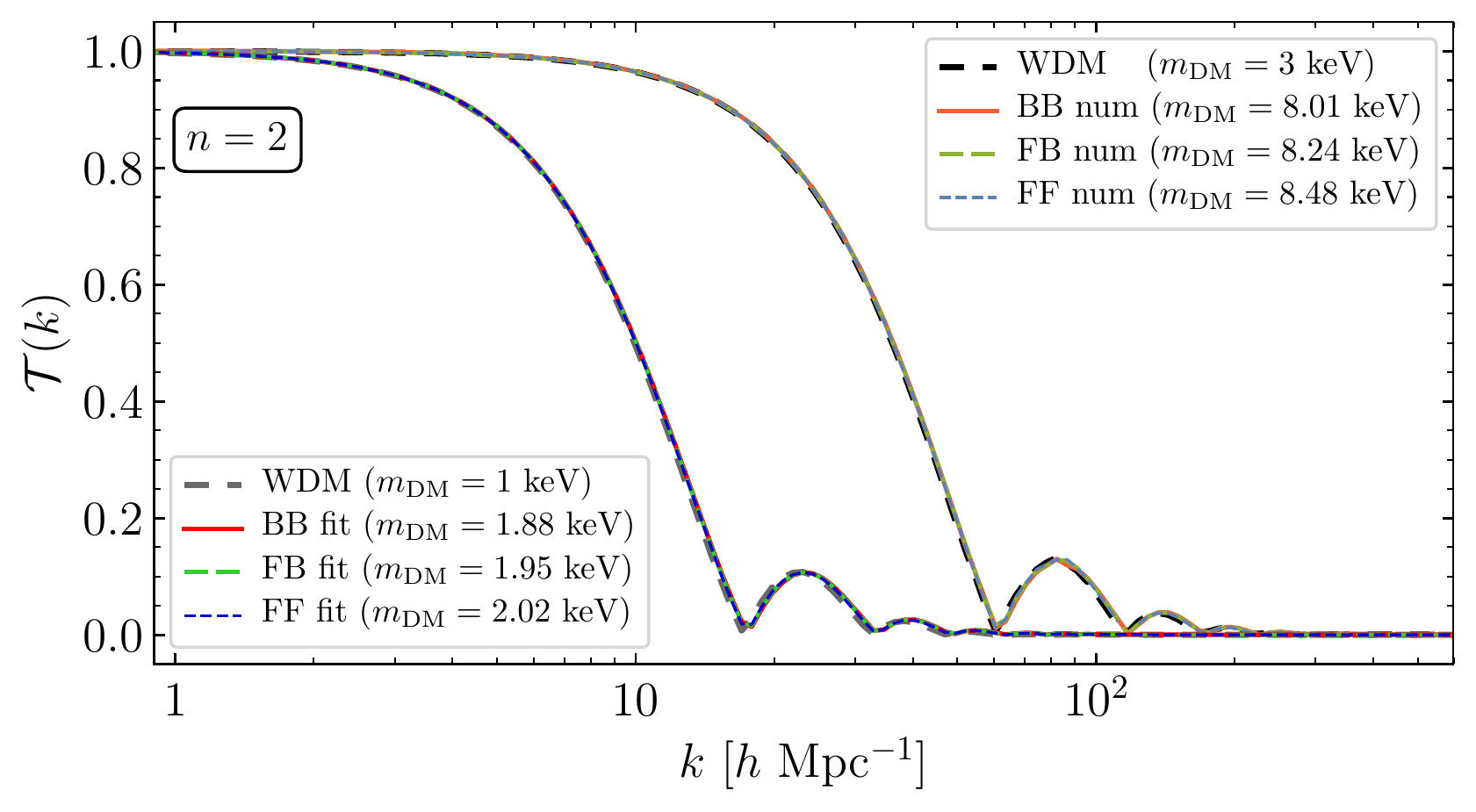}
    \caption{Linear transfer function for DM produced by thermal UV freeze-in. Shown here are the results for the numerical and fit approximations discussed in the text for $n=0$ (top) and $n=2$ (bottom), for initial fermion-fermion (FF), fermion-boson (FB) and boson-boson (BB) states. The DM masses are taken from the rescaled bounds (\ref{eq:mDM_bound_UVFI_n0}) and (\ref{eq:mDM_bound_UVFI_n2}) with $m_{\rm WDM}=1\,{\rm keV}$ (fit) and $m_{\rm WDM}=3\,{\rm keV}$ (numerical). For comparison we show $\mathcal{T}(k)$ for the WDM in each case.}
    \label{fig:TF_UVFI_n_a}
\end{figure}

\begin{figure}[!t]
\centering
    \includegraphics[width=0.75\textwidth]{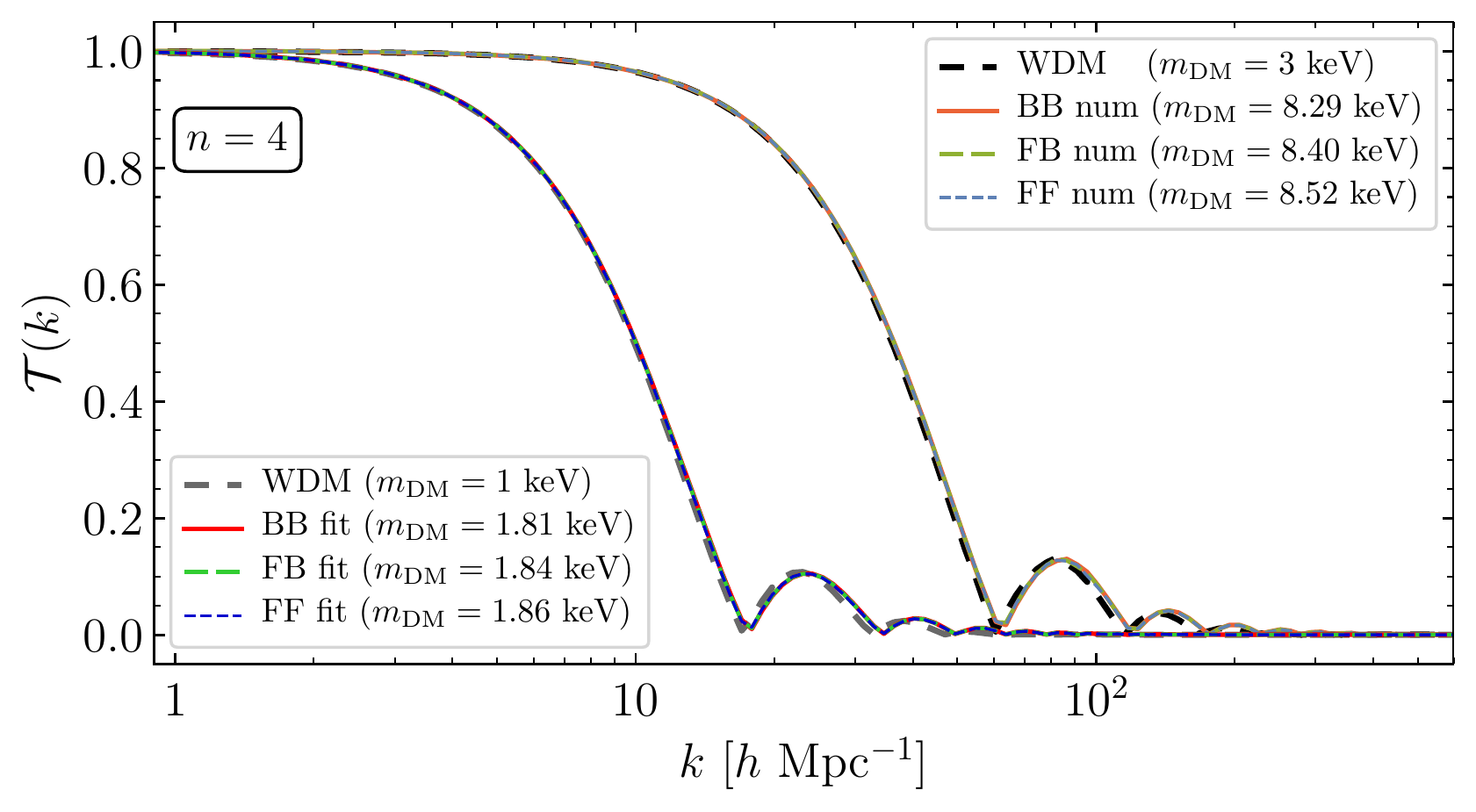}\vspace{-3pt}
    \includegraphics[width=0.75\textwidth]{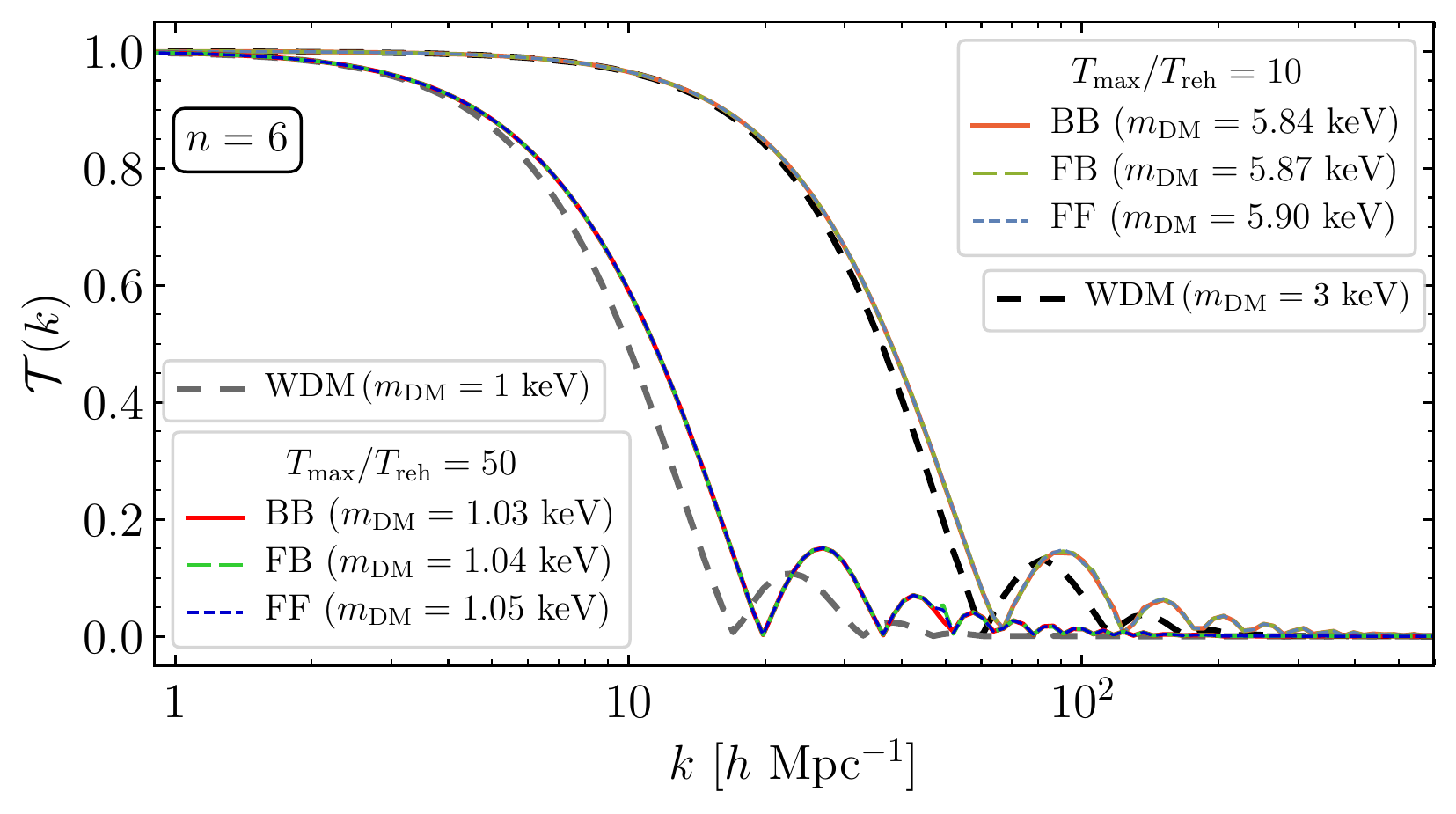}
    \caption{Linear transfer function for DM produced by thermal UV freeze-in. Shown here are the results for the numerical and fit approximations discussed in the text for $n=4$ (top) and the numerical result for $n=6$ (bottom), for initial fermion-fermion (FF), fermion-boson (FB) and boson-boson (BB) states. The DM masses are taken from the rescaled bounds (\ref{eq:mDM_bound_UVFI_n4}), (\ref{eq:mDM_bound_UVFI_n6_1}) and (\ref{eq:mDM_bound_UVFI_n6_2}) with $m_{\rm WDM}=1\,{\rm keV}$ for the fit approximation (top) and $T_{\rm max}/T_{\rm reh}=50$ (bottom), and $m_{\rm WDM}=3\,{\rm keV}$ for the numerical result (top), and $T_{\rm max}/T_{\rm reh}=10$ (bottom). For comparison we show $\mathcal{T}(k)$ for the WDM in each case.}
    \label{fig:TF_UVFI_n_b}
\end{figure}

\noindent
For $n=0$,
\beq
    \hatmncdm \, \gtrsim \,  \,  \left( \dfrac{m_\text{WDM}}{3~\text{keV}} \right)^{4/3} \left( \dfrac{106.75}{g_{*s}^{\rm reh}} \right)^{1/3} \begin{cases}
7.27~(7.17)~\text{keV}\,, \quad & {\rm FF }\quad {\rm Numerical}~{\rm (Fit)}\,,\\
6.41~(6.16)~\text{keV}\,, \quad & {\rm BB}\quad {\rm Numerical}~{\rm (Fit)}\,,\\
6.84~(6.70)~\text{keV}\,, \quad & {\rm FB}\quad {\rm Numerical}~{\rm (Fit)}\,.
\end{cases}
\label{eq:mDM_bound_UVFI_n0}
\eeq
For $n=2$,
\beq
    \hatmncdm \, \gtrsim \,  \,  \left( \dfrac{m_\text{WDM}}{3~\text{keV}} \right)^{4/3} \left( \dfrac{106.75}{g_{*s}^{\rm reh}} \right)^{1/3} \begin{cases}
8.48~(8.73)~\text{keV}\,, \quad & {\rm FF }\quad {\rm Numerical}~{\rm (Fit)}\,,\\
8.01~(8.14)~\text{keV}\,, \quad & {\rm BB}\quad {\rm Numerical}~{\rm (Fit)}\,,\\
8.24~(8.44)~\text{keV}\,, \quad & {\rm FB}\quad {\rm Numerical}~{\rm (Fit)}\,.
\end{cases}
\label{eq:mDM_bound_UVFI_n2}
\eeq
For $n=4$,
\beq
    \hatmncdm \, \gtrsim \,  \,  \left( \dfrac{m_\text{WDM}}{3~\text{keV}} \right)^{4/3} \left( \dfrac{106.75}{g_{*s}^{\rm reh}} \right)^{1/3} \begin{cases}
8.52~(8.05)~\text{keV}\,, \quad & {\rm FF }\quad {\rm Numerical}~{\rm (Fit)}\,,\\
8.29~(7.84)~\text{keV}\,, \quad & {\rm BB}\quad {\rm Numerical}~{\rm (Fit)}\,,\\
8.40~(7.94)~\text{keV}\,, \quad & {\rm FB}\quad {\rm Numerical}~{\rm (Fit)}\,.
\end{cases}
\label{eq:mDM_bound_UVFI_n4}
\eeq
In all cases the agreement between numerics and the fits are evident, since the relative deviations are at most of a few percent. This is further confirmed by the transfer functions shown in Figs.~\ref{fig:TF_UVFI_n_a} and \ref{fig:TF_UVFI_n_b}. There, the three possible initial states for each value of $n$ are shown together with the corresponding WDM transfer function. In these plots, the leftmost set of curves shows $\mathcal{T}(k)$ for $m_{\rm WDM}=1\,{\rm keV}$, built using the fit approximations in each case. The rightmost set of curves, in turn, correspond to $m_{\rm WDM}=3\,{\rm keV}$, and the DM bound derived from the numerically computed distributions. All results confirm our constraint mapping procedure to a precision $\lesssim 3\%$, as shown in Fig.~\ref{fig:transfer_function_residual} for the FF cases. 

As we discussed above, the case with $n=6$ needs to be treated separately, due to its dependence on $T_{\rm max}$. In this case, the mapping relation (\ref{eq:mncdm_fromeos}) gives the following rescaling on the DM mass bound, based on the numerically determined phase space distribution shown in Fig.~\ref{fig:neq6},
\beq
   \hatmncdm \, \gtrsim \,  \,  \left( \dfrac{m_\text{WDM}}{3~\text{keV}} \right)^{4/3} \left( \dfrac{106.75}{g_{*s}^{\rm reh}} \right)^{1/3} 
 \begin{cases}
5.90~\text{keV}\,, \quad & {\rm FF }\,,\\
5.84~\text{keV}\,, \quad & {\rm BB}\,,\\
5.87~\text{keV}\,, \quad & {\rm FB}\,.
\end{cases}
\qquad  {\rm for } \quad \dfrac{T_\text{max}}{T_\text{reh}}\,=\,10\,,\\
\label{eq:mDM_bound_UVFI_n6_1}
\eeq
and
\beq
   \hatmncdm \, \gtrsim \,  \,  \left( \dfrac{m_\text{WDM}}{3~\text{keV}} \right)^{4/3} \left( \dfrac{106.75}{g_{*s}^{\rm reh}} \right)^{1/3} 
 \begin{cases}
4.53~\text{keV}\,, \quad & {\rm FF }\,,\\
4.48~\text{keV}\,, \quad & {\rm BB}\,,\\
4.50~\text{keV}\,, \quad & {\rm FB}\,.
\end{cases}
\qquad  {\rm for } \quad \dfrac{T_\text{max}}{T_\text{reh}}\,=\,50\,,\\
\label{eq:mDM_bound_UVFI_n6_2}
\eeq
Fig.~\ref{fig:TF_UVFI_n_b} shows the form of the linear transfer function for both temperature ratios and all initial configurations, compared to the WDM case. There is a noticeable deviation between the WDM transfer functions and those for this freeze-in scenario, which increases as $T_{\max}$ is increased relative to $T_{\rm reh}$, correlated with the presence of a longer power-like tail in the distribution. For $T_{\rm max}/T_{\rm reh}=10$, this difference is as large as $10\%$ at $k_{\rm 1/2}^{\rm WDM}$, as is shown in Fig.~\ref{fig:transfer_function_residual}. It is worth pointing out that the Maxwell-Boltzmann approximation (\ref{eq:ftfapp}) can be used to estimate the DM mass bound without the need for numerical computations. From it, we obtain $m_{\rm DM} \approx (9\,{\rm keV}) \ln^{-1/2}(T_{\rm max}/T_{\rm reh})$.

\subsubsection{Relic density and phenomenology}\label{sec:phenoFI}

For $-1<n<6$, the DM number density and the abundance can be computed analytically by means of (\ref{eq:boltznchi}), and result in the following expressions,
\begin{align}
n_{\chi} ^{(n)}(T) \;&=\;   \frac{g_Ag_B g_{\psi} g_{\chi}\, g_{*s}^{0} \sqrt{6{ b}}\, 2^{n+3} \Gamma(\frac{n}{2}+3)^2 \zeta(\frac{n}{2}+3)^2 \mathcal{S}(n) M_P T_{\rm reh}^{n+4} }{ (g_{*s}^{\rm reh})^{3/2} \pi^5 (6-n) (n+4) \Lambda^{n+2}}  \left(\frac{T}{T_{\rm reh}}\right)^{3}\,,
\end{align}
and
\begin{align} \notag
\Omega_{\chi}^{(n)}h^2 \;\simeq\;\; &\frac{g_Ag_B g_{\psi} g_{\chi} \sqrt{{ b}}\,2^{n+3} \Gamma(\frac{n}{2}+3)^2 \zeta(\frac{n}{2}+3)^2 \mathcal{S}(n)  }{  (6-n) (n+4) } \left(\frac{106.75}{g_{*s}^{\rm reh}}\right)^{3/2}\\
& \times\left(\frac{T_{\rm reh}}{\Lambda}\right)^{n+1} \left(\frac{10^{16}\,{\rm GeV}}{\Lambda}\right) \left(\frac{\mncdm }{1\,{\rm keV}}\right) \,.
\end{align}
For $n=6$, the late-time DM number density and the present DM relic abundance can be found by integration, and are given by
\beq
n_{\chi}^{(6)}(T) \;=\;  \frac{ g_Ag_B g_{\psi} g_{\chi}\, g_{*s}^{0}  16384 \pi^7 \mathcal{S}(6) M_P T_{\rm reh}^{10} }{ 6615 (g_{*s}^{\rm reh})^{3/2} \Lambda^{8}} \left(\frac{2{ b}}{3}\right)^{1/2}  \left(\frac{T}{T_{\rm reh}}\right)^{3} \ln\left(\frac{T_{\rm max}}{T_{\rm reh}}\right)\,,
\eeq
and
\beq
\Omega_{\chi}^{(6)} h^2 \;=\;  g_Ag_B g_{\psi} g_{\chi}\sqrt{{ b}}\, \mathcal{S}(6)  \left(\frac{106.75}{g_{*s}^{\rm reh}}\right)^{3/2} \left(\frac{m_{\rm DM}}{1.2\,{\rm keV}}\right) \left(\frac{T_{\rm reh}}{10^6\,{\rm GeV}}\right)^7 \left(\frac{10^8\,{\rm GeV}}{\Lambda}\right)^8\ln\left(\frac{T_{\rm max}}{T_{\rm reh}}\right)\,.
\eeq\par\bigskip

\begin{figure}[!t]
\centering
    \includegraphics[width=0.6\textwidth]{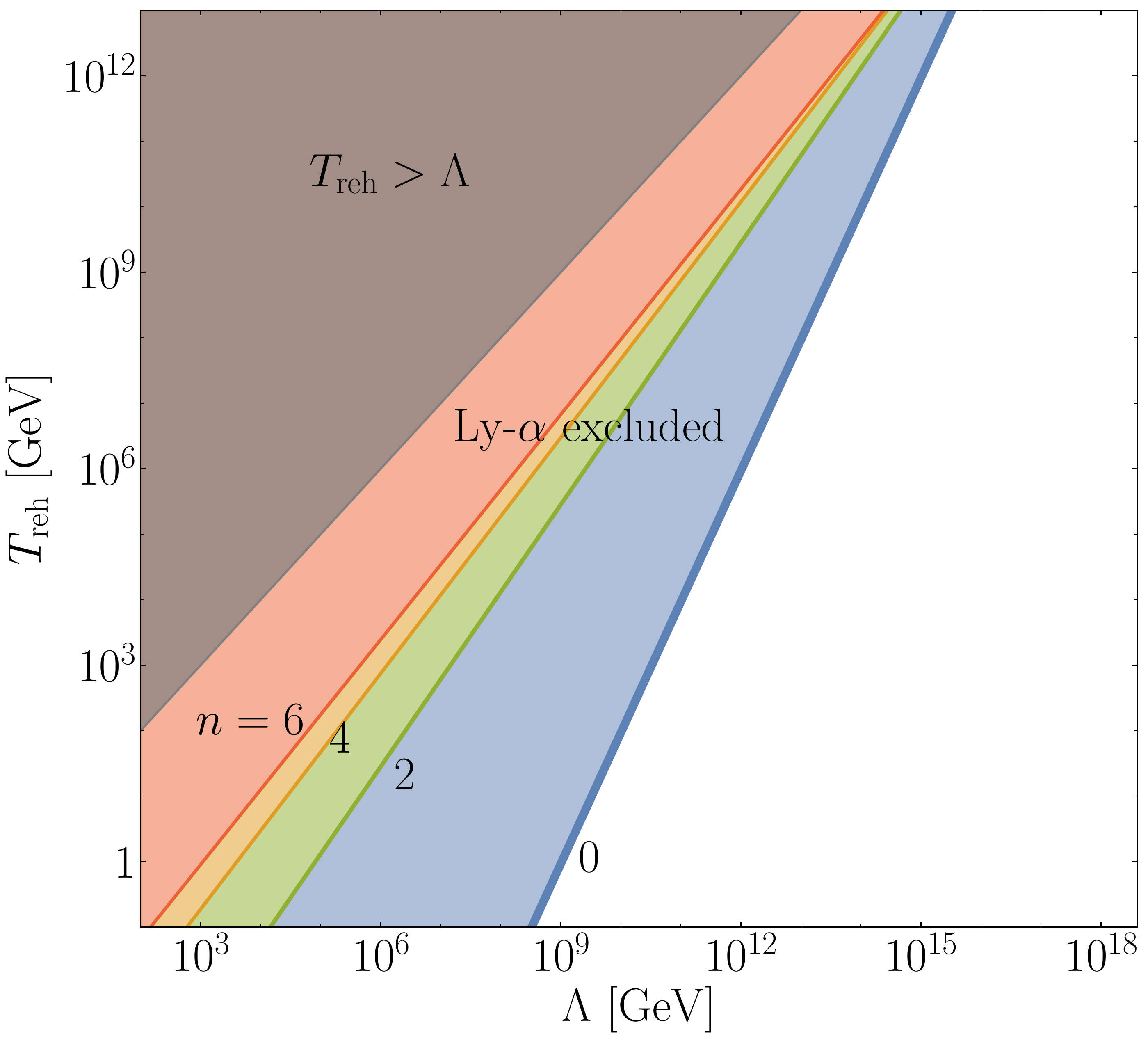}
    \caption{Ly-$\alpha$ constraint mapped on the $(\Lambda,T_{\rm reh})$ plane for thermal freeze-in, for $n=\{0,2,4,6\}$. The width of the solid lines corresponds to the difference in the mass lower bound $m_{\rm DM}$ for fermion-fermion and boson-boson scatterers. Here $g_{*s}^{\rm reh}=106.75$ and $g_A=g_B=g_{\psi}=g_{\chi}=c=1$. Note that the effective description based on (\ref{eq:M2sn}) is valid only if $\Lambda > T_{\rm reh}$.}
    \label{fig:TvsL_th}
\end{figure}
Similarly to the inflaton decay scenario, the Ly-$\alpha$ constraint on the DM mass and the DM relic abundance can be combined, in this case to exclude values of the pairs $(\Lambda,T_{\rm reh})$. Fig.~\ref{fig:TvsL_th} shows the excluded parameter space in the $(\Lambda,T_{\rm reh})$ plane, for $n=\{0,2,4,6\}$. To construct this plot we have taken the parameter $b$ to be 1, and neglected the contribution from the internal degrees of freedom of the annihilating SM particles and the scattering products. The thickness of the boundary lines, at which the bounds are saturated, corresponds to the difference between the possible initial state quantum statistics. As we discussed earlier, this difference is reduced as $n$ is increased. Annihilations with a steeper dependence on the center of mass energy allow for a wider range of values for the scale $\Lambda$ than processes with a low values of $n$. Notice that this scale cannot be taken much below the inflaton mass by construction. Indeed, in this case the effective-field-theory approach used to describe the DM-SM scattering amplitude would cease to be valid for processes involving the most energetic SM particles produced in the early universe.

The bounds tend to converge for larger values of the reheating temperature. For the special case $n=6$ the maximum temperature of the universe is taken to be the temperature at thermalization, which is dependent on the decay rate $\Gamma_{\phi}$ and therefore on the reheating temperature, $T_{\rm max}=T_{\rm max}(T_{\rm reh})$. This functional relation can be obtained by substituting in (\ref{eq:Tvstreh}) the thermalization time-scale (\ref{eq:thts}), which we discuss in more detail in the following section. \par\medskip

It is outside of the scope of our study to provide a detailed account of the implications of our analysis for the many DM models for which the UV thermal freeze-in mechanism is the dominant production channel. Moreover, for many of these constructions, the physics that gives rise to the suppression by the scale $\Lambda$ for the cross section leads to electroweak-scale DM candidates, as in Grand Unification constructions ($n=2$)~\cite{Mambrini:2013iaa,mnoqz1}, or super-heavy DM candidates, as is the case of gravitino DM from high-scale supersymmetry models ($n=6$)~\cite{Benakli:2017whb,grav2,Dudas:2018npp}. Nevertheless, we now identify a few scenarios for which $m_{\rm DM}\gtrsim {\rm keV}$ is viable for the various thermal freeze-in cases that we have discussed above. A well known example of light DM with $n=0$ freeze-in is axino DM, with a production cross section suppressed by the Peccei-Quinn scale, $\sigma\propto f_a^{-2}$~\cite{Kim:1984yn,Rajagopal:1990yx,Goto:1991gq,Chun:1995hc,Brandenburg:2004du,Strumia:2010aa,Bae:2011jb,Bae:2017dpt}.

Light DM produced from UV freeze-in can take the form of a spin-3/2 particle (the ``raritron'')~\cite{Garcia:2020hyo}. In a SM extension that contains a right-handed and/or sterile neutrino $\nu_R$ with mass $m_R$, the following Lagrangian determines the DM interactions,
\beq
{\cal L}_{3/2} \;=\; i\frac{\alpha_1}{2 M_P} \bar \nu_R \gamma^\mu [\gamma^\rho,\gamma^\sigma] \Psi_\mu F_{\rho \sigma}
 + i\frac{\alpha_2}{{2}M_P} i \sigma_2 (D^\mu H)^* \bar L  \Psi_\mu  + {\rm h.c.}
\label{Eq:l32}
\eeq
where $\Psi_{\mu}$ denotes the raritron, $F_{\rho\sigma}$ the $U(1)$ field strength tensor, and $H$ the SM Higgs doublet. When the term proportional to $\alpha_1$ dominates raritron production, the processes $\nu+H\rightarrow B+\Psi$, $H+B\rightarrow \nu+\Psi$ and $\nu+B\rightarrow H+\Psi$ populate the DM energy density during reheating, with $\sigma(s)\propto (\alpha_1 s/m_{3/2}m_R M_P)^2$, i.e.~$n=4$. On the other hand, if the $\alpha_2$ coupling dominates, $\sigma(s)\propto \alpha_2^2 s/m_{3/2}^2M_P^2$, that is $n=2$. We elaborate on the interplay between thermal and non-thermal effects for the $n=4$ case in Section~\ref{sec:phenon4}.

In scenarios inspired by modified gravity, SM-DM interactions can be mediated by a massive spin-2 particle $\tilde{h}_{\mu\nu}$~\cite{Bernal:2018qlk,Anastasopoulos:2020gbu}, 
\beq
\mathcal{L}_2 \;=\; \frac{1}{M}\tilde{h}_{\mu\nu} \left(\alpha_{\rm SM}T_{\rm SM}^{\mu\nu} + \alpha_{\rm DM}T_{\rm DM}^{\mu\nu}\right)\,,
\eeq
where $T_{\rm SM (DM)}^{\mu\nu}$ is the SM (DM) energy-momentum tensor and $M$ some energy scale. Scalar, fermion or vector light DM can be produced through thermal freeze-in during reheating. For a heavy mediator, $m_{\tilde{h}}\gg T_{\rm reh}$, $\sigma \propto \alpha_{\rm SM}^2\alpha_{\rm DM}^2 s^3/(M m_{\tilde{h}})^4$, realizing $n=6$. For a lighter mediator, $m_{\tilde{h}}\ll T_{\rm reh}$, $\sigma \propto \alpha_{\rm SM}^2\alpha_{\rm DM}^2 s/M^4$, i.e.~$n=2$. \par\bigskip

We finish this section by briefly addressing the more exotic $n>6$ scenarios, arising e.g.~in vector non-Abelian DM constructions with SM-DM interactions mediated by a heavy $Z'$~\cite{Bhattacharyya:2018evo}. For these cases, (\ref{eq:ftfapp}) is a good approximation for the phase space distribution. A straightforward computation reveals that the ``plateau'' behaviour observed for $\bar{f}_{\rm TF}^{(6)}(q)$ is also present, with $q^2\bar{f}_{\rm TF}^{(n)}(q)\propto q^{\frac{n}{2}+2}$ and $q^{(13-3n)/5}$ below and at the plateau, respectively. It can be verified that the lower bound on the DM mass is suppressed with respect to the WDM case by powers of the ratio of the reheating temperature and the maximum temperature,
\beq
m_{\rm DM} \;\gtrsim\; \left( \dfrac{m_\text{WDM}}{3~\text{keV}} \right)^{4/3} \left( \dfrac{106.75}{g_{*s}^{\rm reh}} \right)^{1/3} \times \begin{cases}
23\,{\rm keV}\,\left(\dfrac{T_{\rm reh}}{T_{\rm max}}\right)\,, & n=8\,,\\[10pt]
52\,{\rm keV}\,\left(\dfrac{T_{\rm reh}}{T_{\rm max}}\right)^{5/3}\,, & n\geq 10\,.
\end{cases}\label{eq:mDMlargen}
\eeq
These limits must be complemented with the DM density bound, which in these scenarios reads
\begin{align}\notag
\Omega_{\chi}^{(n)}h^2 \;\simeq\;\; &\frac{g_Ag_B g_{\psi} g_{\chi}\sqrt{{ b}}\,2^{n+3} \Gamma(\frac{n+4}{2})\Gamma(\frac{n+6}{2})}{n-6}  \left(\frac{106.75}{g_{*s}^{\rm reh}}\right)^{3/2}\\
&\times\left(\frac{T_{\rm max}}{\Lambda}\right)^{n+1} \left(\frac{T_{\rm reh}}{T_{\rm max}}\right)^7 \left(\frac{10^{16}\,{\rm GeV}}{\Lambda}\right) \left(\frac{m_{\rm DM}}{1.8\,{\rm keV}}\right) \,.
\end{align}
Note the inteplay of the scale $\Lambda$ and the maximum and reheating temperatures. The requirement that $\Lambda \gtrsim T_{\rm max} \gg T_{\rm reh}$ narrows down the available parameter space for a given $n$, disfavoring DM masses near the bound (\ref{eq:mDMlargen}).

\subsection{Non-thermal freeze-in}\label{sec:thenonthfi}

In the previous section we have addressed the production of DM in UV-dominated freeze-in models, exploiting the fact that reheating is not an instantaneous process, as it involves a continuous transfer of inflaton energy density into its relativistic decay products. The thermalization process, during which elastic and inelastic scatterings in the primordial plasma bring it into kinetic and chemical equilibrium is also non-instantaneous. In typical perturbative reheating scenarios, thermalization is reached after the end of inflation, but well before the end of reheating, if it is mediated by (SM) gauge interactions~\cite{Harigaya:2014waa,Harigaya:2013vwa,Mukaida:2015ria}. As it was found in~\cite{Garcia:2018wtq,Harigaya:2019tzu}, even if this non-thermal (or ``pre-thermal'') window may be relatively narrow, the bulk of the DM abundance may be produced during this time interval, provided that the production scattering cross section is a sufficiently steep function of energy. More concretely, if $n>2$ in (\ref{eq:sigmas}), the inflaton decay products, with momenta $p\sim m_{\phi}$, can copiously produce DM particles, which will eventually  dominate the DM density budget despite their dilution by entropy production during the late stages of reheating. In what follows we will consider only the lowest case with even $n$ for which this pre-thermal production can dominate, that is $n=4$. 

\subsubsection{DM phase space distribution}
\label{sec:non-thermal_FI}

Assuming that the initial state particles necessary for DM production are produced directly from inflaton decay, their distribution before thermalization can be approximated by Eq.~(\ref{eq:fpreth}). This distribution is highly non-thermal, peaked at momenta $p\sim m_{\phi}$, with a cutoff at $m_{\phi}/2$. Nevertheless, due to its power-law nature, it allows a closed form computation of the collision term. For the interested reader, this calculation is presented in Appendix~\ref{app:nthfi}. Moreover, integration of the transport equation by means of (\ref{eq:Cgensol}) can also be performed analytically. The resulting phase space distribution at the thermalization time $t_{\rm th}$, when the interactions between the scatterers become sufficiently efficient to bring the plasma into thermal equilibrium, is given by
\beq
f_{\chi}(p,t_{\rm th}) \;=\; \frac{256\pi^2 g_{\psi} \Gamma_{\phi}^3 M_P^4}{15015 \Lambda^6 m_{\phi} (\Gamma_{\phi}t_{\rm th})} \bar{f}_{\rm NF}^{(4)}\left(\frac{2p}{m_{\phi}}\right)\,,
\eeq
where
\begin{align} \notag \displaybreak[0]
q^{3/2} \bar{f}_{\rm NF}^{(4)}(q) \;=\; &\theta(1-q) \bigg[ 4234-\frac{18931}{4 \sqrt{2}}- \frac{4095}{8} \sinh ^{-1}(1) -1716 q^3 -5148 q^{7/2}\\ \notag \displaybreak[0]
& \hspace{40pt} +10010 q^4 + 8008 q^{9/2} -26208 q^5 +(8190+4095 \pi ) q^{11/2} \\ \notag \displaybreak[0]
&\hspace{40pt} -7392 q^6 +990 q^{13/2}\bigg] + \frac{\theta(2-q)\theta(q-1)}{8\sqrt{q}} \bigg[  24064 +41184 q^4\\ \notag \displaybreak[0]
&\hspace{40pt} -64064 q^5 - 65520 q^6 \left(1 + \tan ^{-1}\sqrt{q-1} - \csc ^{-1}\sqrt{q}\right)\\ \notag \displaybreak[0]
&\hspace{40pt} -7920 q^7 - \sqrt{q-1} \Big( 4096 -2047  q -1194 q^2 -904 q^3 \\ \notag \displaybreak[0]
&\hspace{40pt} +40432 q^4 -108192 q^5 -59136 q^6 \Big) \\ \label{eq:fNT}
&\hspace{40pt}  -\sqrt{q} \left( 18931 \sqrt{2} + 4095 \sinh ^{-1}(1) - 4095 \sinh ^{-1}\sqrt{q-1}\right) \bigg]\,.
\end{align}

\begin{figure}[!t]
\centering
    \includegraphics[width=0.7\textwidth]{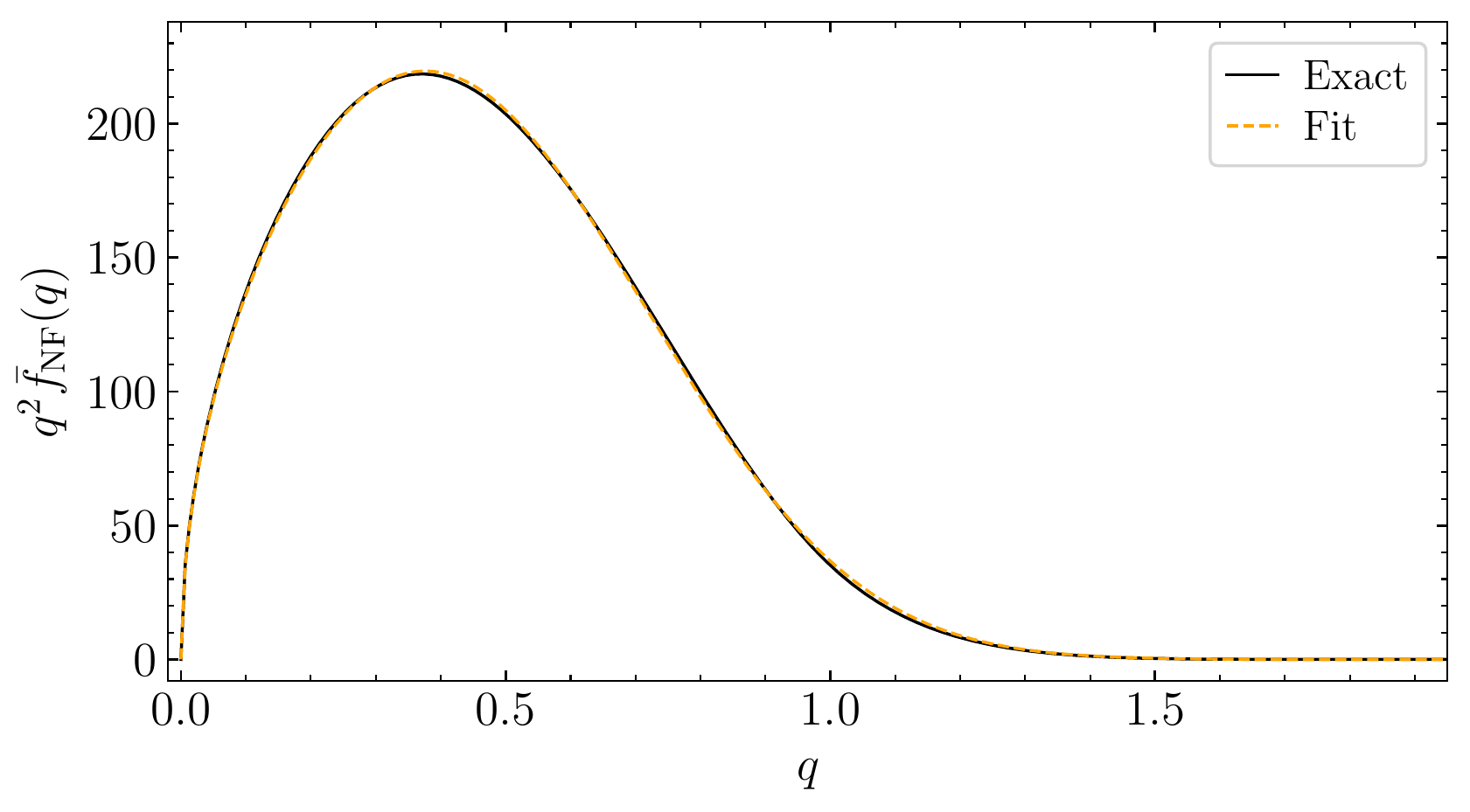}
    \caption{The rescaled distribution function $\bar{f}_{{\rm NF}}$, defined in (\ref{eq:fNT}), as a function of the rescaled momentum $q$, for DM produced from non-thermal freeze-in. Solid, black: the exact result (\ref{eq:fNT}). Dashed, orange: the fit (\ref{eq:fNTfit}).}
    \label{fig:distntfi}
\end{figure}
Fig.~\ref{fig:distntfi} shows the momentum dependence of the rescaled distribution $\bar{f}_{\rm NF}^{(4)}$ for the $n=4$ non-thermal freeze-in scenario. At low momentum the DM inherits the power-law dependence $\sim q^{-3/2}$ of the parent scatterers, c.f.~Eq.~(\ref{eq:fpreth}). At higher momentum, $q\simeq 0.4$, a peak in the distribution appears, and for larger momenta the distribution decays faster than an exponential, finally vanishing at $q=2$. Although we have at our disposal the exact form for the phase space distribution, it is nevertheless instructive to verify that a fit of the form (\ref{eq:allfit}) can be constructed:
\beq\label{eq:fNTfit}
\bar{f}_{\rm NF}^{(4)}(q) \;\approx \; 433.2 q^{-3/2} e^{-2.5 q^{2.6}}\,.
\eeq
This fit is also shown in Fig.~\ref{fig:distntfi}. \par\bigskip

In order to extend the distribution to later times, we require to know the thermalization time-scale and the expansion history from $t_{\rm th}$ to $t_0$. Denoting by $\alpha_{\rm SM}$ the gauge coupling responsible for the thermalization of the (SM) inflaton decay products, the thermalization time-scale can be approximated as follows~\cite{Harigaya:2014waa,Harigaya:2013vwa,Mukaida:2015ria}
\beq\label{eq:thts}
\Gamma_{\phi} t_{\rm th} \;\simeq\; \alpha_{\rm SM}^{-16/5} \left(\frac{\Gamma_{\phi} m_{\phi}^2}{ M_P^3} \right)^{2/5}\,.
\eeq
With the scale factor during reheating being $a(t)\propto t^{2/3}$ we finally have
\beq
f_{\chi}(p,t)\, \diff^3 \bp \;\simeq\; \frac{256\pi^2 g_{\psi}}{15015 \Lambda^6} \left(\frac{\pi^2 { b} g_{*s}^{\rm reh}}{24}\right)^{13/10} \left(\frac{\alpha_{\rm SM}^{16} T_{\rm reh}^{26} M_P^{13}}{m_{\phi}^9}\right)^{1/5}   \left(\frac{a_0}{a(t)}\right)^3 \tncdm^3\, \bar{f}_{\rm NF}^{(4)}(q)\, \diff^3\boldsymbol{q}\,,
\eeq
for $t\gg t_{\rm reh}$. Here
\beq
\tncdm \;=\; \frac{\alpha_{\rm SM}^{-32/15}}{2}\left(\frac{g_{*s}^0}{g_{*s}^{\rm reh}}\right)^{1/3} \left(\frac{\pi^2 { b} g_{*s}^{\rm reh}}{24}\right)^{2/15} \left(\frac{m_{\phi}}{T_{\rm reh}}\right)^{7/15} \left(\frac{m_{\phi}}{M_P}\right)^{16/15}T_0\,,
\eeq
and $b$ was defined in (\ref{eq:Tvstreh}).

\subsubsection{Power spectrum and Ly-$\alpha$ constraints}

The analytical expression for the DM phase space distribution for non-thermal freeze-in allows us to obtain the rescaled bound of the DM mass from Eq.~(\ref{eq:mncdm_fromeos}). It is given by
\begin{align}
    \hatmncdm \, \gtrsim \, &  \,  \left( \dfrac{m_\text{WDM}}{3~\text{keV}} \right)^{4/3} \left( \dfrac{\alpha_\text{SM}}{0.03}\right)^{-32/15} \left( \dfrac{{ b}}{3/5}\right)^{2/15} \left( \dfrac{106.75}{g_{*s}^{\rm reh}} \right)^{1/5}  \nonumber  \\ & \times  \,	\left( \dfrac{10^{10}~\text{GeV}}{T_{\text{reh}}}  \right)^{7/15}  \left( \dfrac{m_\phi}{3\times 10^{13}~\text{GeV}} \right)^{23/15} 
  \begin{cases}
0.44~\text{keV}\,, ~ & {\rm Exact}\,,\\
0.45~\text{keV}\,, ~& {\rm Fit}\,,
\end{cases}
\label{eq:mDM_bound_NTHUVFI}
\end{align}
where, for completeness, we have also included the bound obtained by using the fit approximation (\ref{eq:fNTfit}). Their agreement is excellent. Note here that, for the fiducial values $m_{\rm WDM}=3\,{\rm keV}$ and $T_{\rm reh}=10^{10}\,{\rm GeV}$, the lower bound on the DM mass is one order of magnitude smaller than for WDM, and can be decreased by increasing the reheating temperature. Unlike the case of production from inflaton decay, where a hot spectrum could be obtained for large masses due to a reduced momentum redshift and a large momentum at production, in this case a colder spectrum is obtained due to the redshift that occurs between $t_{\rm th}$ and $t_{\rm reh}$, in addition to the redshift from $t_{\rm reh}$ to $t_0$; despite having $\langle p\rangle \sim m_{\phi}/2$ at production.

\begin{figure}[!t]
\centering
    \includegraphics[width=0.75\textwidth]{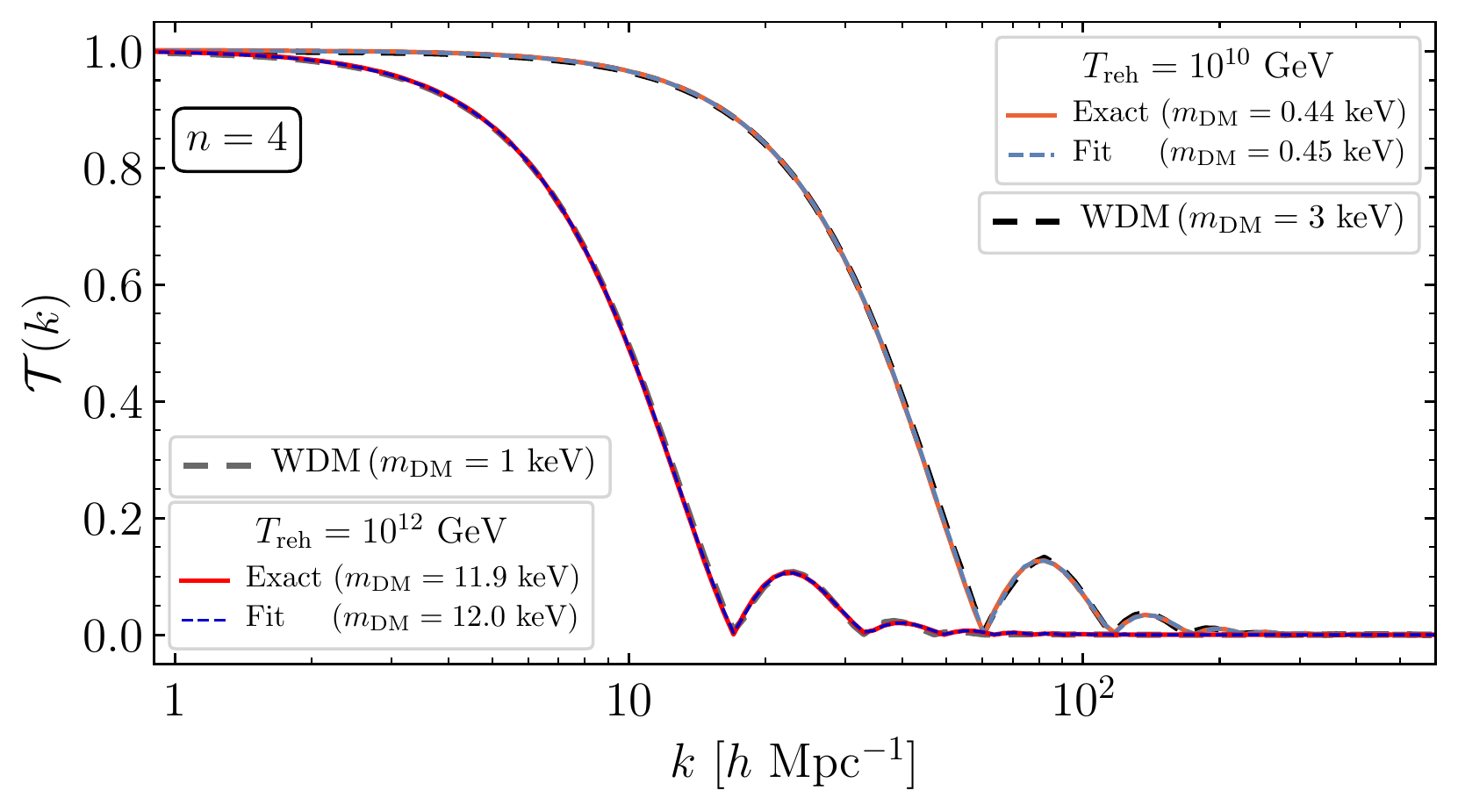}
    \caption{Linear transfer function for DM produced by non-thermal UV freeze-in, for $n=4$. Shown here are the results for the exact and fit approximations (\ref{eq:fNT}) and (\ref{eq:fNTfit}). The DM masses are taken from the rescaled bound (\ref{eq:mDM_bound_NTHUVFI}), with $m_{\rm WDM}=1\,{\rm keV}$ and $T_{\rm reh}=10^{12}\,{\rm GeV}$ (left), and $m_{\rm WDM}=3\,{\rm keV}$ and $T_{\rm reh}=10^{10}\,{\rm GeV}$ (right). For comparison we show $\mathcal{T}(k)$ for WDM in each case.}
    \label{fig:TF_NTHUVFI}
\end{figure}

Fig.~\ref{fig:TF_NTHUVFI} shows the transfer function for non-thermal freeze-in corresponding to the rescaled DM masses (\ref{eq:mDM_bound_NTHUVFI}). Two sets of parameters are explored, one with $m_{\rm WDM}=3\,{\rm keV}$ and $T_{\rm reh}=10^{10}\,{\rm GeV}$ (right), and with $m_{\rm WDM}=1\,{\rm keV}$ and $T_{\rm reh}=10^{12}\,{\rm GeV}$ (left). Both cases overlap with the reference WDM for all the range of scales shown, with a relative difference below the percent level (c.f.~Fig.~\ref{fig:transfer_function_residual}), once again demonstrating the validity of our rescaling program. 
\subsubsection{Relic density and phenomenology}\label{sec:phenon4}

The DM number density at late times can be easily calculated given that 
\beq
\int_0^{\infty} \diff q\, q^2 \bar{f}^{(4)}_{\rm NF}(q) \;=\; \frac{2145}{14}\,.
\eeq
It is given by the following expression, for $T\ll T_{\rm reh}$,
\beq
n_{\chi}(T) \;\simeq\; \frac{8 g_{\psi}g_{\chi}}{49 \Lambda^6}\left(\frac{g_{*s}^0}{g_{*s}^{\rm reh}}\right) \left(\frac{\pi^2 { b} g_{*s}^{\rm reh}}{24}\right)^{17/10} \left(\frac{ m_{\phi}^{14} T_{\rm reh}^{34} }{\alpha_{\rm SM}^{16}M_P^3}\right)^{1/5}   \left(\frac{T}{T_{\rm reh}}\right)^3\,.
\eeq
In turn, the DM abundance has the following form,
\begin{align} \notag
\Omega_{\chi}h^2 \;\simeq\;  (0.72{ b})^{17/10}  g_{\psi}g_{\chi} &\left(\frac{g_{*s}^{\rm reh}}{106.75} \right)^{7/10} \left(\frac{0.03}{\alpha_{\rm SM}}\right)^{16/5} \left(\frac{m_{\phi}}{3\times 10^{13}\,{\rm GeV}}\right)^{14/5}\\ &\times \left(\frac{10^{12}\,{\rm GeV}}{\Lambda}\right)^6 \left(\frac{T_{\rm reh}}{10^{10}\,{\rm GeV}}\right)^{19/5} \left(\frac{ \mncdm  }{1\,{\rm keV}}\right)  \,.
\end{align}
If the decay of the inflaton to the parent scatterers is subdominant with respect to other channels, the previous expression is reduced by the corresponding branching ratios, $\Omega_{\chi} \rightarrow {\rm Br}_A {\rm Br}_B \Omega_{\chi}$. \par\bigskip

\begin{figure}[!t]
\centering
    \includegraphics[width=0.6\textwidth]{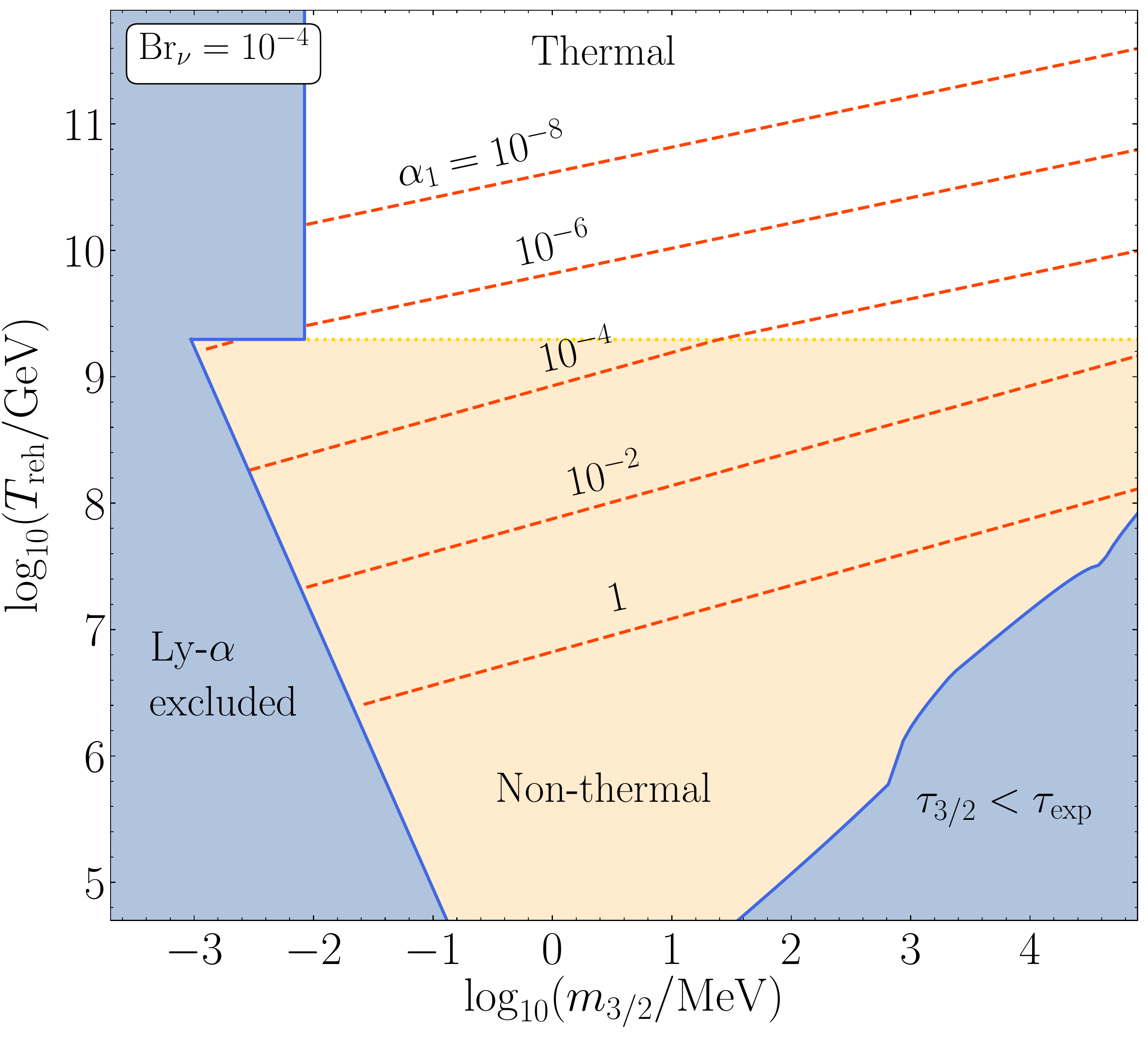}
    \caption{$\Omega_{3/2}h^2=0.1$ curves in the ($m_{3/2}$, $T_{\rm reh}$) plane for the raritron model~(\ref{Eq:l32}), for different values of $\alpha_1$, and ${\rm Br}_{\nu}=10^{-4}$. Shown in blue are the astrophysical constraints on the lifetime from $\gamma$-ray observations and the Ly-$\alpha$ constraint on the mass. In the orange region freeze-in occurs non-thermally. See~\cite{Garcia:2020hyo} for details.}
    \label{fig:raritron}
\end{figure}

Combining the DM abundance and mass constraints, we can obtain a bound on the reheating temperature analogous to that for thermal freeze-in. In this case, the bound takes the form 
\begin{align}
T_{\rm reh} \;\lesssim\; \frac{ 10^{10}\,{\rm GeV} }{(g_{\psi}g_{\chi})^{3/10}} & \left(\frac{3\,{\rm keV}}{m_{\rm WDM}}\right)^{2/5} \left( \dfrac{\alpha_\text{SM}}{0.03}\right)^{8/5} \left( \dfrac{3/5}{{ b}}\right)^{11/20} \left( \dfrac{g_{*s}^{\rm reh}}{106.75} \right)^{3/20}  \nonumber  \\ & \times  \,	 \left( \dfrac{3\times 10^{13}~\text{GeV}}{m_\phi} \right)^{13/10} \left(\frac{\Lambda}{10^{12}\,{\rm GeV}}\right)^{9/5}\,.
\end{align}
As we mentioned in Section~\ref{sec:phenoFI}, freeze-in with $n=4$ is realized for scattering processes involving neutrinos for the spin-3/2 raritron. With the interactions mediated dominantly by the first term of (\ref{Eq:l32}), and under the assumption that the inflaton predominantly decays to Higgs bosons, the thermal and non-thermal relic abundances can be written as 
\begin{align}
\Omega_{3/2}^{\rm thermal} h^2  \;\simeq\;\; &0.1 \left(\frac{\alpha_1}{1.1\times10^{-3}}\right)^{2}\left(\frac{106.9}{g_{*s}^{\rm reh}}\right)^{3 / 2}\left(\frac{T_{\rm reh } }{10^{10} \, \mathrm{GeV}}\right)^{5}  \nonumber \\
& \quad \times \left(\frac{m_{\nu}}{0.15 \, \mathrm{eV}}\right)\left(\frac{10^{14} \, \mathrm{GeV}}{m_{R}}\right)\left(\frac{10^{4} \, \mathrm{GeV}}{m_{3 / 2}}\right),
\label{Eq:omegabis}
\end{align}
and
\begin{align} \notag
\Omega_{3/2}^{{\rm non}\mbox{-}{\rm thermal}}h^2 \;\simeq\;\; &0.1 \left(\frac{\alpha_1}{1.1\times10^{-3}}\right)^2 \left(\frac{g_{*s}^{\rm reh}}{106.5}\right)^{7/10} \left(\frac{0.030}{\alpha_{\rm SM}}\right)^{16/5}
\left(\frac{m_{\phi}}{3\times 10^{13}\,{\rm GeV}}\right)^{14/5}
\\ 
& \times  \left(\frac{m_{\nu}}{0.15\,{\rm eV}}\right)\left(\frac{10^4\,{\rm GeV}}{m_{3/2}}\right) \left(\frac{10^{14}\,{\rm GeV}}{m_R}\right) \left(\frac{T_{\rm reh}}{10^{10}\,{\rm GeV}}\right)^{19/5} \left(\frac{{\rm Br}_{\nu}}{7\times 10^{-4}}\right)\,.
\end{align}
For ${\rm Br}_{\nu} \gtrsim 7\times 10^{-4} (T_{\rm reh}/10^{10}\,{\rm GeV})^{6/5}$ the DM energy density is mostly comprised of non-thermally produced raritrons. Fig.~\ref{fig:raritron} illustrates the different domains in the parameter space where freeze-in production can occur either thermally or non-thermally. For definiteness we have fixed the branching ratio to neutrinos to $10^{-4}$. We observe that for $T_{\rm reh}\gtrsim 2\times 10^9\,{\rm GeV}$ the production is dominated by thermal effects, which are most efficient around $T\sim T_{\rm reh}$. The Ly-$\alpha$ bound in this case is independent of the inflationary parameters, and it is given by (\ref{eq:mDM_bound_UVFI_n4}). On the other hand, for lower reheating temperatures freeze-in occurs before thermalization is complete. The limit on the mass is therefore given by (\ref{eq:mDM_bound_NTHUVFI}) and is $T_{\rm reh}$-dependent. We finally note the region in the botton right corner, in which the unstable raritron would decay faster to photons than experimentally allowed. We refer the interested reader to~\cite{Garcia:2020hyo} for further details. 

\section{Light, but not too light, dark matter} \label{sec:lightbntl}

In our exploration of non-equilibrium DM production scenarios we have focused on WDM mimickers: DM which is sufficiently relativistic during structure formation to leave a detectable imprint in the matter power spectrum at scales below their free-streaming horizon, yet heavy enough to be indistinguishable from CDM at late times. Such a DM candidate may, in principle,  contribute significantly to the number of effective relativistic species, $N_{\rm eff}$, at recombination or BBN. It is for this reason that we evaluate this contribution in this section.
Well after DM decoupling, the total energy density in the Universe can be written as
\beq\label{eq:neffdef}
\rho \;=\; \left[ 1 + \frac{7}{8} \left(\frac{T_{\nu}}{T}\right)^4 \gls{Neff} \right]\rho_{\gamma} + \rho_{\chi} + \cdots
\eeq
where $\rho_{\gamma}$ denotes the energy density of photons, $T_{\nu}$ is the effective neutrino temperature, $T_{\nu}/T = (4/11)^{1/3}$ after electron-positron annihilation, and the dots include all other contributions to $\rho$, such as that of baryons. In the SM, $N_{\rm eff}=3.046$. In order to determine the contribution of DM to $N_{\rm eff}$ one {could naively think of placing} its energy density inside the brackets. However, the equation of state of our DM candidates lies in between that for radiation and pressureless dust {(and varies in time)}. Hence, as it is 
{sometimes done}~\cite{Merle:2015oja,Baumholzer:2019twf,Baldes:2020nuv}, we separate its energy density into relativistic and non-relativistic parts, $\rho_{\chi} = (\rho_{\chi}-\mncdm n_{\chi}) + \mncdm n_{\chi}$, and absorb only the former into the factor multiplying $\rho_{\gamma}$ in (\ref{eq:neffdef}).\footnote{This (artificial) splitting has the effect of suppressing the mass- and time-dependent contribution to $\Delta N_{\rm eff}$, which, if dominant, would signal the non-relativistic nature of the DM.} With this, we can then write
\begin{align}\notag
\Delta N_{\rm eff} \;&=\; \frac{8}{7}\left( \frac{T}{T_{\nu}} \right)^4 \frac{\rho_{\chi} - \mncdm n_{\chi}}{\rho_{\gamma}}\\ \notag
&=\; \frac{8\pi \Omega_{\chi}}{7\Omega_{\gamma}} \left(\frac{g_{*s}(T)}{g_{*s}^0}\right)^{4/3}\left(\frac{T}{T_{\nu}}\right)^4\left(\frac{\tncdm}{\mncdm}\right)\\ 
&\hspace{20pt} \times \left[\left\langle \sqrt{q^2 + \left(\frac{g_{*s}^0}{g_{*s}(T)}\right)^{2/3}\left(\frac{\mncdm}{\tncdm}\right)^2 \left(\frac{T_0}{T}\right)^2 } \right\rangle - \left(\frac{g_{*s}^0}{g_{*s}(T)}\right)^{1/3}\left(\frac{\mncdm}{\tncdm}\right) \left(\frac{T_0}{T}\right) \right]\,.
\end{align}
The contribution to $N_{\rm eff}$ depends on the ratio $\mncdm/\tncdm$. As expected, for a given $T$, decreasing this ratio increases $\Delta N_{\rm eff}$. Hence, the maximal contribution to the number of relativistic degrees of freedom for a given DM scenario is obtained by saturating the Ly-$\alpha$ constraint, which by virtue of (\ref{eq:mncdm_fromeos}) fixes the value of the mass to $\tncdm$ ratio up to $\sqrt{\langle q^2\rangle}$, 
\begin{align}\notag
\Delta N_{\rm eff,max} \;\simeq\; \frac{1.4\times 10^{-4} }{\sqrt{\langle q^2\rangle}} &\left(\frac{g_{*s}(T)}{g_{*s}^0}\right)^{4/3} \left(\frac{\Omega_{\chi}h^2}{0.1}\right) \left(\frac{3\,{\rm keV}}{m_{\rm WDM}}\right)^{4/3} \left(\frac{T}{T_{\nu}}\right)^4\\
&\times \left[ \left\langle \sqrt{q^2 + \mu_*(T)^2} \right\rangle - \mu_*(T) \right]\,,
\end{align}
where
\beq
\mu_*(T) \;\equiv\; \sqrt{\langle q^2\rangle}\left(\frac{g_{*s}^0}{g_{*s}(T)}\right)^{1/3} \left(\frac{3\,{\rm keV}}{m_{\rm WDM}}\right)^{4/3} \left(\frac{7.56\,{\rm keV}}{T}\right)\,.
\eeq

For all the DM production mechanisms that we consider in this work the geometric and arithmetic means of $q$ are approximately $\mathcal{O}(1)$. The largest value of $\sqrt{\langle q^2\rangle}$ ($\langle q\rangle$) is 3.4 (1.7) corresponding to the $n=4$ thermal freeze-in case, while the smallest is 0.3 (0.2) for the non-thermal decay case. With this in mind, we can immediately verify that DM with the lowest allowed mass will be relativistic at $T_{\rm BBN}\simeq 4\,{\rm MeV}$, as $\mu(T_{\rm BBN})\sim 10^{-3}\langle \sqrt{q^2}\rangle$. Hence,
\beq
\Delta N_{\rm eff}(T_{\rm BBN}) \;\lesssim\; 5.4\times 10^{-4} \left(\frac{\langle q\rangle}{\sqrt{\langle q^2\rangle}}\right) \left(\frac{\Omega_{\chi}h^2}{0.1}\right) \left(\frac{3\,{\rm keV}}{m_{\rm WDM}}\right)^{4/3} \,,
\eeq
well below the constraining power of BBN computations, for which $2.3<N_{\rm eff}<3.4$~\cite{Lisi:1999ng,Cyburt:2015mya}. On the other hand, at $T_{\rm CMB}\simeq 0.26\,{\rm eV}$, $\mu(T_{\rm CMB})\sim 3\times10^4\langle \sqrt{q^2}\rangle\gg 1$, {so that} DM is mostly non-relativistic, and contributes negligibly to $N_{\rm eff}$, $\Delta N_{\rm eff}\lesssim 9\times 10^{-9}$, far below the current and projected detectability thresholds~\cite{Aghanim:2018eyx,DiValentino:2019dzu,Abazajian:2019eic}. We can then conclude that for none of the production scenarios explored in this work does DM significantly contributes to the amount of non-photonic relativistic species. 

\section{Conclusions}
\label{sec:conclusion}
We have investigated the imprint on the matter power spectrum at small scales of Non-Cold Dark Matter (NCDM, i.e.\ a DM species with a non-vanishing equation of state parameter $w$) produced in an out-of-equilibrium state. The ratio of the matter spectrum for NCDM to that in $\Lambda$CDM features a cutoff at large Fourier modes. Rather generically, the cutoff scale, i.e.~the free-streaming horizon, depends only on $w$. The Lyman-$\alpha$ forest constraint on the Warm Dark Matter mass $m_\text{WDM}$ can be translated into a constraint on $w$, being $w_\wdm(\rm{today})\lesssim 10^{-15}$. By comparing the theoretical value of $w$ for a given NCDM model with a non-thermal phase space distribution to $w_\wdm$, we can map the constraint on $m_\text{WDM}$ to the DM mass in the NCDM model. The key result of this paper, which we have illustrated with many examples, is that our mapping procedure allows to translate the Lyman-$\alpha$ bound on $m_\text{WDM}$ to non-equlibrium NCDM scenarios without performing a numerically costly computation of the power spectrum for each model. All that is necessary to obtain the (lower) bound on the DM mass is to determine the present DM momentum dispersion and the first two moments of the corresponding distribution function. 

To test our formalism we have considered several NCDM production mechanisms. In all of them, the assumption of DM decoupling immediately after production is implicit, ensuring that its phase space distribution is preserved until later times, up to redshift effects. The scenarios that we have considered correspond to DM production from scalar condensates (inflaton, moduli), thermalized and non-thermalized particles, and thermal and non-thermal freeze-in. In each case we have computed the DM phase space distribution either analytically or numerically by means of the Boltzmann equation and computed numerically the linear matter power spectrum using the \texttt{CLASS} code. We compared our results with the WDM linear power spectrum (assuming a fermionic WDM benchmark candidate), showing a good agreement with our matching procedure on the Lyman-$\alpha$ bound on $m_\text{WDM}$. { Our matching procedure is performed with a relative precision of $\sim 3\%$ for most of the scenarios whereas $\sim 10\%$ for the least precise cases, over the range of Fourier wavenumbers of interest for Lyman-$\alpha$ data. Such a precise matching highlights one of our main conclusion that the linear transfer function, for all the cases considered in this work, is essentially controlled by the single parameter $w$.}

\par\bigskip

\noindent
{\bf Phase space distributions.} For {all but two}
of the scenarios considered in the present work, we have shown that the DM phase space distribution can be well fitted by a generalized distribution of the form
\begin{equation}
f(q)  \, \propto \, q^\alpha\,\exp{ \left(-\beta \, q^\gamma \right)}
\label{eq:PSDalphabetagamma}
\end{equation} 
with constant $\alpha>-3$ and $\beta,\gamma>0$, as required for the DM number density to be finite. The only scenarios in which this fitting fails are for thermal freeze-in production with a SM-DM scattering cross section $\sigma(s)\propto s^{n/2}$ and $n\geq 6$, for DM production from strongly stabilized moduli decay, when the inflaton and modulus decay rates are comparable, and for the decay of a non-thermalized particle with $\mathcal{R}\gg 1$. For the rest of the scenarios the values of the parameters $\alpha, \beta$ and $\gamma$ are summarized in Table~\ref{tab:alphabetagamma}. 
\begin{table}[t]
\centering
\begin{tabular}{|c|c|c|c|c|c|c|}
\hline \multicolumn{2}{|c|}{Scenario} &  $\alpha$ & $\beta$ & $\gamma$ & Figure & Section \\
\hline  \hline \multicolumn{2}{|c|}{Inflaton decay} & -3/2 & 0.74 & 1.00& \ref{fig:distinf}& \ref{sec:inflaton_PSD}\\
\hline  \multirow{2}{*}{Moduli decay} & during reheating & -3/2 & 1.00 & 3/2 & \ref{fig:distmod} & \ref{sec:moduli_PSD}\\
\cline{2-7} & after reheating & -1.00 & 1.00 & 2.00 & \ref{fig:distmod} & \ref{sec:moduli_PSD}\\
\hline  \multicolumn{2}{|c|}{Thermal decay} & -1/2 & 1.00 & 1.00& \ref{fig:distTD}& \ref{sec:thermal_decay_PSD}\\

\hline  \multirow{2}{*}{Non-thermal decay} & non-relativistic & - & - & -& \ref{fig:distNTDNR}& \ref{sec:non-thermal_decay_PSD}\\
\cline{2-7} & relativistic & -5/2 & 0.74 & 2.00& \ref{fig:distNTD}& \ref{sec:non-thermal_decay_PSD}\\
\hline  \multirow{3}{*}{UV Freeze-in ($n=0$)} & BB & -0.70 & 1.13 & 1.00 & \ref{fig:low_n} & \ref{sec:therma_FI_PSD}\\
\cline{2-7} & FB & -0.51 & 1.10 & 1.00 & \ref{fig:low_n} & \ref{sec:therma_FI_PSD}\\
\cline{2-7} & FF &  -0.29 & 1.11 & 1.00 & \ref{fig:low_n} & \ref{sec:therma_FI_PSD}\\
\hline \multirow{3}{*}{UV Freeze-in ($n=2$)} & BB & -0.51 & 0.91 & 1.00 & \ref{fig:low_n} & \ref{sec:therma_FI_PSD}\\
\cline{2-7} & FB & -0.42 & 0.90 & 1.00 & \ref{fig:low_n} & \ref{sec:therma_FI_PSD}\\
\cline{2-7} & FF & -0.33 & 0.90 & 1.00 & \ref{fig:low_n} & \ref{sec:therma_FI_PSD}\\
\hline \multirow{3}{*}{UV Freeze-in ($n=4$)} & BB &  -1.79 & 0.06 & 1.98 & \ref{fig:low_n} & \ref{sec:therma_FI_PSD}\\
\cline{2-7} & FB  & -1.79 & 0.06 & 2.04 & \ref{fig:low_n} & \ref{sec:therma_FI_PSD}\\
\cline{2-7} & FF &  -1.79 & 0.05 & 2.10 & \ref{fig:low_n} & \ref{sec:therma_FI_PSD}\\
\hline \multirow{3}{*}{UV Freeze-in ($n=6$)} & BB & - & - & - & \ref{fig:neq6} & \ref{sec:therma_FI_PSD}\\
\cline{2-7} & FB  & - & - & - & \ref{fig:neq6} & \ref{sec:therma_FI_PSD}\\
\cline{2-7} & FF & - & - & - & \ref{fig:neq6} & \ref{sec:therma_FI_PSD}\\
\hline \multicolumn{2}{|c|}{Non-thermal UV Freeze-in} & -3/2 & 2.5 & 2.6 & \ref{fig:distntfi}& \ref{sec:non-thermal_FI}\\
\hline
\end{tabular}
\caption{
Fit parameters $\alpha, \beta, \gamma$ for a distribution function of the form of Eq.~(\ref{eq:PSDalphabetagamma}) for each scenario considered in this work. For the UV Freeze-in case, BB, FB or FF denote the thermal distributions of the parents B stands for Bose-Einstein and F for Fermi-Dirac. The parameter $\gamma$ is fixed when the fitted expression with $\alpha$ and $\beta$ only allows to accurately describe the numerical phase space distribution. For each distribution, the derivation can be found in the corresponding section and is represented in the corresponding Figure, accompanied by the exact (numerical or analytical) form of the distribution.}
\label{tab:alphabetagamma}
\end{table}

For the decay of a scalar condensate, the form of the distribution is determined by the expansion rate. If the decay occurs during a period with a background equation of state $w_{\rm B}$, then $f(q)\propto q^{\frac{3}{2}(w_{\rm B}-1)}$ at low momentum. On the other hand, the exponential decay of the condensate makes $f(q)\propto \exp(q^{\frac{3}{2}(w_{\rm B}+1)})$ at large momentum. For a stabilized modulus the combination of both regimes results into (\ref{eq:PSDalphabetagamma}).\footnote{{Unless its decay occurs close to the end of reheating.}}  For inflaton and non-stabilized moduli, the low momentum part of the distribution is populated during the matter-like oscillations of the field, with $w_{\rm B}\simeq 0$, while the exponential tail is populated during the subsequent radiation domination, with $w_{\rm B}=1/3$.

In the case of thermal decays, the integrated thermal distribution of the progenitor particles leads to an enhanced distribution at low momentum, $f(q)\propto q^{-1/2}$, while preserving the thermal exponential tail. An analogous effect is present in the decay of a (relativistic) non-thermalized relic produced from the decay of the inflaton. Here the shift to lower momenta corresponds to $f(q)\propto q^{-5/2}$, while the Gaussian tail is preserved. More interesting is the decay of a non-equilibrated non-relativistic relic. In this case we find that the phase space distribution is skewed toward large momentum, with a sharp cutoff dependent on the inflaton and decaying particle masses and widths. 

Thermal freeze-in production during reheating encompasses a variety of scenarios that can be distinguished by the nature of the scattering particles that produce DM, which can be any combination of fermions and bosons, and by the dependence of the production cross section on the center of mass energy. In all cases, the statistics of the scatterers manifests itself mostly as an overall normalization of the distribution function, and only mildly through the shape of the distribution, as it can be seen in Table~\ref{tab:alphabetagamma}. Moreover, the role of statistics is greatly reduced for a steeper energy dependence of the cross section. Regarding this dependence, for $n=0,2$, a shift of the thermal distribution to lower momenta is observed, similar to thermal decays. At $n=4$, the shift is greater, and the thermal exponential tail is lost. For $n\geq 6$, the distribution depends now on two scales: the reheating temperature $T_{\rm reh}$ and the maximum temperature after thermalization $T_{\rm max}$. The result is a distribution with a peak with a location and amplitude inversely dependent on the ratio $T_{\rm max}/T_{\rm reh}\gg 1$, that increases as $q^{n/2}$ below this peak, and decreases as $q^{-\frac{3}{5}(n-1)}$ for momenta larger than the peak and smaller than $q\sim 10$, beyond which the distribution acquires an exponential suppression. 

If the DM parent scatterers are directly produced from inflaton decay, for $n>2$ the DM relic abundance can be produced prior to the thermalization of the primordial plasma. We considered the $n=4$ case, obtaining a closed form expression for the phase space distribution. At low momentum the distribution inherits the power-law increase of the parent scatterers, $f(q)\propto q^{-3/2}$, while for $q\gtrsim 1$ a sharp decrease is observed, with $\gamma\simeq 2.6$.\par\bigskip

\noindent
{\bf Ly-$\alpha$ constraints.} Given the parametrization (\ref{eq:PSDalphabetagamma}), our re-scaling of the WDM Ly-$\alpha$ constraint on the DM mass can be written as
\beq
\hatmncdm \, \gtrsim \, 7.56~\text{keV}  \,  \left( \dfrac{m_\wdm}{3~\text{keV}} \right)^{4/3} \left( \dfrac{\langle p \rangle_{0}}{T_{0}} \right) \, \sqrt{\dfrac{\Gamma \left(\frac{3+\alpha}{\gamma} \right) \, \Gamma \left(\frac{5+\alpha}{\gamma} \right)}{\Gamma^2 \left(\frac{4+\alpha}{\gamma} \right)}}\,,
\eeq
where $\langle p\rangle_0$ denotes the mean DM momentum at present time. The previous expression can be read as follows: the non-thermal shape of the phase space distribution represents, in most cases, an $\mathcal{O}(1)$ correction to the DM mass bound. It is the overall normalization of the distribution what mostly contributes to increasing or decreasing this bound. This normalization depends on the production mechanism, the mean energy of the parent fields, and the expansion history from DM decoupling to the present epoch.

As it can be expected, for all thermal production mechanisms, the resulting Ly-$\alpha$ bound is only mildly corrected with respect to that for WDM. From thermal decays, $m_{\rm DM}\simeq 7.3-7.5\,{\rm keV}$ depending on whether the decaying particle is a fermion or a boson, respectively. For thermal scatterings, we also see a small spread, with $m_{\rm DM}\simeq 6.4-7.3\,{\rm keV}$ for $n=0$, $m_{\rm DM}\simeq 8.0-8.5\,{\rm keV}$ for $n=2$, and $m_{\rm DM}\simeq 8.3-8.5\,{\rm keV}$ for $n=4$. Even for $n=6$, we obtain $m_{\rm DM}\simeq 4.5-5.9\,{\rm keV}$ for $T_{\rm max}/T_{\rm reh}=50-10$. 

For the decay of scalar condensates we find a significant amount of rescaling of the Ly-$\alpha$ bounds. For unstabilized moduli or inflaton decay, the constraint is proportional to the field mass, and inversely proportional to its reheating temperature. If we restrict this ratio to be larger than one, we obtain that $m_{\rm DM}$ can be as low as $1\,{\rm keV}$ or as large as $1\,{\rm EeV}$, as shown in Fig.~\ref{fig:modmass}. For an inflaton of mass $m_{\phi}=3\times 10^{13}\,{\rm GeV}$, $m_{\rm DM}\simeq 4\,{\rm keV} -40\,{\rm EeV} $ for $T_{\rm reh}=10^{13}-10^{-3}\,{\rm GeV}$. For stabilized moduli, which by assumption never dominate the energy density of the Universe, the DM mass bound depends on the ratio of the modulus and the background temperature at the time of its decay. If the modulus decays during reheating the mass bound depends also on the ratio of its decay rate to that of the inflaton. This scenario is model dependent. For the stabilized Polonyi example that we discuss in Section~\ref{sec:modulipheno}, the lower bound on the DM mass ranges from $4\,{\rm MeV}$ if the modulus decays during radiation domination, to $4\,{\rm GeV}$ if it decays during reheating. 

If DM is produced from the decay of a non-thermalized inflaton decay product, $A$, the lower bound on its mass depends on whether the intermediate particle decays while being relativistic or non-relativistic. In the former case, DM inherits a hard distribution with $p\sim m_{\phi}$. The result is a Ly-$\alpha$ bound almost identical to that of the inflaton decay case, except for a reduction of $m_{\rm DM}$ by a factor of 3. On the other hand, if the decaying particle is non-relativistic at the time of decay, the phase space distribution of DM is highly non-thermal, and the Ly-$\alpha$ bound on its mass depends on the ratio of the decaying mass to the temperature $T_{\rm dec}$ at which it decays, $m_{\rm DM} \simeq 1\,{\rm keV} (m_A/T_{\rm dec})$.

When DM is produced from the scattering of non-thermalized inflaton decay products, as in the non-thermal $n=4$ freeze-in case, it also inherits the hard momentum distribution. However, in this case the bound becomes on the thermalization timescale, as the DM momentum can be significantly redshifted away from the thermalization epoch to the end of reheating due to entropy production. Notably, this implies that for this scenario the DM mass bound can be significantly reduced relative to the WDM case. The result is a relatively complicated function of the inflaton mass, the reheating temperature and the gauge coupling that mediates the interactions that thermalize the plasma, as shown in Eq.~(\ref{eq:mDM_bound_NTHUVFI}). This results in $m_{\rm DM}\simeq 18\,{\rm eV} -0.5\,{\rm GeV} $ for $T_{\rm reh}=10^{13}-10^{-3}\,{\rm GeV}$.\par\bigskip

\noindent
{\bf Phenomenological implications.} A lower bound on the DM mass can have far reaching consequences for non-equilibrium DM model building. For thermal freeze-out, only sub-keV particle DM candidates can be ruled out from Ly-$\alpha$ forest observations. For non-equilibrated DM, the constraint can be powerful enough to rule out DM candidates, and hence SM extensions, above the electroweak scale. 

In the case of inflaton decay, the wide range for $m_{\rm DM}$ means that the suppression of structure at small scales can in principle occur for DM masses well above the electroweak scale if the reheating temperature is sufficiently low. Moreover, combining the power spectrum and DM relic abundance constraints, we have found the absolute bound ${\rm Br}_{\chi} \;\lesssim\; 1.5\times 10^{-4}$ for the branching ratio of the decay of the inflaton to DM,
which is independent on the inflaton mass and the reheating temperature. Similar conclusions are found for the decay of an unstable modulus into DM. For the case of stabilized moduli, we do not attempt to arrive at a generic conclusion, due to the model dependence of the DM bounds. Nevertheless, we have explored the parameter space of a strongly stabilized Polonyi modulus, finding that Ly-$\alpha$ observations provide a stronger constraint than entropy production bounds.

For thermal decays, we computed $\Omega_{\chi}$ in general, and briefly discussed axino production from thermal decays. For non-thermal decays, we have computed the DM relic abundance in general, and we have applied it to the particular decay chain inflaton $\rightarrow$ gravitino $\rightarrow$ LSP. For a $10\,{\rm TeV}$ gravitino, its decay occurs while it is relativistic when $T_{\rm reh}<100\,{\rm TeV}$, and non-relativistic for $T_{\rm reh}>100\,{\rm TeV}$. In the former case the limit on the LSP mass grows as the inversely with the reheating temperature, excluding $m_{\rm LSP}\gtrsim 100\,{\rm GeV}$ for $T_{\rm reh}\lesssim 100\,{\rm TeV}$ and bounding the branching ratio of the inflaton $\rightarrow$ gravitino process to ${\rm Br}_{3/2}\lesssim 10^{-3}$. For a non-relativistic decay, the Ly-$\alpha$ bound is independent of the reheating temperature, $m_{\rm DM}\simeq 90\,{\rm GeV}$, while the branching ratio bound is dependent on the reheating temperature, being ${\rm Br}_{3/2}\lesssim 10^{-8}$ for $T_{\rm reh}=10^{10}\,{\rm GeV}$.

Our exploration of UV thermal freeze-in has resulted in precise expressions for the DM abundance for any initial state. We have also recovered the dependence on $T_{\max}$ of the relic density for scattering with $n\geq 6$. Moreover, we have provided provided bounds on the reheating temperature given scale $\Lambda$ that controls the suppression of the scattering cross section. These bounds can be seen in Fig.~\ref{fig:TvsL_th}. We have also briefly discussed extensions of the SM model, such as those containing spin-3/2 and massive spin-2 particles, for which keV NCDM can be produced, for all values of $n$ considered in this work. Finally, for non-thermal DM we found a generic expression for $\Omega_{\chi}$, and applied our formalism to the production of light, non-supersymmetric, spin-3/2 particles. We observed the interplay between thermal and non-thermal effects, which are dependent on the branching ratio of the inflaton to neutrinos. \par\bigskip

\noindent
{\bf $\boldsymbol{N_{\rm eff}}$ constraints.} Dark matter particles that remain warm at late times would have been relativistic at early times, potentially providing a large contribution to the number of non-photonic relativistic species $\Delta N_{\rm eff}$. We have applied our re-scaling formalism to this question, finding that for all scenarios, the Ly-$\alpha$ bound corresponds to an NCDM contribution $\Delta N_{\rm eff}$ of $\mathcal{O}(10^{-4})$ at BBN, and $\mathcal{O}(10^{-8})$ at recombination, well below all current and projected limits.
\par\bigskip

\noindent
{\bf Other WDM bounds.}
The procedure we have presented for mapping Ly-$\alpha$ WDM mass lower bounds onto constraints on the parameters of NCDM models with different shapes of the velocity distribution can also be applied to bounds coming from other data, provided that the dependence on the velocity distribution is encoded in the linear transfer function. As an example, in Section \ref{keysec} we have discussed that this should be the case for WDM bounds inferred from counting the number of Milky Way satellite galaxies and comparing the result with the predictions from N-body simulations. These bounds are sensitive to the total mass of the Milky Way halo and to the physics of reionization. If these are independent of the velocity distribution of dark matter, which is a reasonable assumption, our method also applies to these limits.\par\bigskip

As a final note, we emphasize that this work is not meant to be encyclopedic, and exhaust all possible non-equilibrium DM production mechanisms. Our goal has been to provide a general formalism to rescale the Ly-$\alpha$ WDM mass bound, that is suitable for its application in less conventional scenarios.

\section*{Acknowledgments}
We thank M.\ Garny, T.\ Konstandin, V.\ Poulin and P.\ Quilez for interesting discussions. We thank Ayuki Kamada for pointing out references \cite{Kamada:2019kpe} and \cite{Bae:2017dpt} after the first version of the present work appeared on the arXiv. We also thank L.\ Lopez\ Honorez for pointing out relevant bibliography. The work of GB\ is funded by a Contrato de Atracci\'on de Talento (Modalidad 1) de la Comunidad de Madrid (Spain), with number 2017-T1/TIC-5520, by MINECO (Spain) under contract FPA2016-78022-P, MCIU (Spain) through contract PGC2018-096646-A-I00. The work of MP and MG was supported by the Spanish Agencia Estatal de Investigaci\'{o}n through the grants FPA2015-65929-P (MINECO/FEDER, UE),  PGC2018-095161-B-I00, and Red Consolider MultiDark FPA2017-90566-REDC. All three authors are supported by the IFT Centro de Excelencia Severo Ochoa Grant SEV-2016-0597. MP  thanks the Paris-Saclay Particle Symposium 2019 for support of the P2I and SPU research departments and the P2IO Laboratory of Excellence (program “Investissements d’avenir” ANR-11-IDEX-0003-01 Paris-Saclay and ANR-10-LABX-0038), as well as IPhT. MG would like to thank CNRS and the Laboratoire de Physique des 2 Infinis Ir{\`e}ne Joliot-Curie for their hospitality and financial support while completing this work.

\appendix  

\section{The Boltzmann equation in an expanding universe}
\label{sec:computation_PSD}

\subsection{Generalities}

In a Friedmann-Robertson-Walker Universe the phase space distribution of a given species $i$ is spatially homogeneous and isotropic. The distribution function can be expressed in terms of the norm of the 3-momentum $|\bp|$ and time, i.e. $f_i=f_i(|\bp|,t)$, or equivalently in terms of time and the energy $p_0=\sqrt{|\bp|^2+m_i^2}$ where $m_i$ denotes the mass of the species $i$. 

The evolution equation for the distribution $f_i$ is given by
\beq
\frac{\partial f_i}{\partial t} - H|\bp|\frac{\partial f_i}{\partial |\bp|}\,=\,\mathcal{C}[f_i(|\bp|,t)]
\eeq
where $\mathcal{C}[f_i]$ is the collision term. If we are tracking the phase space density of a particle $\chi$, then the collision term for the process $\chi+a+b+\cdots \longleftrightarrow i+j+\cdots$ is given by
\begin{align}\notag
    \mathcal{C} \,[f_{\chi}] &=  - \frac{1}{2 p_0} \int \frac{g_a \diff ^3 \bp_a}{(2\pi)^3 2p_{a 0}}\frac{g_b \diff ^3 \bp_b}{(2\pi)^3 2p_{b 0}}\cdots \frac{g_i \diff ^3 \bp_i}{(2\pi)^3 2p_{i 0}}\frac{g_j \diff ^3 \bp_j}{(2\pi)^3 2p_{j 0}} \cdots\\ \notag
& \qquad \times (2\pi)^4\,\delta^{(4)}(p_{\chi} + p_a + p_b + \cdots - p_i - p_j - \cdots)\\ \notag
& \qquad \times \Big[ |\mathcal{M}|^2_{\chi+a+b+\cdots \longrightarrow i+j+\cdots}  \,f_a f_b\cdots f_{\chi}(1\pm f_i)(1\pm f_j) \cdots \\ \label{genboltzeq}
& \qquad \quad  - |\mathcal{M}|^2_{i+j+\cdots \longrightarrow \chi+a+b+\cdots} \, f_i f_j\cdots (1\pm f_a)(1\pm f_b) \cdots (1\pm f_{\chi})\Big]\,,
\end{align}
where $f_a$, $f_b$, $f_i$, $f_j$, $\cdots$ and $g_a$, $g_b$, $g_i$, $g_j$, $\cdots$ are the phase space densities of species $a,b,i,j,\cdots$ of 3-momentum $\bp_{a,b,i,j,\cdots}$ and their internal degrees of freedom, respectively; the blocking and stimulated emission factors are $(+)$ for bosons and $(-)$ for fermions. The number density of particle $\chi$ is given by
\beq\label{eq:nfrel}
\gls{nchi}(t) = \frac{\gls{gchi}}{(2\pi)^3}\int \diff ^3\bp\,f_{\chi}(p_0,t)\,.
\eeq 

In the case of an always out-of-equilibrium relic $\chi$, due typically to the combination of a feeble coupling to other fields and a small density, the annihilation process will be negligible with respect to the production process, and the first term inside the brackets of Eq.~(\ref{genboltzeq}) can be disregarded. Moreover, in this limit Bose condensation or Fermi degeneracy will be absent, and one may approximate $1\pm f_{\chi}\simeq 1$. As a consequence, the collision term $\mathcal{C}[f_{\chi}]$ is independent of $f_{\chi}$. Therefore, if the phase space distributions of all other particles involved in the interaction are known, the transport equation can be integrated to yield the following general solution,
\begin{equation}\label{eq:Cgensol}
f_\chi(p_0,t)=\int_{t_i}^{t} \mathcal{C} [ f_\chi ] \left(\dfrac{a(t)}{a(t^\prime)}|\bp|,t^\prime\right) \diff  t^\prime\,,
\end{equation}
where $t_i$ is the initial time at which the number density of $\chi$ vanishes. In this expression, the functional $\mathcal{C}[f_{\chi}]$ is a function of momentum and time.

Consider now the evolution of the distribution function for a particle $\chi$ that becomes effectively non-interacting at the time $t_{\rm dec}$, and which inherits a distribution of the form $\bar{f}(|\bp|)$, whose form is determined by interactions at $t<t_{\rm dec}$. The corresponding Boltzmann equation at $t>t_{\rm dec}$ will have the form
\beq
\frac{\partial f_\chi}{\partial t} - H |\bp|\frac{\partial f_\chi}{\partial |\bp|} \;=\; 0 \,.
\eeq
The general solution for this equation, with initial distribution $\bar{f}(|\bp|)$, is
\beq\label{eq:freestream}
f_\chi(|\bp|,t) \;=\; \bar{f}\left( |\bp| \frac{a(t)}{a_{\rm dec}}, t_{\rm dec} \right)\,.
\eeq
It suffices to find the form of $\bar{f}$ in terms of the comoving momentum $a \bp$ to know the form of the distribution at any later time.

\subsection{Freeze-in via scatterings }\label{app:figen}

In this Appendix we derive a simplified expression for the freeze-in collision term (\ref{eq:CgenFI}) of a DM species $\chi$, namely 
\beq\label{eq:CgenFIA}
\mathcal{C}[f_{\chi}] \;=\; \frac{16\pi g_A g_B g_{\psi}}{\Lambda^{n+2}2 p_0} \int \frac{\diff^3\tilde{\bp}}{2(2\pi)^3 \tilde{p}_0} \frac{\diff^3 \bk}{2(2\pi)^3 k_0} \frac{\diff^3 \tilde{\bk} }{2(2\pi)^3 \tilde{k}_0}  (2\pi)^4 \delta^{(4)}(p+\tilde{p}-k-\tilde{k}) s^{\frac{n}{2}+1} f_A (k_0) f_B(\tilde{k}_0)\,.
\eeq
which allows a straightforward evaluation once the phase space distribution of the parent scatterers is known. This procedure follows the steps presented in~\cite{Bolz:2000fu,GarciaGarcia:2016nhj}. We begin by introducing the auxiliary momentum $\boldsymbol{P}=\tilde{\bk}-\tilde{\bp}$ to write the two innermost integrals in the following way
\begin{align} \notag
I\;&\equiv\; \int \frac{\diff^3 \bk}{2 k_0} \frac{\diff^3 \tilde{\bk} }{2 \tilde{k}_0 }  s^{\frac{n}{2}+1} \delta^{(4)}(p+\tilde{p}-k- \tilde{k})  f_A (k) f_B( \tilde{k} )\\ \notag
& = \; \int \diff^4k\, \diff^4 \tilde{k}\, \diff^3\boldsymbol{P} \, s^{\frac{n}{2}+1} f_A(k_0) f_B(\tilde{k}_0) \delta^{(3)}(\boldsymbol{P}+ \tilde{\bp} -\tilde{\bk}) \delta^{(4)}(p+\tilde{p}-k- \tilde{k} )\delta(k^2)\delta(\tilde{k}^2)\theta(k_0) \theta(\tilde{k}_0)\\ \notag
&=\; \int \diff \tilde{k}_0\, \diff^3 \boldsymbol{P}\, s^{\frac{n}{2}+1} f_A( p_0 + \tilde{p}_0 - \tilde{k}_0 ) f_B(\tilde{k}_0 ) \delta\left(( p_0 + \tilde{p}_0 - \tilde{k}_0 )^2 - |\bp-\boldsymbol{P}|^2 \right)\\
& \hspace{110pt}\times \delta(\tilde{k}_0^2 - |\boldsymbol{P}+\tilde{\bp}|^2) \theta(p_0+\tilde{p}_0 - \tilde{k}_0 ) \theta(\tilde{k}_0)\,.
\end{align}
Next, we specialize to the following coordinate system,
\begin{align}\notag
\boldsymbol{P} \;&=\; P(0,0,1)\,,\\
\bp \;&=\; p_0 (0,\sin\vartheta,\cos\vartheta)\,,\\ \notag
\tilde{\bp} \;&=\; \tilde{p}_0 (\cos\phi\sin\theta, \sin\phi\sin\theta,\cos\theta)\,,
\end{align}
so that
\beq
\begin{aligned} \label{eq:kinvar2}
s \;&=\; (p + \tilde{p})^2 \;=\; 2p_0 \tilde{p}_0 (1- \sin\phi\sin\theta\sin\vartheta- \cos\theta\cos\vartheta)\,,\\
|\bp-\boldsymbol{P}|^2 \;&=\; p_0^2 + P^2 - 2 p_0 P \cos\vartheta\,,\\
|\boldsymbol{P}+\tilde{\bp}|^2 \;&=\; \tilde{p}_0^2 + P^2 + 2\tilde{p}_0 P\cos\theta\,.
\end{aligned}
\eeq
Substitution yields
\begin{align} \notag
I &=\; \int \diff \tilde{k}_0\, \diff^3 \boldsymbol{P}\, s^{\frac{n}{2}+1} f_A( p_0 + \tilde{p}_0 - \tilde{k}_0 ) f_B(\tilde{k}_0) \frac{1}{2p_0 P}\delta\left(\cos\vartheta +  \frac{(p_0+\tilde{p}_0-\tilde{k}_0)^2 - p_0^2 - P^2}{2 p_0 P} \right)\\ \label{eq:Ifin}
& \hspace{110pt}\times \frac{1}{2\tilde{p}_0 P}\delta\left( \cos\theta -  \frac{\tilde{k}_0^2 - \tilde{p}_0^2 - P^2}{2\tilde{p}_0 P} \right) \theta( p_0 + \tilde{p}_0 - \tilde{k}_0 ) \theta(\tilde{k}_0)\,.
\end{align}
From the outermost integral in (\ref{eq:CgenFIA}) we can evaluate the polar angle integral as follows
\begin{align}\notag
\int I\, \diff \cos \vartheta \, \diff \phi \;&=\; \frac{2 \pi}{4 p_0 \tilde{p}_0 }\int \diff \tilde{k}_0 \,\diff \phi \,\diff P\,  f_A( p_0 + \tilde{p}_0 - \tilde{k}_0 ) f_B(\tilde{k}_0) \theta( p_0 + \tilde{p}_0 - \tilde{k}_0 ) \theta(\tilde{k}_0) \theta(\tilde{p}_0+\tilde{k}_0-P)\\ \notag \displaybreak[0]
&\hspace{40pt} \times \theta(2p_0 + \tilde{p}_0 -\tilde{k}_0 - P) \theta(P-|\tilde{k}_0 - \tilde{p}_0|)\, s^{\frac{n}{2}+1}\\ \notag
&=\; \frac{2 \pi}{4 p_0 \tilde{p}_0 }\int \diff \tilde{k}_0 \, f_A( p_0 + \tilde{p}_0 - \tilde{k}_0 ) f_B(\tilde{k}_0) \theta( p_0 + \tilde{p}_0 - \tilde{k}_0 ) \theta(\tilde{k}_0) \\
& \hspace{40pt} \times \left[ \int_{|\tilde{k}_0-\tilde{p}_0|}^{2p_0 + \tilde{p}_0-\tilde{k}_0}\diff P \left(\int s^{\frac{n}{2}+1} \diff\phi\right) - \int_{\tilde{k}_0 + \tilde{p}_0}^{2p_0 + \tilde{p}_0-\tilde{k}_0}\diff P \left(\int s^{\frac{n}{2}+1} \diff\phi\right) \right]\,.
\end{align}
Here $s$ is assumed to be given by (\ref{eq:kinvar2}) after evaluating the Dirac delta distributions
in (\ref{eq:Ifin}). If we denote for simplicity $S^{(n)}(P,\tilde{k}_0,p_0,\tilde{p}_0)\equiv \int s^{\frac{n}{2}+1}\diff \phi$, we can finally write
\begin{align}\notag
\mathcal{C}[f_{\chi}]\;=\; \frac{ g_A g_B g_{\psi}}{2(2\pi)^3\Lambda^{n+2} p_0^2} \Bigg\{ &\int_0^{\infty}\diff \tilde{p}_0 \int_{\tilde{p}_0}^{p_0+\tilde{p}_0}\diff \tilde{k}_0\, f_A(  p_0 + \tilde{p}_0 - \tilde{k}_0 ) f_B(\tilde{k}_0) \int_{\tilde{k}_0-\tilde{p}_0}^{2p_0 + \tilde{p}_0 - \tilde{k}_0} \diff P\, S^{(n)}\\ \notag
&+ \int_0^{\infty}\diff \tilde{p}_0 \int_{0}^{\tilde{p}_0}\diff \tilde{k}_0\, f_A(  p_0 + \tilde{p}_0 - \tilde{k}_0 ) f_B(\tilde{k}_0) \int_{\tilde{p}_0 - \tilde{k}_0}^{2p_0+\tilde{p}_0-\tilde{k}_0} \diff P\, S^{(n)}\\ \label{eq:cfgen}
&- \int_0^{\infty}\diff \tilde{p}_0 \int_{0}^{p_0}\diff \tilde{k}_0 \, f_A(  p_0 + \tilde{p}_0 - \tilde{k}_0 ) f_B(\tilde{k}_0) \int_{\tilde{p}_0+\tilde{k}_0}^{2p_0+\tilde{p}_0-\tilde{k}_0} \diff P\, S^{(n)} \Bigg\}\,.
\end{align}
Given the distribution functions for the scatterers, the previous expression can be integrated, either analytically or numerically.

\subsection{Non-thermal freeze-in}\label{app:nthfi}

In this Appendix we find a closed form for the collision term for non-thermal freeze-in discussed in Section~\ref{sec:thenonthfi}. This requires the knowledge of the initial state distribution functions. For scatterers produced directly from inflaton decay these distributions have the form (\ref{eq:fpreth}) before thermalization. Substitution into (\ref{eq:cfgen}) leads to the following expression for the $\chi$ collision term,
\begin{align}\notag \displaybreak[0]
\mathcal{C}[f_{\chi}] \;=\; &\frac{9\pi g_{\psi} n_A n_B}{2\Lambda^{n+2}m_{\phi}^3 p_0^2} \Bigg\{ \theta \Big( \dfrac{m_{\phi}}{2} - p_0 \Big) \Bigg[ \int_0^{p_0}\diff \tilde{k}_0  \int_{0}^{\tilde{k}_0}\diff \tilde{p}_0 \, (p_0+ \tilde{p}_0 - \tilde{k}_0)^{-3/2} \tilde{k}_0^{-3/2} \int_{\tilde{k}_0 - \tilde{p}_0}^{2p_0+ \tilde{p}_0 - \tilde{k}_0} \diff P\, S^{(n)}\\ \notag \displaybreak[0]
&+ \int_{p_0}^{\frac{m_{\phi}}{2}}\diff \tilde{k}_0  \int_{\tilde{k}_0 - p_0}^{\tilde{k}_0}\diff \tilde{p}_0 \, (p_0 + \tilde{p}_0 - \tilde{k}_0)^{-3/2} \tilde{k}_0^{-3/2} \int_{\tilde{k}_0 - \tilde{p}_0 }^{2p_0 + \tilde{p}_0 - \tilde{k}_0} \diff P\, S^{(n)}\\ \notag \displaybreak[0]
&+ \int_{0}^{\frac{m_{\phi}}{2}}\diff \tilde{k}_0  \int_{\tilde{k}_0}^{ \frac{m_{\phi}}{2} - p_0 +\tilde{k}_0 }\diff \tilde{p}_0 \, (p_0 + \tilde{p}_0 - \tilde{k}_0)^{-3/2} \tilde{k}_0^{-3/2} \int_{\tilde{p}_0 - \tilde{k}_0}^{2p_0 +\tilde{p}_0 -\tilde{k}_0} \diff P\, S^{(n)} \\ \notag \displaybreak[0]
&- \int_{0}^{p_0} \diff \tilde{k}_0  \int_{0}^{ \frac{m_{\phi}}{2} - p_0 + \tilde{k}_0 }\diff \tilde{p}_0 \, (p_0 + \tilde{p}_0 - \tilde{k}_0)^{-3/2} \tilde{k}_0^{-3/2} \int_{\tilde{p}_0 + \tilde{k}_0}^{2p_0 + \tilde{p}_0 - \tilde{k}_0} \diff P\, S^{(n)} \Bigg]\\ \notag \displaybreak[0] 
&+ \theta \Big( m_{\phi} - p_0\Big) \theta \Big(p_0 - \dfrac{m_{\phi}}{2} \Big) \\ \notag
& \hspace{20pt}\times \Bigg[ \int_{p_0 - \frac{m_{\phi}}{2} }^{ \frac{m_{\phi}}{2} }\diff \tilde{k}_0  \int_{0}^{ \frac{m_{\phi}}{2} - p_0 + \tilde{k}_0}\diff \tilde{p}_0 \, (p_0 + \tilde{p}_0 - \tilde{k}_0)^{-3/2} \tilde{k}_0^{-3/2} \int_{\tilde{k}_0 -\tilde{p}_0}^{2p_0 + \tilde{p}_0 - \tilde{k}_0} \diff P\, S^{(n)}  \\ \displaybreak[0] 
& \hspace{40pt} - \int_{p_0- \frac{m_{\phi}}{2} }^{ \frac{m_{\phi}}{2} }\diff \tilde{k}_0  \int_{0}^{ \frac{m_{\phi}}{2} - p_0 + \tilde{k}_0}\diff \tilde{p}_0 \, (p_0 + \tilde{p}_0 - \tilde{k}_0)^{-3/2} \tilde{k}_0^{-3/2} \int_{\tilde{p}_0 +\tilde{k}_0}^{2p_0 + \tilde{p}_0 - \tilde{k}_0} \diff P\, S^{(n)} \Bigg] \Bigg\}\,. \displaybreak[0]
\end{align}
Specializing to $n=4$, and replacing $p_0 \rightarrow p$ for notational simplicity for the ultrarelativistic $\chi$, integration yields the following result
\begin{align}\notag
\mathcal{C}[f_{\chi}]^{(4)} \;=\; & \frac{256\pi^2 g_{\psi} n_A n_B p^4}{5005 \Lambda^{6} m_{\phi}^3} \Bigg\{ \theta \Big(\dfrac{m_{\phi}}{2} - p \Big)\,  \Bigg[  10296 \left(\frac{m_{\phi}}{2 p}\right)^{5/2} +36036 \left(\frac{m_{\phi}}{2 p}\right)^2 \\ \notag
& -80080 \left(\frac{m_{\phi}}{2 p}\right)^{3/2} - 72072 \left(\frac{m_{\phi}}{2 p}\right) +262080 \left(\frac{m_{\phi}}{2 p}\right)^{1/2} -45045 (2+\pi )\\ \notag
&  +88704 \left(\frac{2 p}{m}\right)^{1/2} -12870 \left(\frac{2 p}{m}\right) \Bigg]\\ \notag
& - \theta \Big(m_{\phi} - p\Big)  \theta\Big(p- \dfrac{m_{\phi}}{2} \Big) \, \left(\frac{m_{\phi}}{p}\right)^6 \left(\frac{2p}{m_{\phi}}-1\right)^{-1/2}\Bigg[ 8 - 4\left(\frac{2p}{m_{\phi}}\right) - \left(\frac{2p}{m_{\phi}}\right)^2\\ \notag
&-\frac{1}{2} \left(\frac{2 p}{m_{\phi}}\right)^3 -\frac{4507}{8} \left(\frac{2 p}{m_{\phi}}\right)^4 +\frac{40537}{16} \left(\frac{2 p}{m_{\phi}}\right)^5 - \frac{9387}{16} \left(\frac{2 p}{m_{\phi}}\right)^6 -1386 \left(\frac{2 p}{m_{\phi}}\right)^7\\ \notag
& + \left(\frac{2p}{m}-1\right)^{1/2}\Bigg( 47 - \frac{9009}{16} \left(\frac{2 p}{m_{\phi}}\right)^4 +\frac{9009}{8} \left(\frac{2 p}{m_{\phi}}\right)^5 +\frac{6435}{32} \left(\frac{2 p}{m_{\phi}}\right)^7\\
& +\frac{45045}{32} \left(\frac{2 p}{m_{\phi}}\right)^6 \left(1 - \tan^{-1}\left(\frac{2p}{m}-1\right)^{-1/2} + \tan^{-1} \left(\frac{2p}{m}-1\right)^{1/2} \right)\, \Bigg)\Bigg]\Bigg\}\,.
\end{align}

\printnoidxglossary[type=symbols,
numberedsection=autolabel,
style=long,title={List of main symbols}]

\bibliography{bibfile}

\end{document}